\documentclass[a4paper, 12pt]{article}

\pdfoutput = 1

\usepackage{styleBuding}
\usepackage{bm}
\usepackage{color,listings}
\usepackage[usenames,dvipsnames,svgnames,table]{xcolor}
\usepackage{amsthm, amsmath}
\usepackage{amssymb}
\usepackage{mathrsfs}
\usepackage{graphicx}
\usepackage{slashed}
\usepackage{soul}
\usepackage{multirow}
\usepackage{subfigure}
\usepackage{siunitx} 
\usepackage{url} 
\usepackage[utf8]{inputenc}
\usepackage[figuresright]{rotating}
\definecolor{back}{HTML}{F8F8F8}
\usepackage{epstopdf}

\newcommand{\rom}[1]{\uppercase\expandafter{\romannumeral #1\relax}}

\allowdisplaybreaks
\DeclareUnicodeCharacter{2212}{-}

\let\jnfont=\rm
\def\NPB#1,{{\jnfont Nucl.\ Phys.\ B }{\bf #1},}
\def\PLB#1,{{\jnfont Phys.\ Lett.\ B }{\bf #1},}
\def\EPJC#1,{{\jnfont Eur.\ Phys.\ Jour.\ C }{\bf #1},}
\def\PRD#1,{{\jnfont Phys.\ Rev.\ D }{\bf #1},}
\def\PRL#1,{{\jnfont Phys.\ Rev.\ Lett.\ }{\bf #1},}
\def\MPLA#1,{{\jnfont Mod.\ Phys.\ Lett.\ A }{\bf #1},}
\def\JPG#1,{{\jnfont J.\ Phys.\ G}{\bf #1},}
\def\CTP#1,{{\jnfont Commun.\ Theor.\ Phys.\ }{\bf #1},}
\def\ZPC#1,{{\jnfont Z.\ Phys.\ C }{\bf #1},}
\def\JHEP#1,{{\jnfont JHEP \ }{\bf #1},}

\title{Impact of recent $(g-2)_{\mu}$ measurement on the light CP-even Higgs scenario in general Next-to-Minimal Supersymmetric Standard Model}
\author{Junjie Cao$^{a}$, Jingwei Lian$^a$, Yusi Pan$^a$, Yuanfang Yue$^a$, Di Zhang$^a$}
\affiliation{ $^a$ Department of Physics, Henan Normal University, Xinxiang 453007, China}
\emailAdd{junjiec@alumni.itp.ac.cn}
\emailAdd{ljwfly@hotmail.com}
\emailAdd{panyusi0406@foxmail.com}
\emailAdd{yueyuanfang@htu.edu.cn}
\emailAdd{dz481655@gmail.com}

\abstract{The General Next-to-Minimal Supersymmetric Standard Model (GNMSSM) is an attractive theory that is free from the tadpole problem and the domain-wall problem of $Z_3$-NMSSM, and can form an economic secluded dark matter (DM) sector to naturally predict the DM experimental results. It also provides mechanisms to easily and significantly weaken the constraints from the LHC search for supersymmetric particles. These characteristics enable the theory to explain the recently measured muon anomalous magnetic moment, $(g-2)_\mu$, in a broad parameter space that is consistent with all experimental results and at same time keeps the electroweak symmetry breaking natural. This work focuses on a popular scenario of the GNMSSM in which the next-to-lightest CP-even Higgs boson corresponds to the scalar discovered at the Large Hadron Collider (LHC). Both analytic formulae and a sophisticated numerical study show that in order to predict the scenario without significant tunings of relevant parameters, the Higgsino mass $\mu_{tot} \lesssim 500~{\rm GeV}$ and $\tan \beta \lesssim 30$ are preferred. This character, if combined with the requirement to account for the  $(g-2)_\mu$ anomaly, will entail some light sparticles and make the LHC constraints very tight. As a result, this scenario can explain the muon anomalous magnetic moment in very narrow corners of its parameter space.}

\begin{document}
    \maketitle
    \flushbottom

\section{Introduction}
The latest measurement of the muon anomalous magnetic moment $a_\mu \equiv (g-2)_\mu/2$ announced by the Fermilab National Accelerator Laboratory (FNAL)~\cite{Abi:2021gix} is in full agreement with the Brookhaven National Laboratory (BNL) E821 result~\cite{Bennett:2006fi}. The combined experimental average is given by Equation~(\ref{amu-exp}):
\begin{equation}
    a_\mu^{\rm Exp} = 116592061(41)  \times 10^{-11},  \label{amu-exp}
\end{equation}
which shows a 4.2 $\sigma$ discrepancy from the Standard Model (SM) theoretical prediction $a_\mu^{\rm SM} = 116 591 810(43)  \times 10^{-11}$~\cite{Aoyama:2020ynm,Aoyama:2012wk,Aoyama:2019ryr,Czarnecki:2002nt,Gnendiger:2013pva,Davier:2017zfy,Keshavarzi:2018mgv,Colangelo:2018mtw,Hoferichter:2019gzf,Davier:2019can,
Keshavarzi:2019abf,Kurz:2014wya,Melnikov:2003xd,Masjuan:2017tvw,Colangelo:2017fiz,Hoferichter:2018kwz,Gerardin:2019vio,Bijnens:2019ghy,Colangelo:2019uex,Blum:2019ugy,Colangelo:2014qya}, given by Equation~(\ref{delta-amu}):
\begin{equation}
     \Delta a_\mu = a_\mu^{\rm Exp} - a_\mu^{\rm SM} = (251 \pm 59) \times 10^{-11}.  \label{delta-amu}
\end{equation}
In addition, the Run-1 results in Fermilab also imply that a more thorough analysis in future experiments will most probably substantiate the excess of $a_\mu$ in $5 \sigma$ discovery level. In recent years, this situation has inspired continuous attention to $a_\mu$. In particular, it was widely conjectured that the anomaly may arise from new physics beyond the SM (see, e.g., Ref.~\cite{Athron:2021iuf} and the references therein). Among a variety of theories that can account for the anomalous magnetic moment, supersymmetry (SUSY) is especially promising due to its elegant structure and natural solutions to many puzzles in the SM, such as the hierarchy problem, the unification of different forces, and the dark matter (DM) mystery~\cite{Fayet:1976cr,Haber:1984rc,Martin:1997ns,Jungman:1995df}. Studies of low-energy supersymmetric models have indicated that the source of the significant deviation can be totally or partially contributed to smuon-neutralino or sneutrino-chargino loop effects~\cite{Martin:2001st,Domingo:2008bb,Moroi:1995yh,Hollik:1997vb,Athron:2015rva,Endo:2021zal,
Stockinger:2006zn,Czarnecki:2001pv,Cao:2011sn,Kang:2016iok,Zhu:2016ncq,Yanagida:2017dao, Hagiwara:2017lse,Cox:2018qyi,Tran:2018kxv,Padley:2015uma,
Choudhury:2017fuu,Okada:2016wlm,Du:2017str, Ning:2017dng, Wang:2018vxp,Yang:2018guw,Liu:2020nsm,Cao:2019evo,Cao:2021lmj,Ke:2021kgy,Lamborn:2021snt,Li:2021xmw,Nakai:2021mha,Li:2021koa,Kim:2021suj,Li:2021pnt,Altmannshofer:2021hfu,
Baer:2021aax,Chakraborti:2021bmv,Aboubrahim:2021xfi,Athron:2021iuf,Iwamoto:2021aaf,Chakraborti:2021dli,Cao:2021tuh,Yin:2021mls,Zhang:2021gun,Ibe:2021cvf,
Han:2021ify,Wang:2021bcx,Zheng:2021gug,Chakraborti:2021mbr,Aboubrahim:2021myl,Ali:2021kxa,Wang:2021lwi,Chakraborti:2020vjp,Baum:2021qzx}.

Although SUSY has multiple theoretical advantages, it has been strongly restricted by DM direct detection (DD) experiments, such as XENON-1T~\cite{XENON:2018voc,XENON:2019rxp} and PandaX-4T~\cite{PandaX-II:2021nsg,PandaX-4T:2021bab} experiments, as well as LHC sparticle searches~\cite{Aad:2019vvi,Aad:2019vnb,Aad:2019vvf,Sirunyan:2018nwe,Sirunyan:2018ubx,ATLAS:2019lng,ATLAS:2021moa,CMS:2020bfa}. As a consequence, some of its economical realizations become unnatural for electro-weak symmetry breaking in interpreting the anomalous magnetic moment. In the Minimal Supersymmetric Standard Model (MSSM) with $R$-parity conservation~\cite{Farrar:1978xj,Haber:1984rc,Gunion:1984yn,Djouadi:2005gj}, the lightest neutralino is usually the lightest supersymmetric particle (LSP), and thus a viable DM candidate. In order to fully account for the DM  abundance measured by the Planck experiment~\cite{Planck:2018vyg}, it must be Bino-dominated when it is lighter than $1~\rm TeV$~\cite{Bagnaschi:2017tru}. In this case, the XENON-1T experiment and the LHC experiments prefer that the magnitude of the Higgsino mass parameter, $\mu$, be larger than about $500~\rm GeV$ in explaining the discrepancy in the $2~\rm \sigma$ level~\cite{Chakraborti:2020vjp}. This implies a fine-tuning at the order of $1~\%$ in predicting $m_Z$~\cite{Baer:2012uy}. This conclusion may be understood from the features of the DM annihilation mechanisms in the MSSM:
\begin{itemize}
\item In the case that the DM co-annihilates with Wino-dominated particles to obtain the measured abundance, the Higgsino mass prefers to be much larger than the DM mass, which was explained in Appendix A of this work. In addition, the SUSY explanation of the deviation together with the results of the LHC search for SUSY can further restrict $\mu$ values in a significant way\footnote{In carrying out this study, we also investigated the characteristics of MSSM and $Z_3$-NMSSM in a similar way to this work. We found that the co-annihilations with Wino- and Slepton-dominated particles are the main annihilation mechanisms of the bino-dominated DM , i.e., their corresponding Bayesian evidences are the largest in comparison with the other annihilation mechanisms. We also found that the DM preferred to be heavier than about $300~{\rm GeV}$ and $|\mu| \gtrsim 500~{\rm GeV}$  when we implemented detailed Monte Carlo simulations for the constraints from the latest LHC searches for electroweakinos. These observations significantly improve the conclusions of ~\cite{Chakraborti:2020vjp} and~\cite{Baum:2021qzx}, since more LHC analyses were considered carefully. }.
\item In the case that the DM co-annihilates with Higgsino-dominated particles to obtain the measured abundance, the DM DD experiments have required $|\mu|$ to be as large as several TeV, because DM-nucleon scattering rates are enhanced by a factor of $1/(1-m_{\tilde{\chi}_1^0}^2/\mu^2)^2$ as $m_{\tilde{\chi}_1^0}^2/\mu^2 \to 1$ (see discussions in Appendix A).
\item In the case that the DM co-annihilates with Sleptons to obtain the measured abundance, the LHC's searches for electroweakinos require Higgsinos to be massive because Wino- and Higgsino-dominated electroweakinos can decay into Sleptons and thus enhance the production rate of lepton signals at the LHC~\cite{Cao:2021tuh}.
\item In the case that the DM co-annihilates with Squarks or Gluinos to obtain the measured abundance, the LHC's searches for colored sparticles require the DM mass to be heavier than $1~{\rm TeV}$~\cite{CMS:2019ybf}.
\item In the case that the DM obtains the measured abundance by the SM-like Higgs funnel or Z funnel, the LHC's searches for eletroweakinos prefer massive Higgsinos because the DM is relatively light and the LHC's constraints on sparticle mass spectrum are rather strong~\cite{Cao:2018rix}.
\item The case in which the DM obtains the measured abundance by the resonance of heavy doublet Higgs bosons is rare. One reason for this is that this case requires significant tuning of SUSY parameters to realize the correlation $|m_{\tilde{\chi}_1^0}| \simeq m_A/2$, where $m_A$ denotes the mass of CP-odd Higgs bosons in MSSM. Another reason is that the LHC's searches for exotic Higgs bosons prefer the bosons to be very massive~\cite{ATLAS:2021upq}. Consequently, the DM is also massive.
\end{itemize}

Next, we consider the Next-to-Minimal Supersymmetric Standard Model with a $Z_3$ symmetry ($Z_3$-NMSSM)~\cite{Ellwanger:2009dp,Maniatis:2009re}. This model
extends MSSM by a gauge-singlet Higgs superfield $\hat{S}$, and has the advantage that either a Bino-dominated (in most physical cases) or a Singlino-dominated
neutralino can act as a viable DM candidate~\cite{Cao:2016nix,Ellwanger:2016sur,Xiang:2016ndq,Baum:2017enm,Cao:2018rix,Ellwanger:2018zxt,Domingo:2018ykx, Baum:2019uzg,vanBeekveld:2019tqp, Abdallah:2019znp,Cao:2019qng,Guchait:2020wqn,Abdallah:2020yag}. The Bino-dominated DM candidate differs from the MSSM prediction mainly in that it could co-annihilate with a Singlino-dominated neutralino to obtain the measured abundance~\cite{Baum:2017enm}. This situation, however, occurrs in very narrow parameter space characterized by $|2 \kappa \mu/\lambda| \simeq |M_1|$, moderately large $\lambda$ and $\kappa$, and $|\mu| \gtrsim 300~{\rm GeV}$~\cite{Baum:2017enm,Cao:2019qng}. In addition, $a_\mu^{\rm SUSY}$ is insensitive to Yukawa coupling $\lambda$ because the Singlino field has no mixing with Wino and Bino fields, and it does not couple directly to the muon lepton. As a result, the formulae to calculate $a_\mu^{\rm SUSY}$ in NMSSM are same as those  at the lowest order of the mass-insertion approximation in MSSM~\cite{Cao:2021tuh}. Considering these features, we expected that the Bino-dominated DM case in the $Z_3$-NMSSM and MSSM would not show significant differences in explaining the discrepancy.
The properties of the Singlino-dominated DM are determined by $\lambda$, the Higgsino mass $\mu$ (denoted by $\mu_{\rm tot}$ in this work), and $\rm tan\beta$ for a given DM mass~\cite{Zhou:2021pit}.  A relatively large $\lambda$ can increase the DM-nucleon scattering cross-sections, and so far, $\lambda \gtrsim 0.3$ is disfavored by the XENON-1T experiments~\cite{Cao:2019qng,Zhou:2021pit}. This conclusion implies that the traditional DM annihilation channels, $\tilde{\chi}_1^0 \tilde{\chi}_1^0 \to t \bar{t}, h_s A_s, h A_s$, where $t$, $h$, $h_s$, and $A_s$ denote the top quark, SM-like Higgs boson, and singlet-dominated CP-even and CP-odd Higgs bosons, respectively, can not be fully responsible for the measured abundance~\cite{Cao:2019qng}. As a result, the DM is more likely to obtain the abundance by means of the co-annihilation with the Higgsino-dominated particles, which corresponds to a correlated parameter space of $2|\kappa| \simeq \lambda$ with $\lambda \lesssim 0.1$. The Bayesian evidence in this case is heavily suppressed owing to the very narrow parameter space, which entails a certain degree of fine-tuning to meet the DM experiments~\cite{Zhou:2021pit}. In addition, the interpretation of the magnetic moment causes the $\rm Z_3$-NMSSM to be further restricted by the updated searches for SUSY at the LHC with $139 fb^{-1}$ data. In particular, the region of $\tan \beta \lesssim 30$ in Figure 7 of~\cite{Cao:2018rix} has been excluded because both the DM and Higgsinos are relatively light. Such a situation, as we will show below, was frequently encountered in this work.

The dilemma of MSSM and $Z_3$-NMSSM inspired us to study the general Next-to-Minimal Supersymmetric Standard Model (GNMSSM)~\cite{Cao:2021ljw}. Unlike $Z_3$-NMSSM, GNMSSM usually predicts the Singlino-dominated neutralino as a viable DM candidate due to its following specific theoretical feature: the properties of the Singlino-dominated DM are described by $\lambda$, $\mu_{tot}$, $m_{{\tilde\chi}^0_1}$, $\rm tan \beta$, and $\kappa$, among which the first four parameters determine the DM couplings to nucleon, and $\kappa$ mainly dominates the DM couplings to singlet-dominated Higgs bosons~\cite{Cao:2021ljw}. Consequently, singlet-dominated particles $\tilde{\chi}_1^0$, $h_s$, and $A_s$ can constitute a secluded DM sector,  where  the measured DM abundance can be achieved by the $h_s/A_s$-mediated resonant annihilation into SM particles or through the annihilation process of $\tilde{\chi}_1^0 \tilde{\chi}_1^0 \to h_s A_s$ by adjusting the value of $\kappa$.  Given that this sector interacts with SM matters only through weak singlet-doublet Higgs field mixing, the DM-nucleon scattering rate can be naturally suppressed by $\lambda v/\mu_{tot}$ when $\lambda$ is small~\cite{Cao:2021ljw}. Since the parameters need no significant tuning to be consistent with the constraints from the DM experiments, the corresponding Bayesian evidence is significantly larger than that for the Bino-dominated DM case~\cite{Cao:2021ljw}. Other characteristics of the theory include that, due to the very weak couplings of the Singlino-dominated DM to other sparticles, heavy sparticles initially prefer to decay into next-to-LSP (NLSP) or next-next-to-LSP (NNLSP). As a result, their decay chains are lengthened and their signals become complicated. In addition, the DM as LSP may be moderately heavy, since the annihilation $\tilde{\chi}_1^0 \tilde{\chi}_1^0 \to h_s A_s$ requires $m_{\rm DM} > (m_{h_s}+m_{A_s})/2$. These features weaken significantly the limitations from the LHC's searches for SUSY. Specifically, in our recent work, we studied $a_\mu^{\rm SUSY}$ in a simplified version of GNMSSM, which we called $\mu$-extended NMSSM ($\mu$NMSSM)~\cite{Cao:2021tuh}. We found that, by presuming the DM and LHC experiments are satisfied, $\mu$NMSSM can explain the discrepancy in a broad parameter space where Higgsinos are lighter than about $500~\rm GeV$.

In our previous work~\cite{Cao:2021tuh}, we considered only the $h_1$ scenario, in which the lightest CP-even Higgs boson corresponds to the SM-like Higgs boson discovered at the LHC. A typical feature of NMSSM is that the next lightest CP-even Higgs boson may also act as the SM-like Higgs boson, which has been dubbed the $h_2$ scenario in the literature. Thus, a full understanding of GNMSSM necessitates
the study of the discrepancy in the $h_2$ scenario. In particular, given that specific configurations of Higgs parameters are needed to predict $m_{h_1} < (m_{h_2} \simeq 125~{\rm GeV})$ and $h_2$ to be SM-like, it is conceivable that the $h_2$ scenario suffers tighter experimental constraints than the $h_1$ scenario\footnote{This conclusion may also be understood intuitively as follows: the lightness of $h_1$ and the premise that $h_1$ and $h_2$ are singlet-dominated and SM-like, respectively, lead to the tendency that some parameters in the Higgs sector are relatively small. Consequently, light sparticles are usually predicted. This phenomenon is similar to the well-known fact that the natural result for electroweak-symmetry breaking prefers $|\mu| \lesssim 500~{\rm GeV}$~\cite{Baer:2012uy}.}. This leaves in doubt the idea that the $h_2$ scenario can explain the discrepancy. As a result, a careful examination of the experimental constraints on the $h_2$ scenario is needed, which is the focus of this work.

This work is organized as follows. In Section \ref{theory-section}, we briefly introduce the basics of GNMSSM and the SUSY contribution to the moment. In Section \ref{numerical study1}, we perform a sophisticated scan over the broad parameter space of $\mu$NMSSM, and show the features of the theory in explaining the discrepancy. By using specific Monte Carlo simulations, we also comprehensively study the constraints from the LHC's searches for SUSY. In Section \ref{numerical study2}, we concentrate on the GNMSSM, which has much broader parameter space than $\mu$NMSSM, and perform a similar study to those in Section \ref{numerical study1}. Lastly, we draw conclusions in Section \ref{conclusion}.

\section{\label{theory-section}Theoretical preliminaries}

It is well-known that the superpotential of the popular $Z_3$-NMSSM is given by~\cite{Ellwanger:2009dp,Maniatis:2009re}
\begin{eqnarray}
	W_{\rm Z_3-NMSSM} = W_{\rm Yukawa} + \lambda \hat{S} \hat{H_u} \cdot \hat{H_d} + \frac{1}{3} \kappa \hat{S}^3,
\end{eqnarray}
where the Yukawa terms $W_{\rm Yukawa}$ are the same as those in MSSM, $\hat{H}_u=(\hat{H}_u^+,\hat{H}_u^0)^T$ and $\hat{H}_d=(\hat{H}_d^0,\hat{H}_d^-)^T$ are $SU(2)_L$ doublet Higgs superfields,
and $\lambda$, $\kappa$ are dimensionless couplings coefficient parameterizing the $Z_3$-invariant trilinear terms. GNMSSM differs from $Z_3$-MSSM in that its superpotential
does not respect the $Z_3$ symmetry, and thus it contains the following most general renormalizable couplings:
\begin{eqnarray}
W_{\rm GNMSSM} = W_{\rm Z_3-NMSSM} + \mu \hat{H_u} \cdot \hat{H_d} + \frac{1}{2} \mu^\prime \hat{S}^2+\xi \hat{S}.  \label{Superpotential}
\end{eqnarray}
Historically, the terms characterized by the bilinear mass parameters $\mu$, $\mu^\prime$ and the singlet tadpole parameter $\xi$ were introduced to solve the tadpole problem~\cite{Ellwanger:1983mg, Ellwanger:2009dp} and the cosmological domain-wall problem of $Z_3$-NMSSM~\cite{Abel:1996cr, Kolda:1998rm, Panagiotakopoulos:1998yw}, and the
$\xi$-term can be eliminated by shifting the $\hat{S}$ field and redefining the $\mu$ parameter~\cite{Ross:2011xv}\footnote{Throughout this work, we adopt this convention consistently.}. The bilinear terms could stem from an underlying discrete R symmetry, $Z^R_4$ or $Z^R_8$, after supersymmetry breaking, and be naturally at the electroweak scale~\cite{Abel:1996cr,Lee:2010gv,Lee:2011dya,Ross:2011xv,Ross:2012nr}. Note that these extra terms can change the properties of Higgs bosons and neutralinos (in comparison with the $Z_3$-NMSSM prediction) and significantly alter the phenomenology of the theory. As emphasized in the introduction, this is one of main motivations of this work.

\subsection{Higgs sector of GNMSSM }

Corresponding to the potential in Eq.~(\ref{Superpotential}), the soft-breaking terms of the GNMSSM are given by~\cite{Ellwanger:2009dp,Maniatis:2009re}
\begin{align}
-\mathcal{L}_{soft} = &\Bigg[\lambda A_{\lambda} S H_u \cdot H_d + \frac{1}{3} A_{\kappa} \kappa S^3+ m_3^2 H_u\cdot H_d + \frac{1}{2} m_S^{\prime \ 2} S^2 + h.c.\Bigg] \nonumber \\
& + m^2_{H_u}|H_u|^2 + m^2_{H_d}|H_d|^2 + m^2_{S}|S|^2 .
\end{align}
where $H_u$, $H_d$ and $S$ denote the scalar components of the Higgs superfields. The soft-breaking mass parameters $m^2_{H_u}$, $m^2_{H_d}$ and $m^2_{S}$ can be fixed by solving the conditional equations for minimizing the scalar potential and then expressing them in terms of the vacuum expectation values (vevs) of the scalar fields:
$\left\langle H_u^0 \right\rangle = v_u/\sqrt{2}$, $\left\langle H_d^0 \right\rangle = v_d/\sqrt{2}$ and $\left\langle S \right\rangle = v_s/\sqrt{2}$
with $v = \sqrt{v_u^2+v_d^2}\simeq 246~\mathrm{GeV}$. As usual, the ratio of the two Higgs doublet vevs is defined as $\tan{\beta} \equiv v_u/v_d$, and an effective $\mu$-parameter of MSSM is generated by $\mu_{\rm eff} \equiv \lambda v_s/\sqrt{2}$. Consequently, the Higgs sector is described by ten free parameters: $\tan \beta$, $\mu_{eff}$, the Yukawa couplings $\lambda$ and $\kappa$, the soft-breaking trilinear coefficients $A_\lambda$ and $A_\kappa$, the bilinear mass parameters $\mu$ and $\mu^\prime$, and their soft-breaking parameters $m_3^2$ and $m_S^{\prime\ 2}$.

The GNMSSM predicts three CP-even Higgs bosons $h_i=\{h,H,h_{s}\}$, two CP-odd Higgs bosons $a_i=\{A_H, A_{s}\}$, and a pair of charged Higgs bosons $H^\pm = \cos \beta H_u^\pm + \sin \beta H_d^\pm$. In the field convention that $H_{\rm SM} \equiv \sin\beta {\rm Re}(H_u^0) + \cos\beta {\rm Re} (H_d^0)$, $H_{\rm NSM} \equiv \cos\beta {\rm Re}(H_u^0) - \sin\beta {\rm Re}(H_d^0)$, and $A_{\rm NSM} \equiv \cos\beta {\rm Im}(H_u^0) - \sin\beta  {\rm Im}(H_d^0)$~\cite{Cao:2012fz}, the elements of the $CP$-even Higgs boson mass matrix $\mathcal{M}_S^2$ in the bases $\left(H_{\rm NSM}, H_{\rm SM}, {\rm Re}[S]\right)$ are read as follows in Equations~(\ref{CP-even Hisggs Mass})~\cite{Ellwanger:2009dp}:
\begin{eqnarray}\label{CP-even Hisggs Mass}
  {\cal M}^2_{S, 11}&=& \frac{2 \left [ \mu_{eff} (\lambda A_\lambda + \kappa \mu_{eff} + \lambda \mu^\prime ) + \lambda m_3^2 \right ] }{\lambda \sin 2 \beta} + \frac{1}{2} (2 m_Z^2- \lambda^2v^2)\sin^22\beta, \nonumber \\
  {\cal M}^2_{S, 12}&=&-\frac{1}{4}(2 m_Z^2-\lambda^2v^2)\sin4\beta, \nonumber \\
  {\cal M}^2_{S, 13}&=&-\frac{1}{\sqrt{2}} ( \lambda A_\lambda + 2 \kappa \mu_{eff} + \lambda \mu^\prime ) v \cos 2 \beta, \nonumber \\
  {\cal M}^2_{S, 22}&=&m_Z^2\cos^22\beta+ \frac{1}{2} \lambda^2v^2\sin^22\beta,\nonumber  \\
  {\cal M}^2_{S, 23}&=& \frac{v}{\sqrt{2}} \left[2 \lambda (\mu_{eff} + \mu) - (\lambda A_\lambda + 2 \kappa \mu_{eff} + \lambda \mu^\prime ) \sin2\beta \right], \nonumber \\
  {\cal M}^2_{S, 33}&=& \frac{\lambda (A_\lambda + \mu^\prime) \sin 2 \beta}{4 \mu_{eff}} \lambda v^2   + \frac{\mu_{eff}}{\lambda} (\kappa A_\kappa +  \frac{4 \kappa^2 \mu_{eff}}{\lambda} + 3 \kappa \mu^\prime ) - \frac{\mu}{2 \mu_{eff}} \lambda^2 v^2, \quad \label{Mass-CP-even-Higgs}
\end{eqnarray}
and those for $CP$-odd Higgs fields in the bases $\left( A_{\rm NSM}, {\rm Im}(S)\right)$ are given by Equations~(\ref{Mass-CP-odd-Higgs}):
\begin{eqnarray}
{\cal M}^2_{P,11}&=& \frac{2 \left [ \mu_{eff} (\lambda A_\lambda + \kappa \mu_{eff} + \lambda \mu^\prime ) + \lambda m_3^2 \right ] }{\lambda \sin 2 \beta}, \nonumber  \\
{\cal M}^2_{P,22}&=& \frac{(\lambda A_\lambda + 4 \kappa \mu_{eff} + \lambda \mu^\prime ) \sin 2 \beta }{4 \mu_{eff}} \lambda v^2  - \frac{\kappa \mu_{eff}}{\lambda} (3 A_\kappa + \mu^\prime) - \frac{\mu}{2 \mu_{eff}} \lambda^2 v^2 - 2 m_S^{\prime\ 2}, \nonumber  \\
{\cal M}^2_{P,12}&=& \frac{v}{\sqrt{2}} ( \lambda A_\lambda - 2 \kappa \mu_{eff} - \lambda \mu^\prime ). \label{Mass-CP-odd-Higgs}
\end{eqnarray}
The mass eigenstates $h_i=\{h, H, h_s\}$ and $a_i=\{A_H, A_s\}$ are achieved by unitary rotations $V$ and $V_P$ to diagonalize ${\cal{M}}_S^2$ and ${\cal{M}}_P^2$, respectively, as given by Equations~(\ref{Mass-eigenstates}):
\begin{eqnarray} \label{Mass-eigenstates}
h_i & = & V_{h_i}^{\rm NSM} H_{\rm NSM}+V_{h_i}^{\rm SM} H_{\rm SM}+V_{h_i}^{\rm S} Re[S], \nonumber \\
a_i & = & V_{P, a_i}^{\rm NSM} A_{\rm NSM}+ V_{P, a_i}^{\rm S} Im [S].
\end{eqnarray}
Among these states, $h$ is defined as the scalar state discovered at the LHC, $H$ and $A_H$ represent the doublet-dominated states which prefer to be heavy in the LHC's search for extra Higgs bosons~\cite{ATLAS:2021upq}, and $h_s$ and $A_s$ denote the singlet-dominated states. For the sake of discussion, these states are also labelled in an ascending mass order, i.e. $m_{h_1} < m_{h_2} < m_{h_3}$, and $m_{A_1} < m_{A_2}$. Thus, $h_s \equiv h_1$ and $h \equiv h_2$ for the $h_2$ scenario. The mass of the charged Higgs state $H^\pm$ is expressed as Equation~(\ref{charged-Higgs}):
\begin{eqnarray}  \label{charged-Higgs}
m^2_{H^{\pm}} = \frac{2 \left [ \mu_{eff} (\lambda A_\lambda + \kappa \mu_{eff} + \lambda \mu^\prime ) + \lambda m_3^2 \right ] }{\lambda \sin 2 \beta} + m^2_W -\lambda^2 v^2.
\end{eqnarray}

Regarding the input parameters in the Higgs sector, we note that they have been tightly constrained by the LHC Higgs data for the $h_2$ scenario, especially $\mu_{tot} \equiv \mu + \mu_{eff}$ does not prefer to be excessively large. To illustrate this point, we assume $m_{H^\pm}$ and $A_\lambda$ to be at most several TeV for natural electroweak symmetry breaking (see the equations to minimize the Higgs potential in~\cite{Ellwanger:2009dp}), $\tan \beta \gg 1$ to predict a sizable $a_\mu^{\rm SUSY}$, and $\lambda \leq 0.1$ to suppress the DM-nucleon scattering for Singlino-dominated DM case. Then, after integrating out the heavy Higgs fields $H_{\rm NSM}$ and $A_{\rm NSM}$~\cite{Cao:2012fz,Baum:2017enm}, we obtain the effective mass matrix for CP-even Higgs bosons in the bases $(H_{\rm SM}, Re[S])$, as shown in Equations~(\ref{CP-even Hisggs Mass1}):
\begin{eqnarray}\label{CP-even Hisggs Mass1}
  {\bar {\cal M}}^2_{H_{\rm SM} H_{\rm SM}}& \simeq & m_Z^2 + \delta_t, \quad \quad  {\bar {\cal M}}^2_{H_{\rm SM} Re[S]} \simeq {\cal{M}}_{S,23}^2 \simeq \sqrt{2} \lambda v \mu_{tot}, \nonumber \\
  {\bar {\cal M}}^2_{Re[S] Re[S]} & \simeq & {\cal M}^2_{S, 33} - \frac{{\cal M}^4_{S, 13}}{{\cal M}^2_{S, 11} - {\cal M}^2_{S, 33}} \nonumber \\
  & \simeq & \frac{\mu_{eff}}{\lambda} (\kappa A_\kappa +  \frac{4 \kappa^2 \mu_{eff}}{\lambda} + 3 \kappa \mu^\prime ) - \frac{\mu_{tot}}{2 \mu_{eff}} \lambda^2 v^2 + \frac{1}{2} \lambda^2 v^2,  \label{m_hs}
\end{eqnarray}
where $\delta_t$ denotes top/stop loop correction to Higgs boson mass, and the singlet-dominated CP-odd Higgs boson mass is given by Equation~(\ref{CP-odd Hisggs Mass1}):
\begin{eqnarray} \label{CP-odd Hisggs Mass1}
m_{A_s}^2 & \simeq &  {\cal M}^2_{P, 22} - \frac{{\cal M}^4_{P, 12}}{{\cal M}^2_{P, 11} - {\cal M}^2_{P, 22}} \nonumber \\
& \simeq &  - \frac{\kappa \mu_{eff}}{\lambda} (3 A_\kappa + \mu^\prime) - \frac{\mu_{tot}}{2 \mu_{eff}} \lambda^2 v^2 + \frac{1}{2} \lambda^2 v^2 - 2 m_S^{\prime\ 2}.
\end{eqnarray}
We also obtain the following approximations as shown in Equation~(\ref{Approximations-Higgs}):
\begin{eqnarray}  \label{Approximations-Higgs}
\frac{V_{h}^{\rm S}}{V_h^{\rm SM}} & \simeq &  \frac{{\bar {\cal M}}^2_{H_{\rm SM} Re[S]}}{m_h^2 - m_{h_s}^2} \simeq \frac{\sqrt{2} \lambda \mu_{tot}}{m_h^2 - m_{h_s}^2}, \quad V_{h}^{\rm NSM} \sim 0, \quad V_h^{\rm SM} \simeq \left [ 1 + \left ( \frac{V_{h}^{\rm S}}{V_h^{\rm SM}} \right )^2 \right ]^{-1/2}  \sim 1, \nonumber \\
V_{P, A_s}^{\rm NSM} & \simeq & 0, \quad \quad \quad V_{P, A_s}^{\rm S} \simeq 1.
\end{eqnarray}
These formulae reveal the following facts:
\begin{itemize}
\item Parameters $A_\lambda$ and $m_3$ mainly determine the heavy Higgs boson masses, and they have little impact on the other Higgs bosons' mass spectrums.
\item $m_{h_s}$ and $m_{A_s}$ depend on parameters $\lambda$, $\kappa$, $\mu_{eff}$, $\mu_{tot}$, $A_\kappa$ and $\mu^\prime$. In addition, $m_{A_s}$ also depends on $m_S^{\prime}$. This implies that, even when $\lambda$, $\kappa$, $\mu_{eff}$, $\mu_{tot}$, and $\mu^\prime$ are fixed, $m_{h_s}$ and $m_{A_s}$ can still vary freely by the adjustment of $A_\kappa$ and $m_S^{\prime}$, respectively. This situation is different from that of $Z_3$-NMSSM, where $\mu_{tot} \equiv \mu_{eff}$, $\mu^\prime = 0$, and $m_S^{\prime}=0$, and consequently, the masses of singlet fields are correlated~\cite{Cao:2018rix}.
\item The most important feature is that the latest LHC Higgs data have imposed an upper limit of about $40~{\rm GeV}$  on $|\lambda \mu_{tot} |$ in the tremendously large $\tan \beta$ limit, since  $|\lambda \mu_{tot} |$ may induce a sizable $V_h^S$. Furthermore, since a small $\lambda$ is preferred by DM DD experiments in the Singlino-dominated DM case, $|\mu_{eff}|$, $|\mu_{tot}|$ and  $|\mu_{tot}/\mu_{eff}|$ in Eq.(\ref{m_hs}) are unlikely to be exceedingly large. Otherwise, strong cancellations among the different terms on the right side of Eq.(\ref{m_hs}) are needed to predict $m_{h_s} < 125~{\rm GeV}$, which makes the theory fine-tuned.
\end{itemize}

Given that too many parameters are involved in the Higgs sector, the $h_2$ scenario is studied using the following strategy. First, we assume the charged Higgs bosons to be very massive by setting $A_\lambda = 2~{\rm TeV}$ and $m_3 = 1~{\rm TeV}$, following the discussion above. Second, we investigate the characteristics of $\mu$NMSSM, where $\mu^\prime$ and $m_S^\prime$ are taken to be zero\footnote{Note that $\mu$NMSSM as the most economical realization of GNMSSM could arise from the $Z_3$-NMSSM when it was embedded into canonical superconformal supergravity in the Jordan frame, and had applications to the inflation in the early universe~\cite{Ferrara:2010yw,Ferrara:2010in,Einhorn:2009bh,Hollik:2018yek,Hollik:2020plc}. This is an interesting realization of supersymmetry in particle physics.}. This model contains most of the key features of the GNMSSM~\cite{Cao:2021ljw}, and thus has pedagogical significance. Finally, we concentrate on the GNMSSM by treating $\mu$, $\mu^\prime$ and $m_S^\prime$ as variables, and investigate its features in explaining the discrepancy.

\subsection{Neutralino sector of GNMSSM }

The neutralino sector in the GNMSSM consists of the mixtures among the Bino field $\tilde{B}$, the Wino field $\tilde{W}$, the Higgsino fields $\tilde{H}_d^0$, $\tilde{H}_u^0$ and the Singlino field $\tilde{S}$. Its mass matrix in the basis $(-i \tilde{B}, -i \tilde{W},\tilde{H}_d^0,\tilde{H}_u^0,\tilde{S})$ takes the the following form ~\cite{Ellwanger:2009dp}, as shown in Equation~(\ref{eq:mmn}):
\begin{equation}
    M_{\tilde{\chi}^0} = \left(
    \begin{array}{ccccc}
    M_1 & 0 & -m_Z \sin \theta_W \cos \beta & m_Z \sin \theta_W \sin \beta & 0 \\
      & M_2 & m_Z \cos \theta_W \cos \beta & - m_Z \cos \theta_W \sin \beta &0 \\
    & & 0 & -\mu_{tot} & - \frac{1}{\sqrt{2}} \lambda v \sin \beta \\
    & & & 0 & -\frac{1}{\sqrt{2}} \lambda v \cos \beta \\
    & & & & \frac{2 \kappa}{\lambda} \mu_{\rm eff} + \mu^\prime
    \end{array}
    \right), \label{eq:mmn}
\end{equation}
where $M_1$ and $M_2$ are gaugino soft-breaking masses, and $\mu_{tot}$ represents the Higgsino mass. This matrix can be diagonalized by a rotation matrix $N$, and subsequently the mass eigenstates are expressed by Equation~(\ref{Mass-eigenstate-neutralino}):
\begin{eqnarray}  \label{Mass-eigenstate-neutralino}
\tilde{\chi}_i^0 = N_{i1} \psi^0_1 +   N_{i2} \psi^0_2 +   N_{i3} \psi^0_3 +   N_{i4} \psi^0_4 +   N_{i5} \psi^0_5.
\end{eqnarray}
where $\tilde{\chi}_i^0\,(i=1,2,3,4,5)$ are labeled in an ascending mass order. $N_{i3}$ and $N_{i4}$ characterize the $\tilde{H}_d^0$ and $\tilde{H}_u^0$ components in $\tilde{\chi}_i^0$, and $N_{i5}$ denotes the Singlino component.

In the case of very massive gauginos and $|m_{\tilde{\chi}_1^0}^2 - \mu_{tot}^2 | \gg \lambda^2 v^2$, the following approximations are obtained for the  Singlino-dominated $\tilde{\chi}_1^0$~\cite{Cheung:2014lqa,Badziak:2015exr,Badziak_2017}, given by Equations~(\ref{Approximation-neutralinos}):
\begin{eqnarray} \label{Approximation-neutralinos}
\mu^\prime & \simeq & m_{\tilde{\chi}_1^0} - \frac{1}{2} \frac{\lambda^2 v^2 ( m_{\tilde{\chi}_1^0} - \mu_{tot} \sin 2 \beta )}{m_{\tilde{\chi}_1^0}^2 - \mu_{tot}^2} - \frac{2 \kappa}{\lambda} \mu_{eff}, \quad N_{11} \sim 0, \quad N_{12} \sim 0, \label{Neutralino-Mixing}  \\
\frac{N_{13}}{N_{15}} &= & \frac{\lambda v}{\sqrt{2} \mu_{\rm tot}} \frac{(m_{\tilde{\chi}_1^0}/\mu_{\rm tot})\sin\beta-\cos\beta} {1-\left(m_{\tilde{\chi}_1^0}/\mu_{\rm tot}\right)^2}, \quad \quad  \frac{N_{14}}{N_{15}} =  \frac{\lambda v}{\sqrt{2} \mu_{\rm tot}} \frac{(m_{\tilde{\chi}_1^0}/\mu_{\rm tot})\cos\beta-\sin\beta} {1-\left(m_{\tilde{\chi}_1^0}/\mu_{\rm tot}\right)^2}, \nonumber \\
N_{15}^2 & \simeq & \left(1+ \frac{N^2_{13}}{N^2_{15}}+\frac{N^2_{14}}{N^2_{15}}\right)^{-1} \nonumber \\
&= & \frac{\left[1-(m_{\tilde{\chi}_1^0}/\mu_{\rm tot})^2\right]^2}{\left[(m_{\tilde{\chi}_1^0}/\mu_{\rm tot})^2
-2(m_{\tilde{\chi}_1^0}/\mu_{\rm tot})\sin2\beta+1 \right]\left(\frac{\lambda v}{\sqrt{2}\mu_{\rm tot}}\right)^2
+\left[1-(m_{\tilde{\chi}_1^0}/\mu_{\rm tot})^2\right]^2}. \nonumber
\end{eqnarray}
These approximations indicate that the mass of the Singlino-dominated DM is determined by the parameters $\lambda$, $\kappa$, $\mu_{eff}$, $\mu_{tot}$, and $\mu^\prime$. In particular, $\lambda$ and $\kappa$ are two independent parameters in predicting $|m_{\tilde{\chi}_1^0}| < |\mu_{tot}|$. This situation is different from that of the $Z_3$-NMSSM, where $\mu^\prime \equiv 0$, $\mu_{tot} \equiv \mu_{eff}$, and consequently, $|\kappa|$ must be less than $\lambda/2$ to predict the Singlino-dominated neutralino as the LSP~\cite{Ellwanger:2009dp}. They also indicate that, for fixed $\tan \beta$, the Higgsino compositions in $\tilde{\chi}_1^0$ depend only on $\lambda$, $\mu_{tot}$, and $m_{\tilde{\chi}_1^0}$. Therefore,  it is convenient to take the three parameters and $\kappa$ as theoretical inputs in studying the $\tilde{\chi}_1^0$'s properties, where $\kappa$ determines the interactions among the singlet-dominated particles. This characteristic contrasts with that of the $Z_3$-NMSSM, which only needs the three input parameters of $\lambda$, $\mu_{tot}$, and any of $m_{\tilde{\chi}_1^0}$ or $\kappa$ to describe $\tilde{\chi}_1^0$ properties~\cite{Zhou:2021pit}. These differences imply that the singlet-dominated particles may form a secluded DM sector~\cite{Pospelov:2007mp}, which has the following salient features:
\begin{itemize}
\item The Singlino-dominated DM can achieve the correct abundance by the process $\tilde{\chi}_1^0 \tilde{\chi}_1^0 \to h_s A_s$, through adjusting the value of $\kappa$, or by the $h_s/A_s$-mediated resonant annihilation into SM particles.
\item Since the secluded sector communicates with the SM sector only through the weak singlet-doublet Higgs mixing, the interaction between the DM and nucleus is naturally feeble when $\lambda$ is small.
\end{itemize}

We added that, even when $\lambda$, $\kappa$, $\mu_{eff}$ and $\mu_{tot}$ are fixed, $m_{\tilde{\chi}_1}^0$ can still vary freely through the tuning of $\mu^\prime$. In addition to the process $\tilde{\chi}_1^0 \tilde{\chi}_1^0 \to h_s A_s$ and $h_s/A_s$ resonant annihilation, the DM has other annihilation channels for obtaining the measured abundance~\cite{Cao:2021tuh}, e.g., co-annihilation with Higgsino-dominated electroweakinos and/or sleptons, and resonant $Z/h$ annihilations. Owing to these features, the GNMSSM has a broad parameter space consistent with the current DM experimental results. As a result, it is the Singlino-dominated LSP, instead of the Bino-dominated LSP, that is most favored to be a viable DM candidate.

\subsection{\label{DMRD}Muon g-2 in the GNMSSM }

The SUSY source of the muon anomalous magnetic moment $a^{\rm SUSY}_{\mu}$ mainly includes loops with a smuon and a neutralino, as well as those with a muon-type sneutrino and a chargino~\cite{Moroi:1995yh,Domingo:2008bb,Hollik:1997vb,Martin:2001st}. The one-loop contributions to $a^{\rm SUSY}_{\mu}$ in GNMSSM are given by~\cite{Domingo:2008bb,Cao:2021tuh} as Equations~(\ref{amuon}):
\begin{small}\begin{equation}\begin{split}
	&a_{\mu}^{\rm SUSY} = a_{\mu}^{\tilde{\chi}^0 \tilde{\mu}} + a_{\mu}^{\tilde{\chi}^{\pm} \tilde{\nu}},\\
    a_{\mu}^{\tilde{\chi}^0 \tilde{\mu}} &= \frac{m_{\mu}}{16 \pi^2}\sum_{i,l}\left\{
    -\frac{m_{\mu}}{12 m_{\tilde{\mu}_l}^2} \left( |n_{il}^{\rm L}|^2 + |n_{il}^{\rm R}|^2 \right) F_1^{\rm N}(x_{il}) + \frac{m_{\tilde{\chi}_i^0}}{3 m_{\tilde{\mu}_l}^2} {\rm Re}(n_{il}^{\rm L} n_{il}^{\rm R}) F_2^{\rm N}(x_{il})
    \right\}, \\
    a_{\mu}^{\tilde{\chi}^\pm \tilde{\nu}} &= \frac{m_{\mu}}{16 \pi^2}\sum_{k}\left\{
    \frac{m_{\mu}}{12 m_{\tilde{\nu}_{\mu}}^2} \left( |c_{k}^{\rm L}|^2 + |c_{k}^{\rm R}|^2 \right) F_1^{\rm C}(x_{k}) + \frac{2 m_{\tilde{\chi}_k^\pm}}{3 m_{\tilde{\nu}_{\mu}}^2} {\rm Re}(c_{k}^{\rm L}c_{k}^{ \rm R}) F_2^{\rm C}(x_{k})
    \right\}, \label{amuon}
\end{split}
\end{equation}\end{small}
where $i=1,\cdots,5$, $j=1,2$ and $l=1,2$ denote the neutralino, chargino and smuon index, respectively.This gives us Equations~(\ref{Mixing-notations}):
\begin{equation}
    \begin{split}
        n_{il}^{\rm L} 	= \frac{1}{\sqrt{2}}\left( g_2 N_{i2} + g_1 N_{i1} \right)X^*_{l1} -y_{\mu} N_{i3}X^*_{l2}, \quad
        &n_{il}^{\rm R} = \sqrt{2} g_1 N_{i1} X_{l2} + y_{\mu} N_{i3} X_{l1},\\
        c_{k}^{\rm L}  	= -g_2 V^{\rm c}_{k1}, \quad
        &c_{k}^{\rm R} 	= y_{\mu} U^{\rm c}_{k2}, \\
    \end{split}   \label{Mixing-notations}
\end{equation}
where $N$ is the neutralino mass rotation matrix, $X$ the smuon mass rotation matrix, and $U^{\rm c}$ and $V^{\rm c}$ the chargino mass rotation matrices defined by ${U^{\rm c}}^* M_{C} {V^{\rm c}}^\dag = m_{\tilde{\chi}^\pm}^{\rm diag}$. $F(x)$s are the loop functions of the kinematic variables defined as $x_{il} \equiv m_{\tilde{\chi}_i^0}^2 / m_{\tilde{\mu}_l}^2$ and $x_{k} \equiv m_{\tilde{\chi}_k^\pm}^2 / m_{\tilde{\nu}_{\mu}}^2$, and take the following form given by Equations~(\ref{Function-1})--(\ref{Function-4}):
\begin{align}
    F^N_1(x) & = \frac{2}{(1-x)^4}\left[ 1-6x+3x^2+2x^3-6x^2\ln x\right]  \label{Function-1} \\
    F^N_2(x) & = \frac{3}{(1-x)^3}\left[ 1-x^2+2x\ln x\right] \\
    F^C_1(x) & = \frac{2}{(1-x)^4}\left[ 2+ 3x - 6x^2 + x^3 +6x\ln x\right] \\
    F^C_2(x) & = -\frac{3}{2(1-x)^3}\left[ 3-4x+x^2 +2\ln x\right] \label{Function-4},
\end{align}
They satisfy $F^N_1(1) = F^N_2(1) = F^C_1(1) = F^C_2(1) = 1$ for the mass-degenerate sparticle case.

In practice, it is instructive to understand the features of $a_\mu^{\rm SUSY}$  through the mass insertion approximation~\cite{Moroi:1995yh}. Specifically, for the lowest order of the approximation, the contributions to $a_\mu^{\rm SUSY}$ can be classified into four types: "WHL", "BHL", "BHR", and "BLR", where $W$, $B$, $H$, $L$, and $R$ stands for Wino, Bino, Higgsino, and left-handed and right-handed Smuon fields, respectively. These are from the Feynman diagrams involving $\tilde{W}-\tilde{H}_d$, $\tilde{B}-\tilde{H}_d^0$, $\tilde{B}-\tilde{H}_d^0$, and $\tilde{\mu}_L-\tilde{\mu}_R$ transitions, respectively, and take the following form~\cite{Athron:2015rva, Moroi:1995yh,Endo:2021zal} given by Equations~(\ref{eq:WHL})--(\ref{eq:BLR}):
\begin{eqnarray}
a_{\mu, \rm WHL}^{\rm SUSY}
    &=&\frac{\alpha_2}{8 \pi} \frac{m_{\mu}^2 M_2 \mu_{tot} \tan \beta}{m_{\tilde{\nu}_\mu}^4} \left \{ 2 f_C\left(\frac{M_2^2}{m_{\tilde{\nu}_{\mu}}^2}, \frac{\mu_{tot}^2}{m_{\tilde{\nu}_{\mu}}^2} \right) - \frac{m_{\tilde{\nu}_\mu}^4}{m_{\tilde{\mu}_L}^4} f_N\left(\frac{M_2^2}{m_{\tilde{\mu}_L}^2}, \frac{\mu_{tot}^2}{m_{\tilde{\mu}_L}^2} \right) \right \}\,, \quad \quad
    \label{eq:WHL} \\
a_{\mu, \rm BHL}^{\rm SUSY}
  &=& \frac{\alpha_Y}{8 \pi} \frac{m_\mu^2 M_1 \mu_{tot}  \tan \beta}{m_{\tilde{\mu}_L}^4} f_N\left(\frac{M_1^2}{m_{\tilde{\mu}_L}^2}, \frac{\mu_{tot}^2}{m_{\tilde{\mu}_L}^2} \right)\,,
    \label{eq:BHL} \\
a_{\mu, \rm BHR}^{\rm SUSY}
  &=& - \frac{\alpha_Y}{4\pi} \frac{m_{\mu}^2 M_1 \mu_{tot} \tan \beta}{m_{\tilde{\mu}_R}^4} f_N\left(\frac{M_1^2}{m_{\tilde{\mu}_R}^2}, \frac{\mu_{tot}^2}{m_{\tilde{\mu}_R}^2} \right)\,,
    \label{eq:BHR} \\
a_{\mu \rm BLR}^{\rm SUSY}
  &=& \frac{\alpha_Y}{4\pi} \frac{m_{\mu}^2  M_1 \mu_{tot} \tan \beta}{M_1^4}
    f_N\left(\frac{m_{\tilde{\mu}_L}^2}{M_1^2}, \frac{m_{\tilde{\mu}_R}^2}{M_1^2} \right)\,,
    \label{eq:BLR}
\end{eqnarray}
where the loop functions are given by
\begin{eqnarray}
    \label{eq:loop-aprox}
    f_C(x,y)
    &=&  \frac{5-3(x+y)+xy}{(x-1)^2(y-1)^2} - \frac{2\ln x}{(x-y)(x-1)^3}+\frac{2\ln y}{(x-y)(y-1)^3} \,,
      \\
    f_N(x,y)
    &=&
      \frac{-3+x+y+xy}{(x-1)^2(y-1)^2} + \frac{2x\ln x}{(x-y)(x-1)^3}-\frac{2y\ln y}{(x-y)(y-1)^3} \,,
\end{eqnarray}
and they satisfy $f_C(1,1) = 1/2$ and $f_N(1,1) = 1/6$. Note that the Singlino field $\tilde{S}$ can also enter the insertions. Because both the $\tilde{W}-\tilde{S}$ and $\tilde{B}^0-\tilde{S}$ transitions and the $\bar{\mu} \tilde{S} \tilde{\mu}_{L,R}$ couplings vanish, the Singlino field only appears in the "WHL", "BHL"
and "BHR" loops by two more insertions at the lowest order, which correspond to the $\tilde{H}_d^0-\tilde{S}$ and $\tilde{S}-\tilde{H}_d^0$ transitions in the
neutralino mass matrix in Eq.~(\ref{eq:mmn}). Since a small $\lambda$ is preferred by the DM physics, the Singlino-induced contributions are never significant~\cite{Cao:2021tuh}. Although in this case the GNMSSM prediction of $a_\mu^{\rm SUSY}$ is roughly the same as that of the MSSM, except that the $\mu$
parameter of the MSSM should be replaced by $\mu_{tot}$, the two models predict different DM physics and different sparticle signals at the LHC. Thus, they are
subject to different theoretical and experimental constraints. It should also be noted that, although there is a prefactor of the Higgsino mass $\mu$ in the expression of
the "WHL", "BHL", and "BHR" contributions, the involved loop functions approach zero with the increase of $|\mu|$, and consequently these contributions depend on $\mu$ in a complex way.
By focusing on several typical patterns of sparticle mass spectrum with a positive $\mu$, we found that the "WHL" contribution decreases monotonously as $\mu$ increases, while the magnitude of the "BHL" and "BHR" contributions increases when $\mu$ is significantly smaller than the slepton mass and decreases when $\mu$ is larger than the slepton mass. In addition, the "WHL" contribution is usually much larger than the other contributions if $\tilde{\mu}_L$ is not significantly heavier than $\tilde{\mu}_R$.

\section{\label{numerical study1} Explaining $\Delta a_\mu$ in the $h_2$ scenario of $\mu$NMSSM}

\subsection{\label{scan} Research strategy}

This sector focuses on the $h_2$ scenario of $\mu$NMSSM. In order to analyze its characteristics in explaining the discrepancy, the following parameter space was scanned by MultiNest algorithm~\cite{Feroz:2008xx}:
\begin{equation}\begin{split}
    &0 \leq \lambda \leq 0.75, \quad |\kappa| \leq 0.75, \quad 1 \leq \tan{\beta} \leq 60,  \\
	|M_1| \leq 1.5&~{\rm TeV}, \quad
	100~{\rm GeV} \leq M_2 \leq 1.5~{\rm TeV}, \quad
	-5~{\rm TeV} \leq A_{t} \leq 5~{\rm TeV}, \\
	|\mu| \leq 1&~{\rm TeV}, \quad
	100~{\rm GeV} \leq \mu_{\rm tot} \leq 1~{\rm TeV},\quad
	|A_\kappa| \leq 1~{\rm TeV}, \\
	100~&{\rm GeV} \leq m_{\tilde{\mu}_L} \leq 1~{\rm TeV}, \quad
	100~{\rm GeV} \leq m_{\tilde{\mu}_R} \leq 1~{\rm TeV},   \label{Parameter-region}
\end{split}\end{equation}
where the flat prior distribution was chosen for all input parameters and the active point number, $n_{\rm live}$, was set to be 6000\footnote{In the MultiNest algorithm, the active points are used to determine the iso-likelihood contours in each iteration of sampling. Note that \bf{the results obtained by this algorithm has statistical significance.}}.  Other dimensional parameters that are unimportant to this study were fixed at $2~{\rm TeV}$, and include SUSY parameters for the first and third generation sleptons, three generation squarks, and gluinos. In numerical calculations, the model file of GNMSSM was constructed using the package \texttt{SARAH-4.14.3}~\cite{Staub:2008uz, Staub:2012pb, Staub:2013tta, Staub:2015kfa}. Particle mass spectra and low-energy observables, such as $a_\mu^{\rm SUSY}$ and B-physics observables, were generated by the codes \texttt{SPheno-4.0.4}~\cite{Porod:2003um, Porod:2011nf} and \texttt{FlavorKit}~\cite{Porod:2014xia}.  The DM abundance and direct/indirect detection cross-sections were calculated by the package~\texttt{MicrOMEGAs-5.0.4}~\cite{Belanger:2001fz, Belanger:2005kh, Belanger:2006is, Belanger:2010pz, Belanger:2013oya, Barducci:2016pcb}. The likelihood function used to guide the scan mainly contains the value of $a_\mu^{\rm SUSY}$~\cite{Cao:2021tuh}, and it takes the following form, given by Equation~(\ref{eq:likeliamm}):
\begin{eqnarray}
\cal{L} = \left \{ \begin{aligned} &\exp\left[-\frac{1}{2} \left( \frac{a_{\mu}^{\rm SUSY}- 2.51\times 10^{-9}}{5.9\times 10^{-10} }\right)^2\right],\ & &{\rm if\ restrictions\ satisfied}; \\ &\exp\left[-100\right], & &{\rm if\ restrictions\ unsatisfied}, \end{aligned} \right . \label{eq:likeliamm}
 \end{eqnarray}
where the restrictions on each sample include:
\begin{enumerate}
\item DM relic abundance, $0.096 \leq \Omega h^2 \leq 0.144$. In implementing this constraint, the central value of the Planck-2018 data, $\Omega h^2 = 0.120$~\cite{Aghanim:2018eyx}, was used, and a theoretical uncertainty of $10\%$ in abundance calculation was assumed.
\item DM direct and indirect detections. Specifically, the SI and SD DM-nucleon scattering cross-sections should be lower than the bounds from the XENON-1T experiments~\cite{Aprile:2018dbl,Aprile:2019dbj}, and the DM annihilation rate at present time should be consistent with dwarf galaxies observations from Fermi-LAT collaboration \cite{Ackermann:2015zua}. The method suggested in~\cite{Carpenter:2016thc} was adopted in studying the latter constraint.
\item Higgs data fit. The properties of the next lightest CP-even Higgs boson $h_2$ (also denoted by $h$ throughout this work) should be consistent at the $95\%$ confidence level with corresponding data obtained by ATLAS and CMS collaborations. This condition was checked with the code~\texttt{HiggsSignal-2.2.3}~\cite{Bechtle:2014ewa} by requiring the sample's $p$ value to be larger than 0.05.
\item Direct searches for extra Higgs bosons at LEP, Tevatron and LHC. This requirement was examined by the code \texttt{HiggsBounds-5.3.2}~\cite{Bechtle:2015pma}.
\item Some B-physics observations. Specifically, the branching ratios of $B_s \to \mu^+ \mu^-$ and $B \to X_s \gamma$ should agree with their experimental measurements, which were summarized in~\cite{PhysRevD.98.030001} at the $2 \sigma$ level.
\item LHC searches for SUSY. In order to explain the discrepancy, the electoweakinos and sleptons in the GNMSSM can not be excessively heavy. Thus, they will be produced at the LHC to generate multi-lepton signals. The code \texttt{SModelS-2.1.1}~\cite{Khosa:2020zar} was used to set limits on the signals in some simple topology cases.
    Sophisticated study of the constraints will be carried out in subsection \ref{LHCMC}, using the package \texttt{CheckMATE-2.0.29}~\cite{Drees:2013wra,Dercks:2016npn, Kim:2015wza}.
\item Vacuum stability for the scalar potential consisting of the Higgs fields and the last two generation slepton fields. This condition was checked by the  Vevacious code~\cite{Camargo-Molina:2013qva,Camargo-Molina:2014pwa}, and its effect on the GNMSSM was recently discussed in~\cite{Cao:2021tuh}.
\end{enumerate}

\begin{figure}[t]
	\centering
	\includegraphics[width=0.505\textwidth]{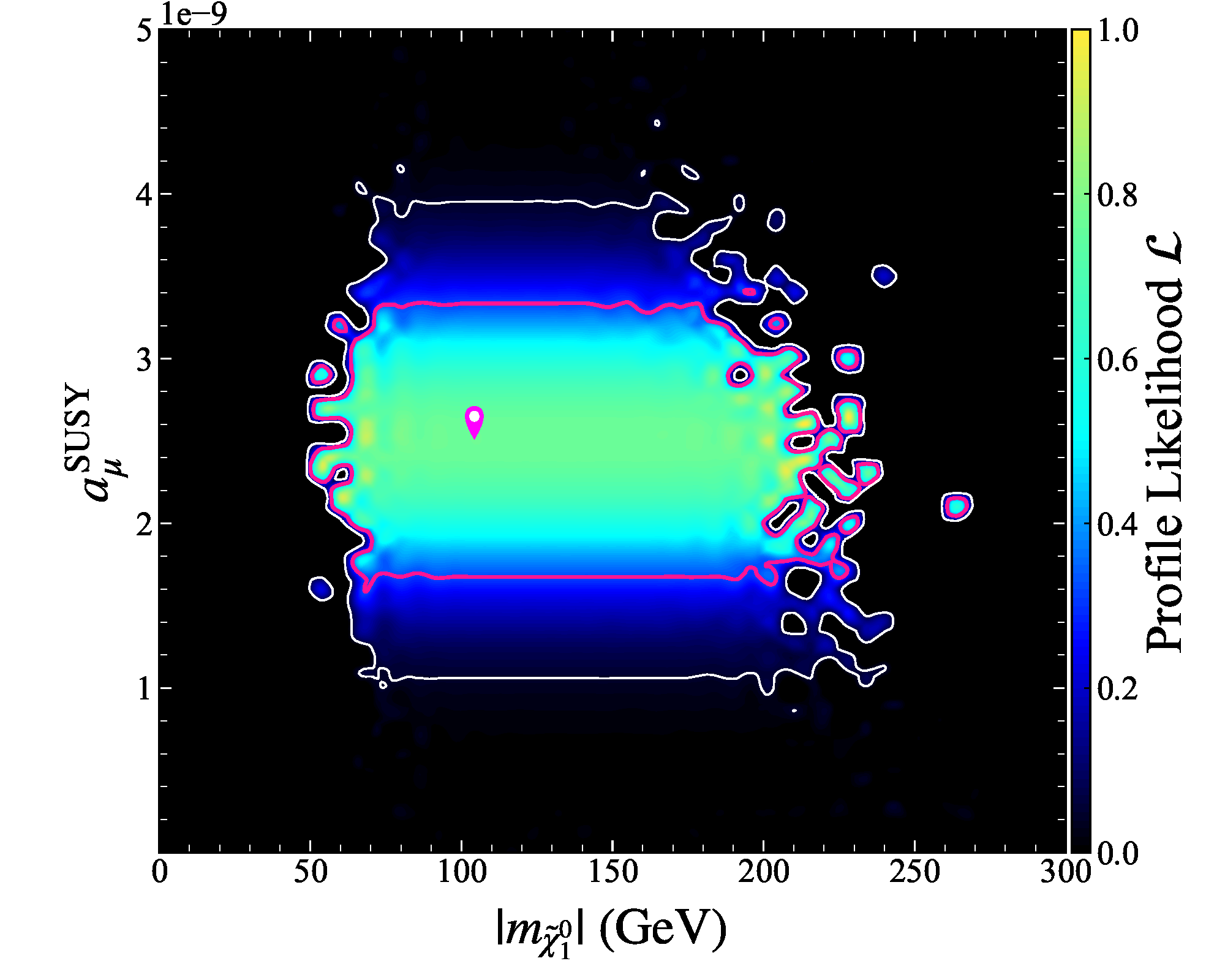}\hspace{-0.3cm}
     \includegraphics[width=0.505\textwidth]{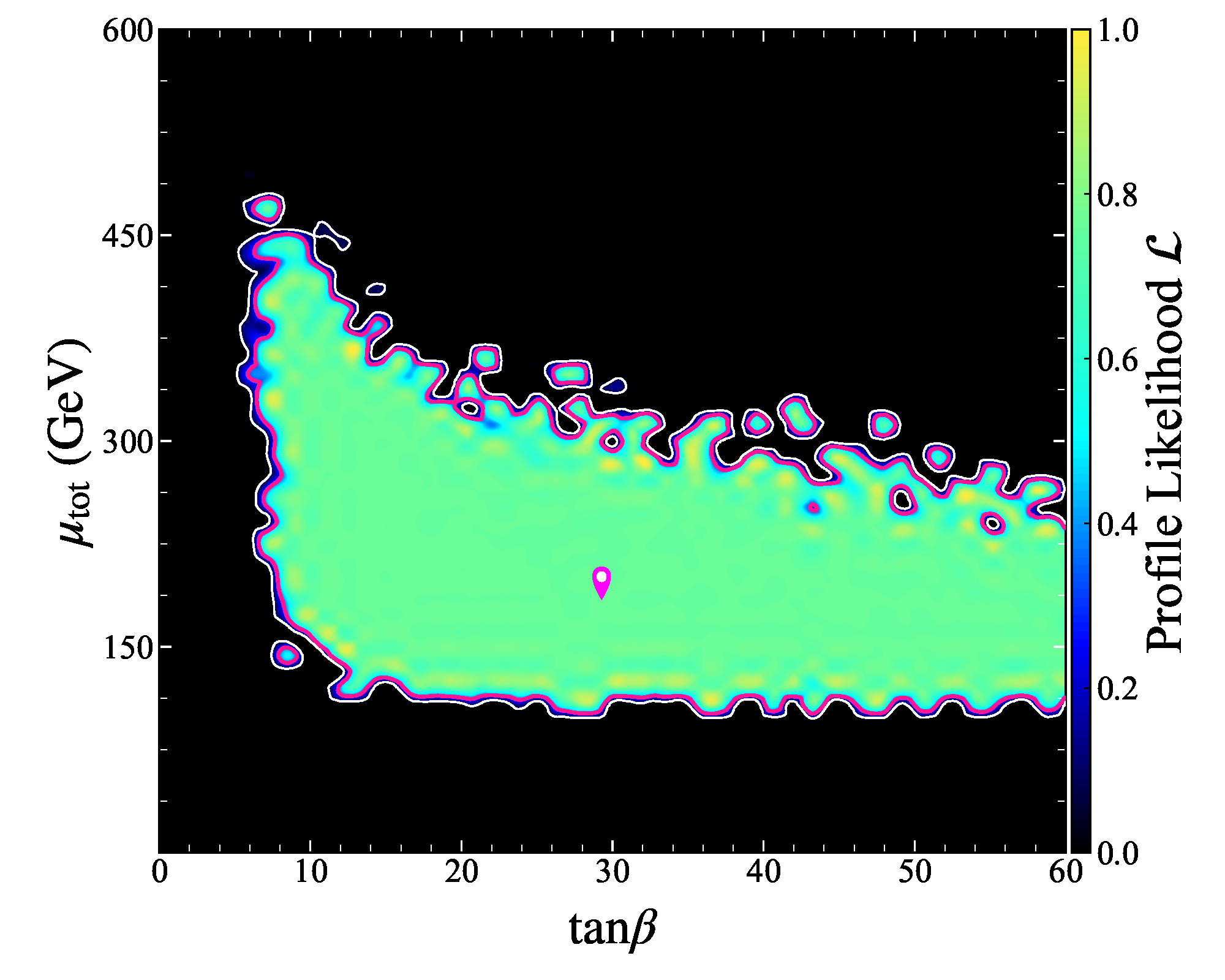} \\
	  \includegraphics[width=0.505\textwidth]{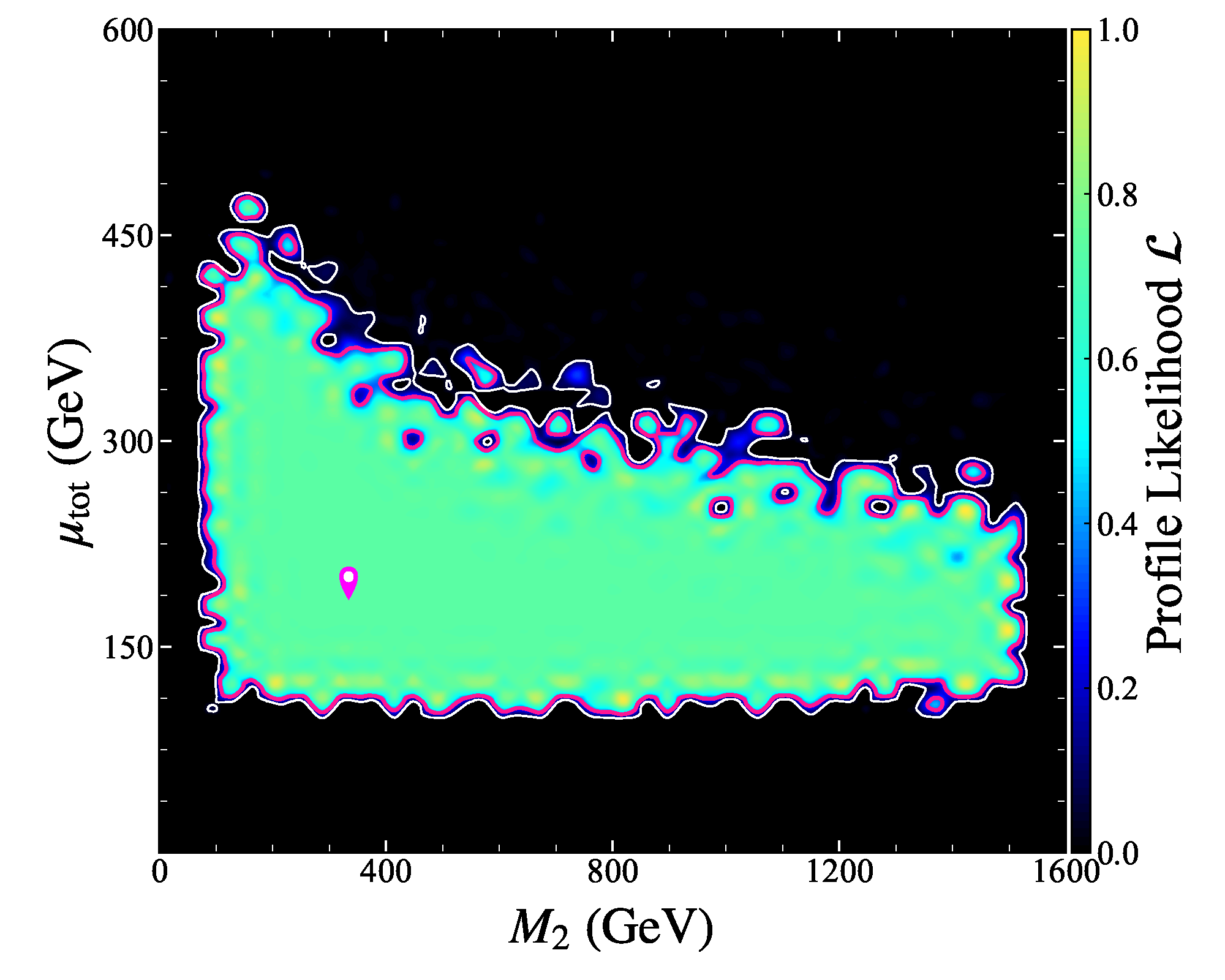}\hspace{-0.3cm}
      \includegraphics[width=0.505\textwidth]{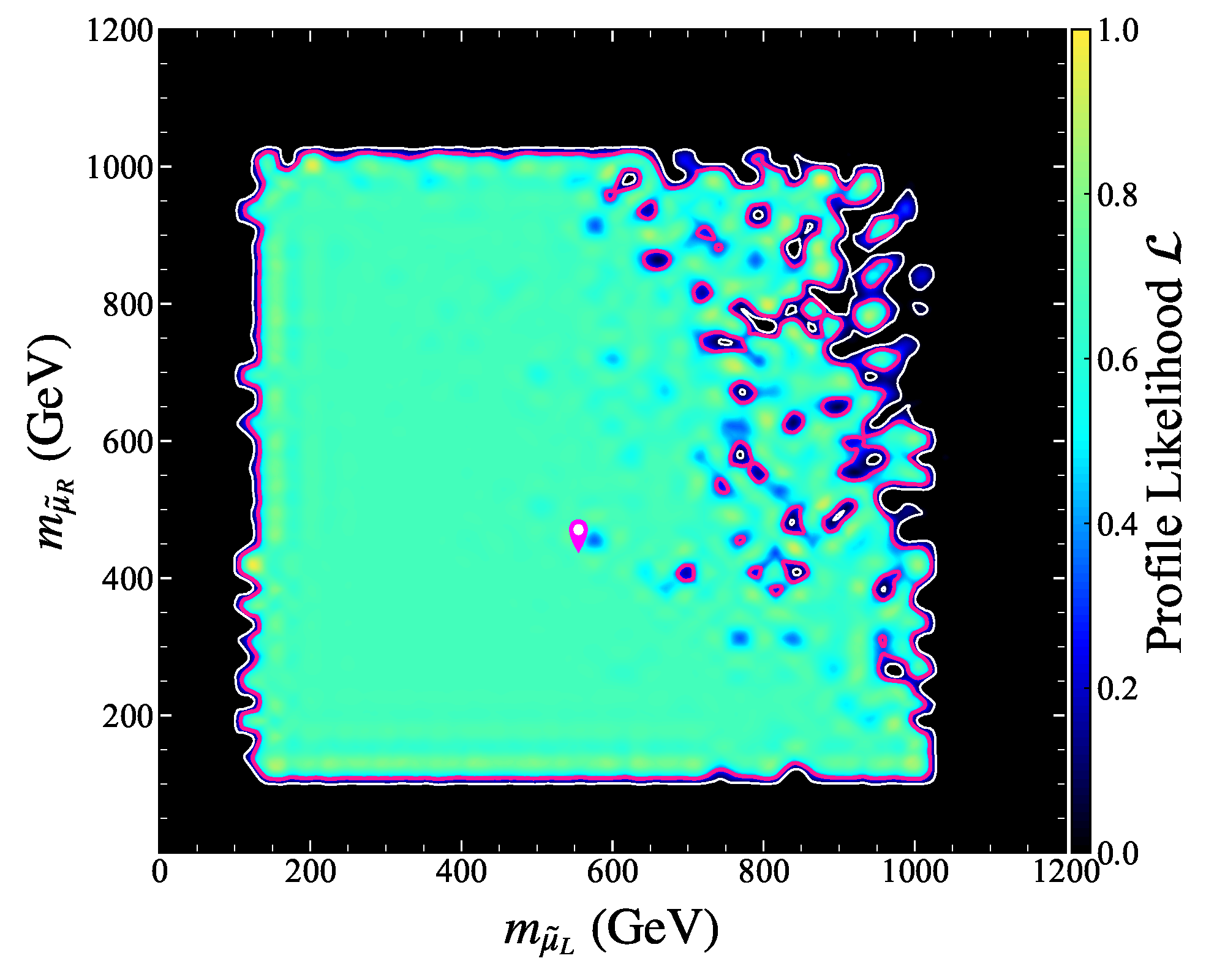}
	\caption{\label{fig:plamm}Two-dimensional profile likelihood maps of the function $\mathcal{L}$ in Eq.~(\ref{eq:likeliamm}) projected onto $|m_{\tilde{\chi}_1^0}|-a_{\mu}^{\rm SUSY}$,  $\tan{\beta}-\mu_{tot}$, $M_2-\mu_{tot}$, and $m_{\tilde{\mu}_L} - m_{\tilde{\mu}_R}$ planes. Pink and white contour lines enclose $1\sigma$ and $2 \sigma$ confidence regions, respectively. The best point is marked by the pin symbol, and it is located at $\tan \beta \simeq 30$, $m_{\tilde{\chi}_1^0} \simeq 103~{\rm GeV}$, $\mu_{tot} \simeq 210~{\rm GeV}$, $M_2 \simeq 330~{\rm GeV}$, $m_{\tilde{\mu}_L} \simeq 470~{\rm GeV}$, and $m_{\tilde{\mu}_R} \simeq 550~{\rm GeV}$.}
\end{figure}

\begin{figure}[t]
	\centering
	\includegraphics[width=0.505\textwidth]{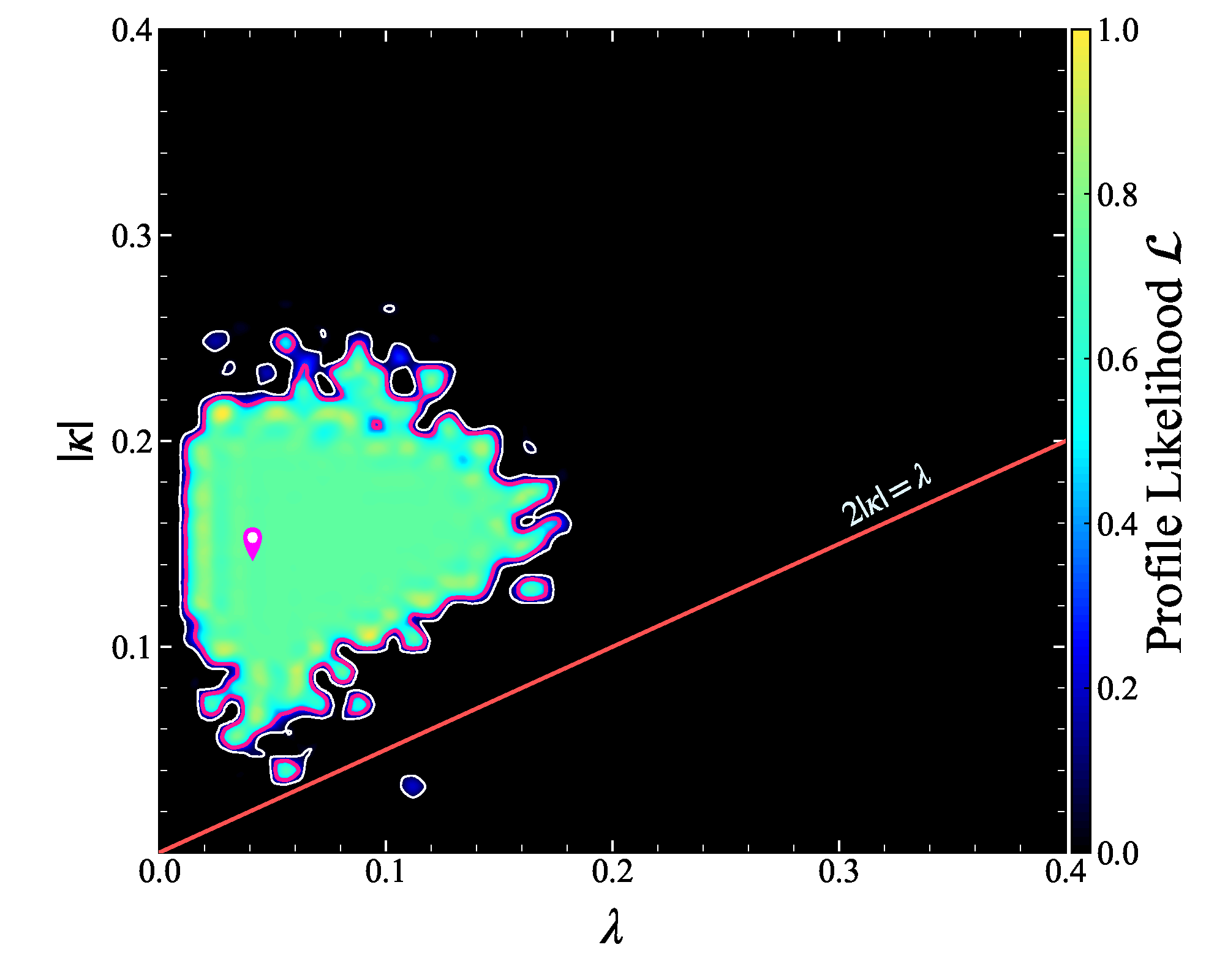}\hspace{-0.3cm}
	\includegraphics[width=0.505\textwidth]{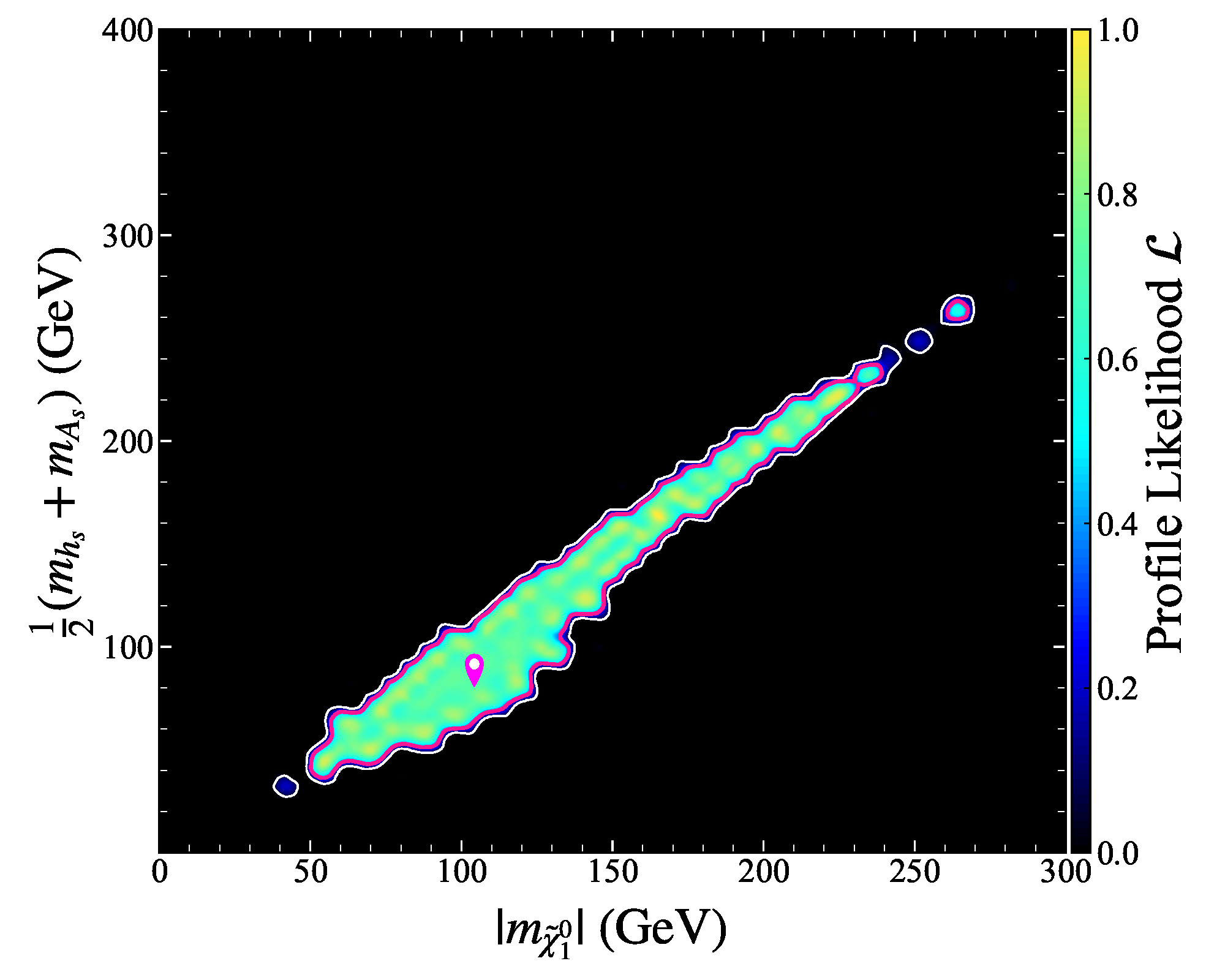}\\
	\includegraphics[width=0.505\textwidth]{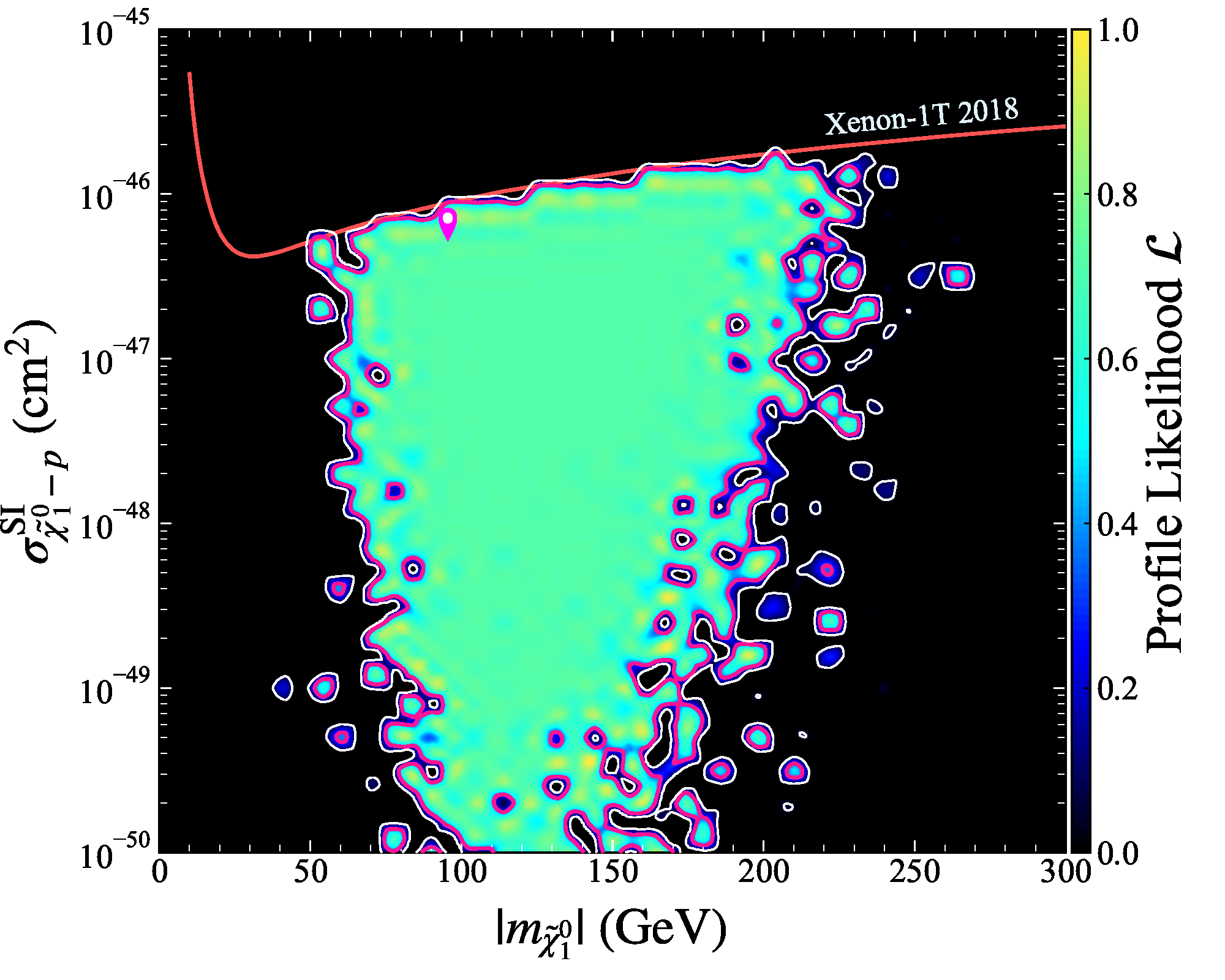}\hspace{-0.3cm}
	\includegraphics[width=0.505\textwidth]{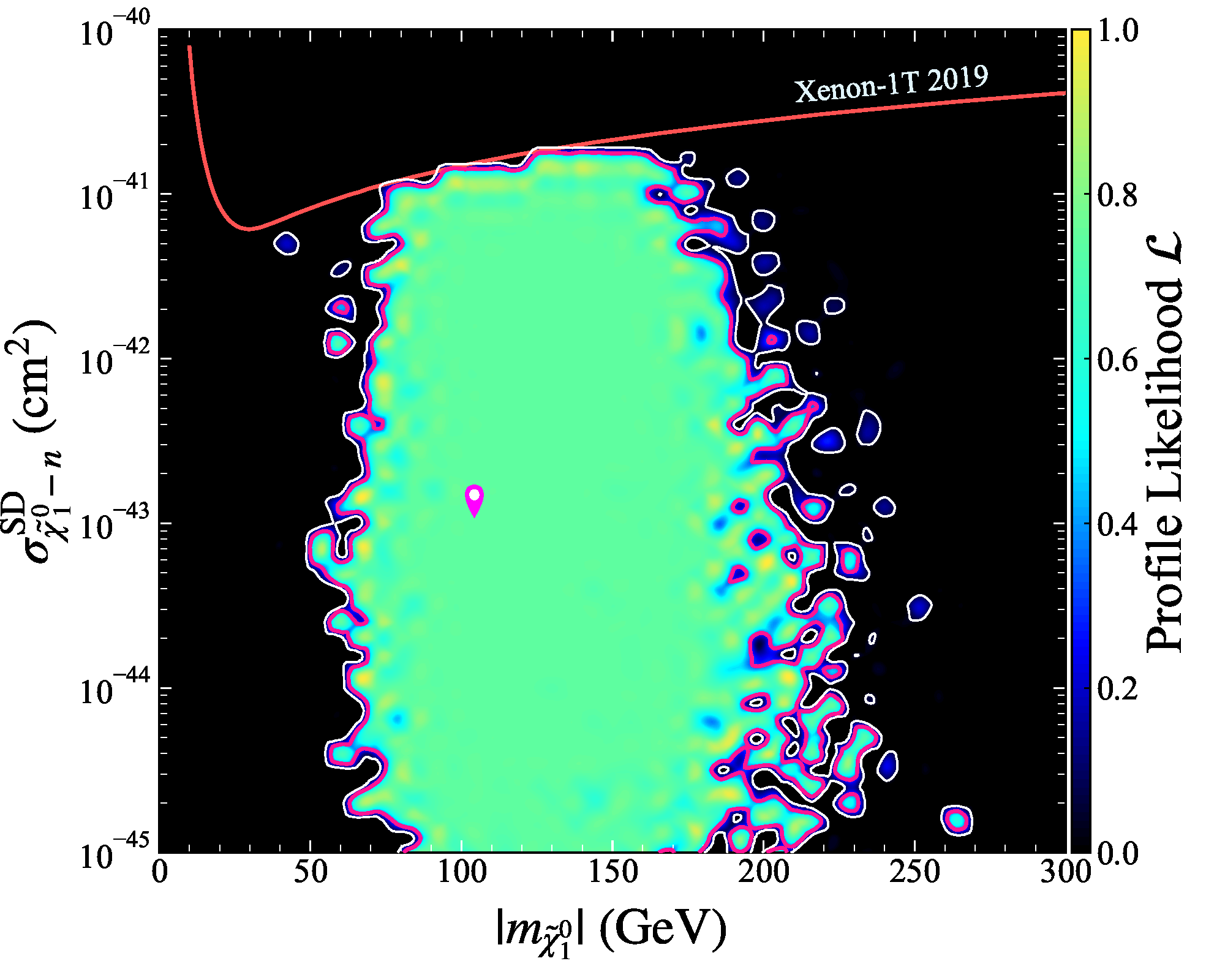}
	\caption{\label{fig:dm}Same as Fig.~\ref{fig:plamm}, but on the $\lambda-|\kappa|$,  $|m_{\tilde{\chi}_1^0}| -\frac{1}{2}(m_{h_s} + m_{A_s})$, $|m_{\tilde{\chi}_1^0}| - \sigma_{n}^{\rm SD}$ and $|m_{\tilde{\chi}_1^0}| - \sigma_{p}^{\rm SI}$ planes, respectively. }
\end{figure}

In presenting the results, two-dimensional profile likelihood (PL) for the function $\cal{L}$ in Eq.(\ref{eq:likeliamm}) was used. It is defined by what follows in Equation~(\ref{PL-DM}):~\cite{Fowlie:2016hew}
\begin{eqnarray}
\mathcal{L} (\Theta_A,\Theta_B)=\mathop{\max}_{\Theta_1,\cdots,\Theta_{A-1},\Theta_{A+1},\cdots, \Theta_{B-1}, \Theta_{B+1},\cdots} \mathcal{L} (\Theta),  \label{PL-DM}
\end{eqnarray}
where $\Theta_i (i=1, 2, ...)$ denote the input parameters, $\Theta_{A,B}$ are the variables of interest, and the maximization of $\mathcal{L} (\Theta_A,\Theta_B)$ is achieved by scanning the parameters other than $\Theta_{A}$ and $\Theta_{B}$. Related quantities includes $1\sigma$ and $2\sigma$ confidence intervals (CI), and the $\chi^2$ function defined by $\chi^2 \equiv -2 \rm ln \mathcal{L} (\Theta_A,\Theta_B)$. These statistical measures were briefly introduced in~\cite{Fowlie:2016hew}, and they reflect the capability of the theory
to explain the discrepancy.

\subsection{\label{region} Key features of the interpretation}

All samples obtained in the scan were projected onto different parameter planes to show two-dimensional PLs, which could reveal the underlying physics of the $h_2$ scenario. Fig.~{\ref{fig:plamm}} illustrates that the scenario can interpret the discrepancy in a broad parameter space. Specifically, the upper left panel indicates that the best point predicts $a_\mu^{\rm SUSY} = 25.1 \times 10^{-10}$,  which means $\chi^2_{\rm Best} = 0$, $\chi^2 \le 2.3$ for $1\sigma$ CI and $\chi^2 \le 6.18$ for $2 \sigma$ CIs. In term of $a_\mu^{\rm SUSY}$, the $\chi^2$ ranges correspond to $ 16.2 \times 10^{-10} \leq a_\mu^{\rm SUSY} \leq 34.0 \times 10^{-10}$  and $ 10.4 \times 10^{-10} \leq a_\mu^{\rm SUSY} \leq 39.8 \times 10^{-10}$, respectively. The upper right panel shows that the maximum reach of $\mu_{tot}$ decreases monotonously with the increase of $\tan \beta$, and it is about $500~{\rm GeV}$ ($260~{\rm GeV}$) for $\tan \beta = 10$ ($\tan \beta = 60$). The reason for such a behavior is that, in the case of a relatively small $\tan \beta$, the second term in ${\cal M}^2_{S, 23}$ of Eq.~(\ref{Mass-CP-even-Higgs}) is sizable, and can cancel the first term to suppress $V_h^S$, which is preferred by LHC Higgs data. As $\tan \beta$ increases, the cancellation effect becomes weak since the second term is suppressed by $\sin 2 \beta$, and tighter constraints are set on $\mu_{tot}$\footnote{Throughout this work, $A_\lambda$ is fixed at $2~{\rm TeV}$. If a larger $A_\lambda$, e.g., $A_\lambda = 10~{\rm TeV}$, was taken, it was found that  $\tan \beta$ tended to become larger, while $|\mu_{tot}|$ and $\lambda$ tended to be smaller~\cite{Cao:2018iyk}. This tendency is needed to suppress ${\cal M}^2_{S, 23}$ and ${\cal M}^2_{S, 33}$ in Eq.(\ref{Mass-CP-even-Higgs}) simultaneously. In addition, the Bayesian evidence of the scenario decreases significantly as $A_\lambda$ increases~\cite{Cao:2018iyk}, which means that setting a large $A_\lambda$ will cause a more subtle parameter tuning to obtain $m_{h_1} \lesssim 125~{\rm GeV}$ and correct electroweak symmetry breaking. In brief, even when $A_\lambda$ is treated as a variable in studying the parameter space, the natural realization of the $h_2$ scenario to interpret the anomaly in the GNMSSM, as suggested by this work, has been tightly limited. This conclusion was verified by our alternative scans.}.
Moreover, analyzing the posterior probability of the scan results indicates that the scenario prefers small $\tan \beta$ region. Thus, most samples obtained in the scan predict $\tan \beta \lesssim 30$.

The lower left and right panels of Fig.~{\ref{fig:plamm}} depict the ranges of $M_2$, $\tilde{\mu}_L$, and $\tilde{\mu}_R$, which are determined by $a_\mu^{\rm SUSY}$ in Eqs.~(\ref{eq:WHL}-\ref{eq:BLR}). They show that $M_2$ may be as large as $1.5~{\rm TeV}$, and $\tilde{\mu}_L$ and $\tilde{\mu}_R$ may be as large as $1~{\rm TeV}$. The lower left panel also exhibits that the mass of chargino $\tilde{\chi}_1^\pm$ is less than about $350~{\rm GeV}$. It should be noted that the ranges of $M_2$ and $\tilde{\mu}_L$ depend strongly on the value of $\tan \beta$. For example, assuming that the theory explains the discrepancy of $\Delta a_\mu$ at $1\sigma$ level, it was found that $M_2$ and $\tilde{\mu}_L$ must be less than about $400~{\rm GeV}$ and $350~{\rm GeV}$, respectively, for $\tan \beta = 10$. The upper bounds become $1.2~{\rm TeV}$ and $700~{\rm GeV}$ for $\tan \beta = 20$, and $1.4~{\rm TeV}$ and $1~{\rm TeV}$ for $\tan \beta = 27$. By contrast, $m_{\tilde{\chi}_1^0}$ and $\tilde{\mu}_R$ are not sensitive to $\tan \beta$, e.g., $m_{\tilde{\chi}_1^0}$ and $\tilde{\mu}_R$ may vary in the range of $50~{\rm GeV} \lesssim m_{\tilde{\chi}_1^0} \lesssim 250~{\rm GeV}$ and $100~{\rm GeV} \lesssim \tilde{\mu}_R \lesssim 1~{\rm TeV}$ for any value of $\tan \beta$. The basic reason for the phenomenon is that the WHL contribution to $a_\mu^{\rm SUSY}$ is usually the dominant one. It depends on $M_2$, $\mu_{tot}$, and $\tilde{\mu}_L$, and is in particular proportional to $\tan \beta$. Therefore, when $\tan \beta$ is relatively small, the invovled SUSY particles must be moderately light to predict a sizable $a_\mu^{\rm SUSY}$. As a result, the left sides of the lower panels usually correspond to a relatively small $\tan \beta$, and the right sides correspond to a large $\tan \beta$.

Fig.~{\ref{fig:dm}}  focuses on the DM physics of the $h_2$ scenario, which involves the parameters $\lambda$, $\kappa$, $\mu_{tot}$, and the masses of singlet-dominated particles, i.e., $m_{\tilde{\chi}_1^0}$, $m_{h_s}$ and $m_{A_s}$. It reveals the following features:
\begin{itemize}
\item $2m_{\tilde\chi^0_1} > m_{h_s} + m_{A_s}$ for most of the parameter areas (see the upper right panel), which implies that  in the early universe, the Singlino-dominated DM might annihilate into the singlet-dominated Higgs bosons $h_s$ and $A_s$. As pointed out in~\cite{Cao:2021ljw}, this annihilation proceeded by the s-channel exchange of $Z$ boson and CP-odd Higgs bosons and the $t$-channel exchange of neutralinos. If the $t$-channel contribution to the annihilation rate was much larger than the $s$-channel contribution, $|\kappa| \simeq 0.15 \times (m_{\tilde{\chi}_1^0}/300~{\rm GeV})^{1/2}$ could predict the measured abundance, while if the interference of the two contributions was significantly constructive/deconstructive, a smaller/larger $|\kappa|$ could be fully responsible for the abundance. Given that $ 0.05 \lesssim |\kappa| \lesssim 0.25$ on the upper left panel, we infer that the process played an important role in determining the abundance. In addition, it was verified in fewer cases that $2m_{\tilde\chi^0_1} < m_{h_s} + m_{A_s}$ and/or $|\kappa| \lesssim 0.1$, so that the DM obtained the measured abundance mainly by co-annihilating with the Higgsino-dominated electroweakinos or $\mu$-type sleptons.
\item The SI and SD cross-sections of DM-nucleon scattering may be as low as $10^{-50}~{\rm cm}$ and $10^{-45}~{\rm cm}$, respectively (see the lower left and right panels). The SD scattering proceeds only through the $Z$-mediated Feynman diagram, and the rate is proportional to $(\lambda v/\mu_{tot})^4$~\cite{Cao:2021ljw}. Thus, it is a small $\lambda$, e.g., $\lambda \sim {\cal O}(0.01)$, that is responsible for the low SD cross-section (see the upper left panel). By contrast, the SI scattering is induced by three CP-even Higgs bosons, and it is the cancellation of $h$- and $h_s$-mediated contributions that mainly accounts for the small SI cross section~\cite{Zhou:2021pit}.
\end{itemize}

\begin{figure}[t]
	\centering
	\includegraphics[width=0.9\textwidth]{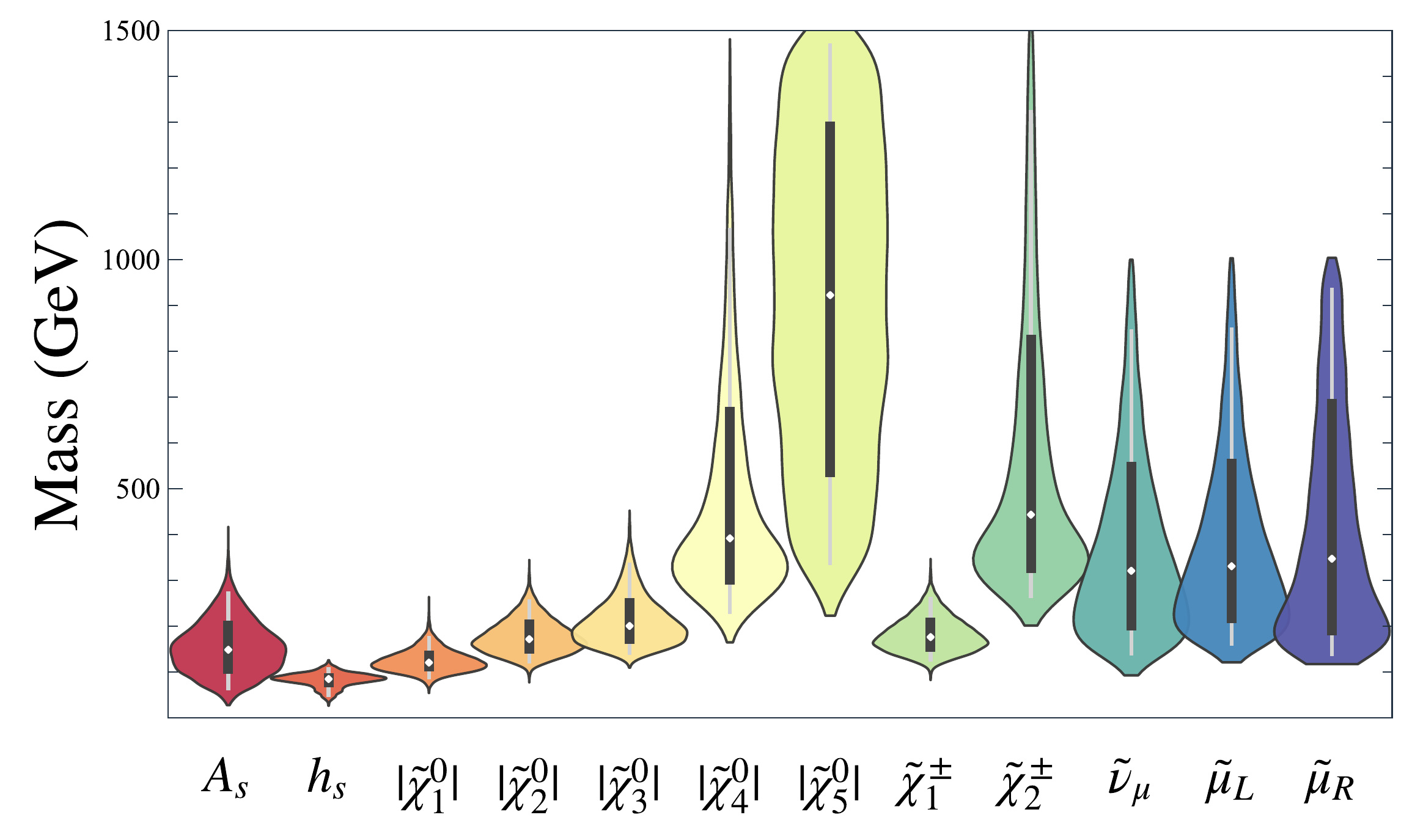}
	\caption{\label{fig:smassviolin} Violin diagrams showing the mass distributions of the singlet Higgs bosons, the electoweakinos, and $\mu$-type sparticles. Smuons are labeled by their dominated component. The violins are scaled by count. The thick vertical bar in the center represents the interquartile range with the white dot denoting the median, and the long vertical line indicates the $95\%$ confidence interval.}
\end{figure}

The DM physics in the $h_2$ scenario differs from those of the $h_1$ scenario, which were presented in Fig.2 of~\cite{Cao:2021tuh}, in three aspects. The first is that the DM is relatively light, i.e., $50~{\rm GeV} \lesssim m_{\tilde{\chi}_1^0} \lesssim 250~{\rm GeV}$ for the $h_2$ scenario and $150~{\rm GeV} \lesssim m_{\tilde{\chi}_1^0} \lesssim 650~{\rm GeV}$ for the $h_1$ scenario. Two reasons may explain this phenomenon. One is that $|m_{\tilde{\chi}_1^0}|$ must be less than $\mu_{tot}$, and as shown in Fig.~{\ref{fig:plamm}}, a moderately small $\mu_{tot}$ is experimentally preferred for the $h_2$ scenario. The other reason is that $|m_{\tilde{\chi}_1^0}|$ must be larger than $(m_{h_s} + m_{A_s})/2$ for most cases to proceed to the annihilation $\tilde{\chi}_1^0 \tilde{\chi}_1^0 \to h_s A_s$. A relatively light $\tilde{\chi}_1^0$ can meet this condition in the $h_2$ scenario (see the previous discussions). The second one is that $|\kappa|$ is less than 0.25 in the $h_2$ scenario, while it is less than 0.4 in the $h_1$ scenario. The underlying reason for this is that a smaller $|\kappa|$ can be fully responsible for the measured abundance in the $h_2$ scenario. The last aspect is that $\lambda$ in the $h_2$ scenario may reach about 0.2, while it is at most 0.1 in the $h_1$ scenario. This is because the cancellation effect in the SI scattering is usually significant in the $h_2$ scenario, and consequently, a larger $\lambda$ is still allowed by DM DD experiments.

In Fig.~\ref{fig:smassviolin}, the mass distributions of the singlet-dominated Higgs states and the SUSY particles relevant to $a_\mu^{\rm SUSY}$ are shown by a series of violin plots, which combines the advantages of the box plot and probability density distribution plot~\cite{hintze1998violin}. This figure shows that all SUSY particles except for $\tilde{\chi}_5^0$ tend to be lighter than $500~{\rm GeV}$, and in particular, $\tilde{\chi}_2^0$, $\tilde{\chi}_3^0$, and $\tilde{\chi}_1^\pm$ are lighter than $500~{\rm GeV}$ for nearly all samples obtained in the scan.  The fundamental reason for the phenomenon, besides the explanation presented before, arises from the fact that a low $\tan \beta$ is preferred to predict the $h_2$ scenario. This tendency, once combined with the requirement of a sizable $a_\mu^{\rm SUSY}$, will necessitate light SUSY particles\footnote{Without the $a_\mu^{\rm SUSY}$ requirement, $\tilde{\chi}_1^0$ may be very massive (e.g., $|m_{\tilde{\chi}_1^0}| > 300~{\rm GeV}$~\cite{Cao:2021ljw}). In this case, the LHC constraints are significantly weakened.}.  Given that these electroweakinos can be richly produced at the LHC, they have been restricted by searching for multi-lepton signals. This issue will be intensively studied in the following.

\subsection{\label{LHCMC}LHC constraints}

To comprehensively study the constraints from the LHC search for sparticles on the obtained parameter points, the following processes were analyzed in the Monte Carlo (MC) event simulation as given by Equations~(\ref{MC-processes})\footnote{In the compressed spectrum case, relevant processes with additional jets were also considered in the simulations.}:
\begin{equation}  \label{MC-processes} \begin{split}
pp \to \tilde{\chi}_i^0\tilde{\chi}_j^{\pm} &, \quad i = 2, 3, 4, 5; \quad j = 1, 2 \\
pp \to \tilde{\chi}_i^{\pm}\tilde{\chi}_j^{\mp} &, \quad i,j = 1, 2; \\
pp \to \tilde{\chi}_i^{0}\tilde{\chi}_j^{0} &, \quad i,j = 2, 3, 4, 5; \\
pp \to \tilde{\mu}_i \tilde{\mu}_j &,\quad i,j = 1, 2;
\end{split}\end{equation}
In the calculation, the cross-sections of $\sqrt{s}$ = 13 TeV were obtained at the next-to-leading order (NLO) by the package \texttt{Prospino2}~\cite{Beenakker:1996ed}. The MC events were generated by the package \texttt{MadGraph\_aMC@NLO}~\cite{Alwall:2011uj, Conte:2012fm} with the code \texttt{PYTHIA8}~\cite{Sjostrand:2014zea} for parton showers, hadronizations, and sparticle decays. The event files were finally input into the package \texttt{CheckMATE-2.0.29} with the code \texttt{Delphes}~\cite{deFavereau:2013fsa} for detector simulation.

\begin{table}[]
	\caption{Experimental analyses considered in this work. Some of them were implemented in \texttt{CheckMATE-2.0.29} by us. In particular, the validation of the very recent analysis, ATLAS-2106-01676~\cite{ATLAS:2021moa}, was presented in Appendix B of this work.}
	\label{tab:analyses}
	\vspace{0.3cm}
	\resizebox{1.0\textwidth}{!}{
		\begin{tabular}{llll}
			\hline\hline
			\texttt{Name} & \texttt{Scenario} &\texttt{Final State} &$\texttt{Luminosity} (\texttt{fb}^{\texttt{-1}})$ \\\hline
			\texttt{ATLAS-1909-09226}~\cite{Aad:2019vvf}                  & $\tilde{\chi}_{2}^0\tilde{\chi}_1^{\pm}\rightarrow Wh\tilde{\chi}_1^0\tilde{\chi}_1^0$ & $1\ell + h(h\rightarrow bb) + \text{E}_\text{T}^{\text{miss}}$                  & 139                  \\\\
			\texttt{CMS-SUS-20-001}~\cite{CMS:2020bfa}& $\tilde{\chi}_{2}^0\tilde{\chi}_1^{\pm}\rightarrow WZ\tilde{\chi}_1^0\tilde{\chi}_1^0$                            & $2\ell + nj(n\textgreater{}0) + \text{E}_\text{T}^{\text{miss}}$    & 137                  \\\\
			\texttt{ATLAS-1912-08479}~\cite{Aad:2019vvi}&$\tilde{\chi}_2^0\tilde{\chi}_1^{\pm}\rightarrow W\tilde{\chi}_1^0Z\tilde{\chi}_1^0$& $3\ell + \text{E}_\text{T}^{\text{miss}}$                           & 139 \\\\
			\multirow{2}{*}{\texttt{ATLAS-1908-08215}~\cite{Aad:2019vnb}} &$\tilde{\ell}\tilde{\ell}\rightarrow \ell\tilde{\chi}_1^0\ell\tilde{\chi}_1^0$& \multirow{2}{*}{$2\ell + \text{E}_{\text{T}}^{\text{miss}}$} & \multirow{2}{*}{139} \\
			&$\tilde{\chi}_1^{\pm}\tilde{\chi}_1^{\mp}(\tilde{\chi}_1^{\pm}\rightarrow \tilde{\ell}\nu/\tilde{\nu}\ell)$&                  &                    \\\\
			\texttt{ATLAS-2106-01676}~\cite{ATLAS:2021moa}                &$\tilde{\chi}_2^0\tilde{\chi}_1^{\pm}\rightarrow W^{(*)}Z^{(*)}\tilde{\chi}_1^0\tilde{\chi}_1^0$,$Wh\tilde{\chi}_1^0\tilde{\chi}_1^0$&$3\ell + \text{E}_\text{T}^{\text{miss}}$ & 139 \\\\
			\multirow{3}{*}{\texttt{ATLAS-1803-02762}~\cite{ATLAS:2018ojr}}                &$\tilde{\chi}_2^0\tilde{\chi}_1^{\pm}\rightarrow WZ\tilde{\chi}_1^0\tilde{\chi}_1^0$,$\nu\tilde{\ell}l\tilde{\ell}$&\multirow{3}{*}{n$ \ell$ (n\textgreater{}=2) + $E_{\rm T}^{\rm miss}$}&\multirow{3}{*}{36.1} \\&$\tilde{\chi}_1^{\pm}\tilde{\chi}_1^{\mp}\rightarrow \nu\tilde{\ell}\nu\tilde{\ell}$&\\& $ \tilde{\ell}\tilde{\ell}\rightarrow\ell\tilde{\chi}_1^0\ell\tilde{\chi}_1^0$  \\\\
			\multirow{5}{*}{\texttt{ATLAS-1802-03158}~\cite{ATLAS:2018nud}}                &$\tilde{g}\tilde{g}\rightarrow 2q\tilde{\chi}_1^0 2q\tilde{\chi}_1^0(\rightarrow \gamma \tilde{G})$&\multirow{5}{*}{n$ \gamma$ (n\textgreater{}=1) + nj(n\textgreater{}=0) + $E_{\rm T}^{\rm miss}$}&\multirow{5}{*}{36.1} \\&$\tilde{g}\tilde{g}\rightarrow 2q\tilde{\chi}_1^0(\rightarrow \gamma \tilde{G}) 2q\tilde{\chi}_1^0(\rightarrow Z \tilde{G})$\\&$\tilde{q}\tilde{q}\rightarrow q\tilde{\chi}_1^0(\rightarrow \gamma \tilde{G}) q\tilde{\chi}_1^0(\rightarrow \gamma \tilde{G})$&\\&$\tilde{\chi}_2^0\tilde{\chi}_1^{\pm}\rightarrow Z/h\tilde{\chi}_1^0 W\tilde{\chi}_1^0$\\& $\tilde{\chi}_1^{\pm}\tilde{\chi}_1^{\pm}\rightarrow W\tilde{\chi}_1^0 W\tilde{\chi}_1^0$  \\\\
			\multirow{3}{*}{\texttt{ATLAS-1712-08119}~\cite{ATLAS:2017vat}} &$\tilde{\ell}\tilde{\ell}\rightarrow \ell\tilde{\chi}_1^0 \ell\tilde{\chi}_1^0$& \multirow{3}{*}{$2\ell + nj(n\textgreater{}=0) + \text{E}_\text{T}^{\text{miss}}$} & \multirow{3}{*}{36.1} \\
			&$(\text{Wino})\tilde{\chi}_2^0\tilde{\chi}_1^{\pm}\rightarrow WZ\tilde{\chi}_1^0\tilde{\chi}_1^0$&                    &                    \\
			&$(\text{Higgsino})\tilde{\chi}_2^0\tilde{\chi}_1^{\pm}$ + $\tilde{\chi}_1^{+}\tilde{\chi}_1^{-}$ + $\tilde{\chi}_2^0\tilde{\chi}_1^0$&                    &                    \\\\
			\multirow{2}{*}{\texttt{CMS-SUS-17-004}~\cite{Sirunyan:2018ubx}} &$\tilde{\chi}_2^0\tilde{\chi}_1^{\pm}\rightarrow WZ\tilde{\chi}_1^0\tilde{\chi}_1^0$,$ WH\tilde{\chi}_1^0\tilde{\chi}_1^0$& \multirow{2}{*}{$n\ell(n\textgreater{}0) + \text{E}_\text{T}^{\text{miss}}$}&\multirow{2}{*}{35.9} \\
			&$\tilde{\chi}_1^0\tilde{\chi}_1^{0}\rightarrow ZZ\tilde{G}\tilde{G}$,$HZ\tilde{G}\tilde{G}$,$HH\tilde{G}\tilde{G}$& &\\\\
			\multirow{3}{*}{\texttt{CMS-SUS-16-039}~\cite{CMS:2017moi}}&$\tilde{\chi}_2^0\tilde{\chi}_1^{\pm}\rightarrow \nu\tilde{\ell}\ell\tilde{\ell}$,$\nu\tilde{\nu}\ell\tilde{\nu}$,$\tilde{\tau}\nu\tilde{\ell}\ell$,$\tilde{\tau}\nu\tilde{\tau}\tau$&\multirow{3}{*}{$n\ell(n\textgreater{}0) + n\tau(n\textgreater{}=0) + \text{E}_\text{T}^{\text{miss}}$}&\multirow{3}{*}{35.9} \\ &$\tilde{\chi}_2^0\tilde{\chi}_1^{\pm}\rightarrow WZ\tilde{\chi}_1^0\tilde{\chi}_1^0$,$WH\tilde{\chi}_1^0\tilde{\chi}_1^0$\\&$\tilde{\chi}_1^0\tilde{\chi}_1^{0}\rightarrow ZZ\tilde{G}\tilde{G}$,$HZ\tilde{G}\tilde{G}$,$HH\tilde{G}\tilde{G}$& \\\\
			\multirow{3}{*}{\texttt{CMS-SUS-16-048}~\cite{CMS:2018kag}} &$\tilde{t}\tilde{t}\rightarrow b\tilde{\chi}_1^{\pm}b\tilde{\chi}_1^{\pm}$&\multirow{3}{*}{$n\ell(n\textgreater{}=0)+nb(n\textgreater{}=0)+ \text{E}_\text{T}^{\text{miss}}$} & \multirow{3}{*}{35.9} \\
			&$\tilde{\chi}_2^0\tilde{\chi}_1^{\pm}\rightarrow W^{*}Z^{*}\tilde{\chi}_1^0\tilde{\chi}_1^0$&                  & \\
			&$(\text{Higgsino})\tilde{\chi}_2^0\tilde{\chi}_1^{\pm}/\tilde{\chi}_1^0$&&    \\\\
			\multirow{3}{*}{\texttt{CMS-SUS-PAS-16-025}~\cite{CMS:2016zvj}} &$\tilde{t}\tilde{t}\rightarrow b\tilde{\chi}_1^{\pm}b\tilde{\chi}_1^{\pm}$& \multirow{1}{*}{$n\ell(n\textgreater{}=0) + nb(n\textgreater{}=0)$} & \multirow{3}{*}{12.9} \\
			&$\tilde{\chi}_2^0\tilde{\chi}_1^{\pm}\rightarrow W^{*}Z^{*}\tilde{\chi}_1^0\tilde{\chi}_1^0$&\multirow{1}{*}{$ + nj(n\textgreater{}=0) + \text{E}_\text{T}^{\text{miss}}$}&                    \\\\
			&$(\text{Higgsino})\tilde{\chi}_2^0\tilde{\chi}_1^{\pm}/\tilde{\chi}_1^0$&&    \\\\
			\multirow{2}{*}{\texttt{ATLAS-CONF-2016-096}~\cite{ATLAS:2016uwq}} &$\tilde{\chi}_1^{\pm}\tilde{\chi}_1^{\mp}(\tilde{\chi}_1^{\pm}\rightarrow \tilde{\ell}\nu/\tilde{\nu}\ell)$& \multirow{2}{*}{$n\ell(n\textgreater{}=2) + \text{E}_{\text{T}}^{\text{miss}}$} & \multirow{2}{*}{13.3} \\\\
			&$\tilde{\chi}_1^{\pm}\tilde{\chi}_2^{0}(\tilde{\chi}_1^{\pm}\rightarrow \tilde{\ell}\nu/\tilde{\nu}\ell, \tilde{\chi}_2^{0}\rightarrow \tilde{\ell}\ell/\tilde{\nu}\nu)$&                  &                    \\\\
			\hline\hline
	\end{tabular}}
\end{table}

\begin{figure}[t]
	\centering
	\includegraphics[width=0.495\textwidth]{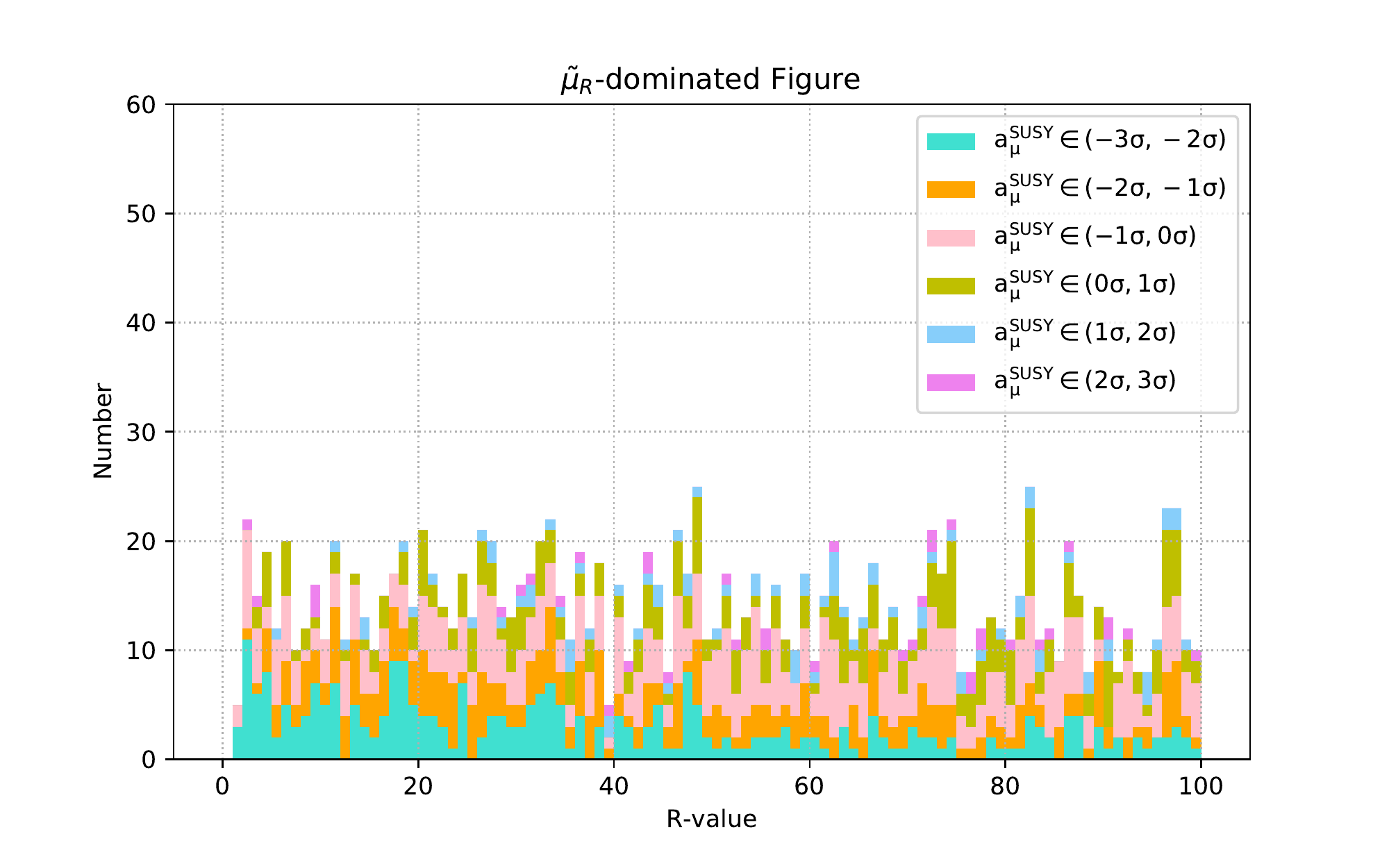}
	\includegraphics[width=0.495\textwidth]{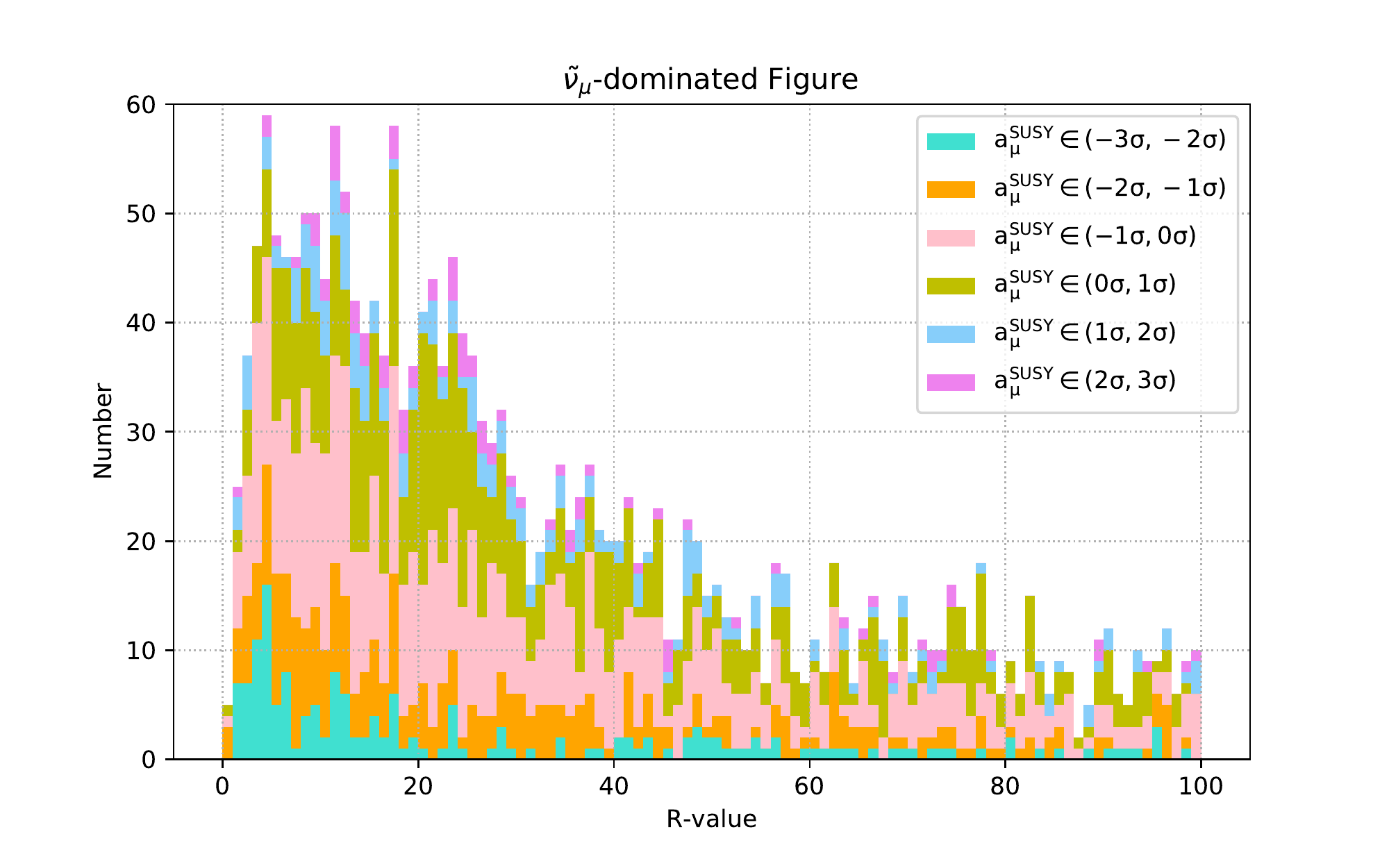}\\
    \includegraphics[width=0.495\textwidth]{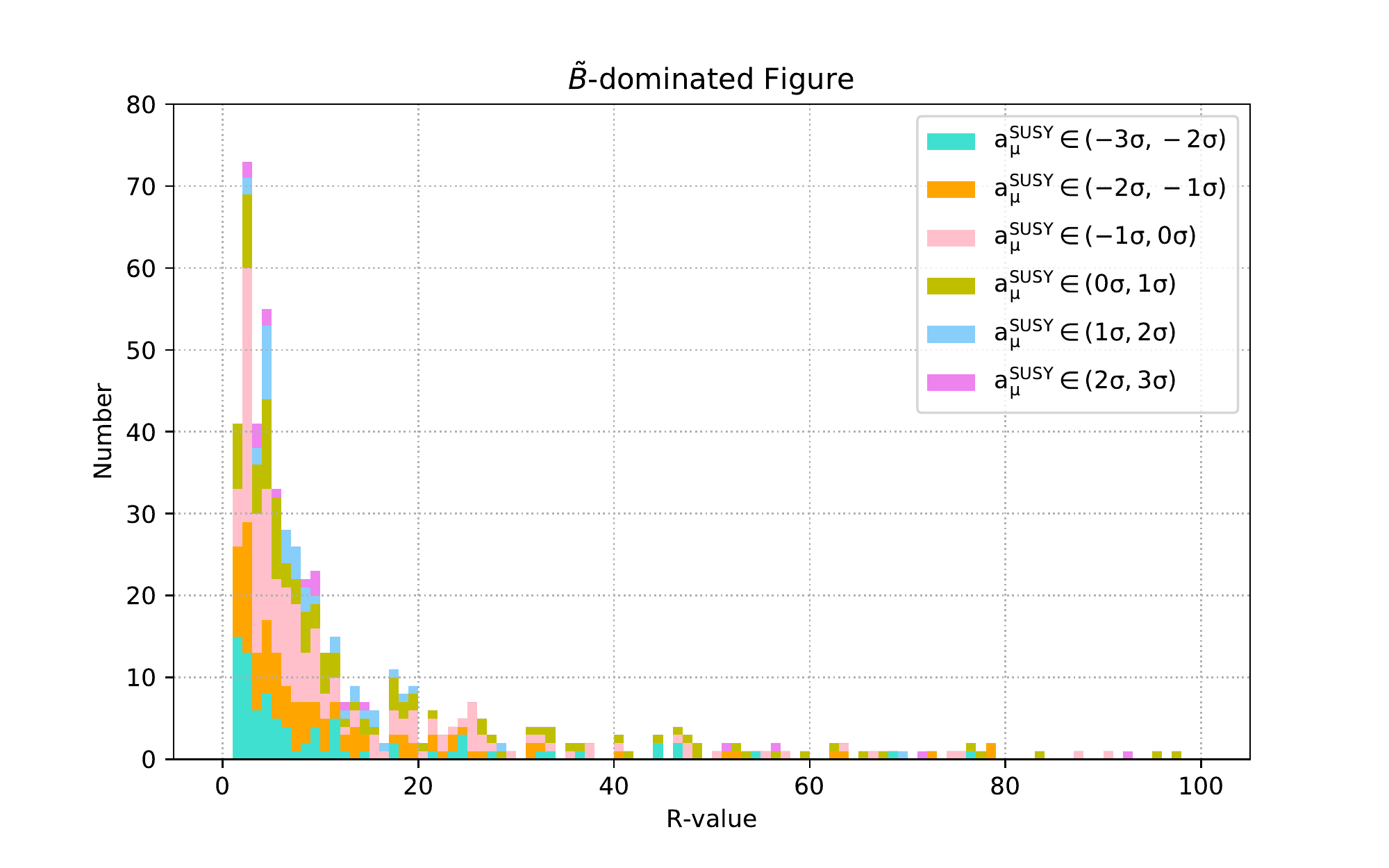}
	\includegraphics[width=0.495\textwidth]{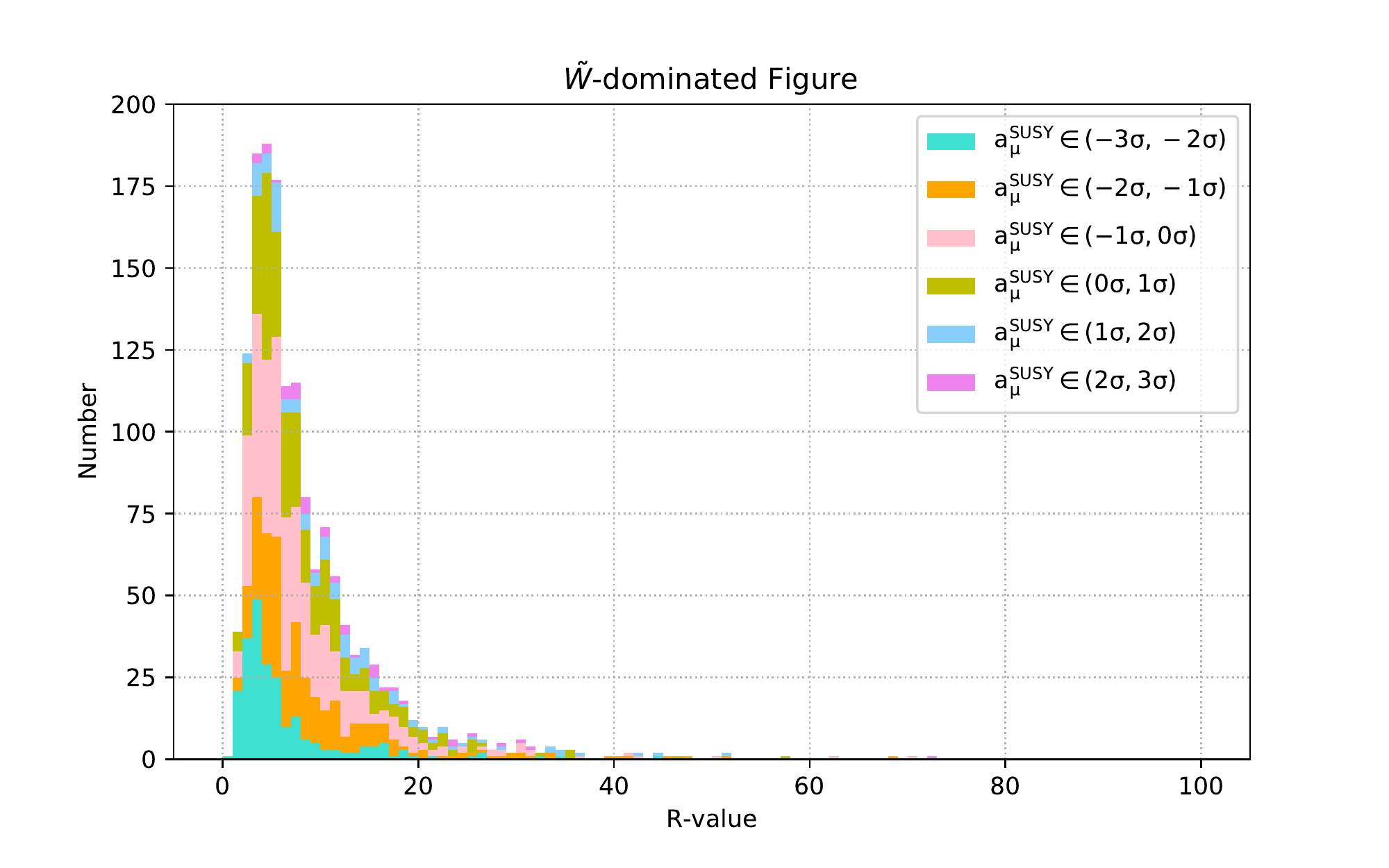}\\
	\includegraphics[width=0.495\textwidth]{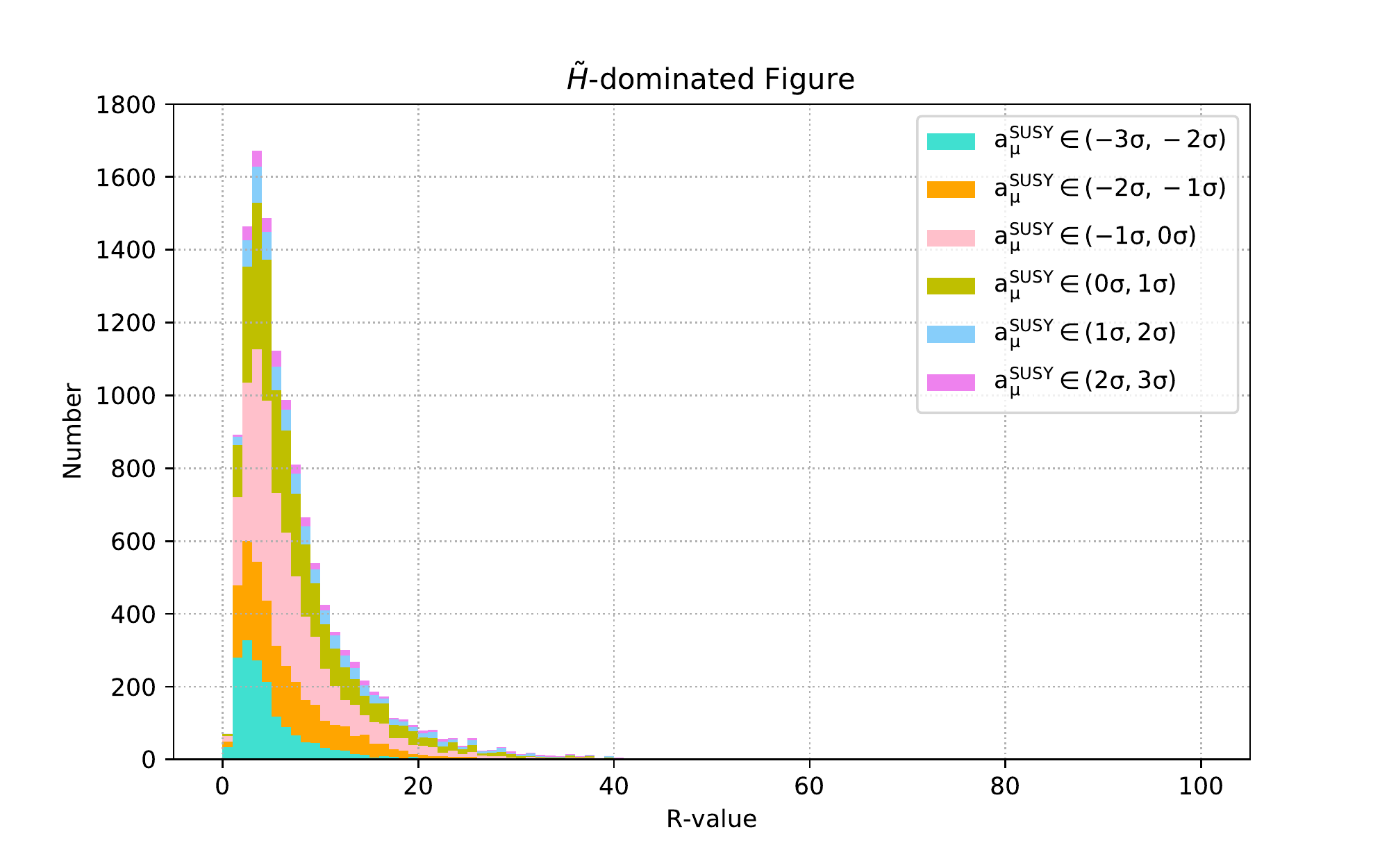}
	\includegraphics[width=0.495\textwidth]{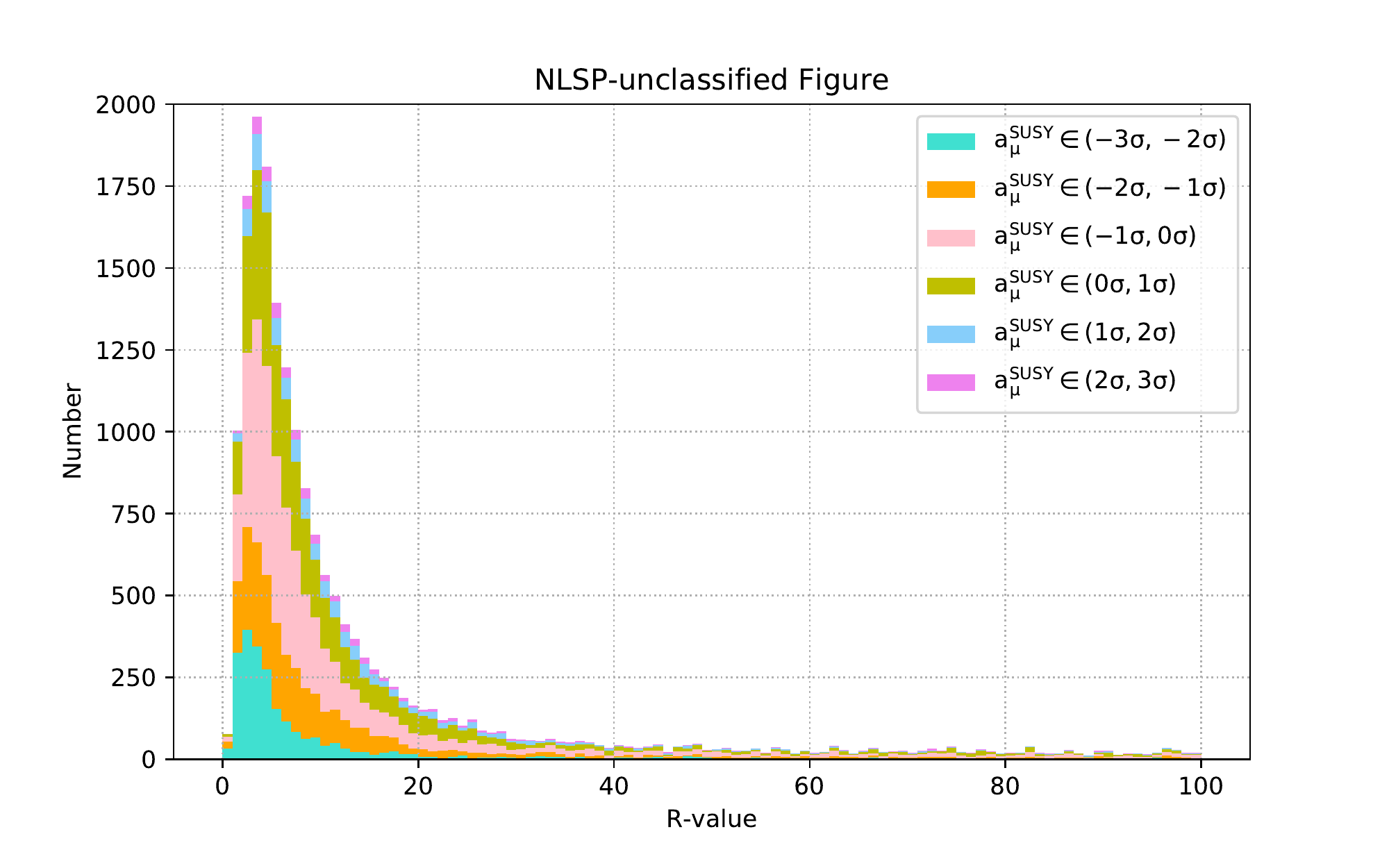} \\
	\caption{\label{fig:Rck}The Histograms of R-value distribution for the different NLSP types. Panels from left to right and top to bottom correspond to the results for the points that NLSP are $\tilde{\mu}_R$,  $\tilde{\nu}_\mu$-, $\tilde{B}$-, $\tilde{W}$-, and $\tilde{H}$-dominated, respectively. The last panel shows the distribution for all the points. The magnitude of $a^{\rm SUSY}_{\mu}$ is expressed in terms of its deviation from the FNAL measurement center value and is shown by different colors: turquoise for $(-3 \sigma, -2 \sigma)$, orange for $(-2 \sigma, -1 \sigma)$, pink for $(-1 \sigma, 0 \sigma)$, green for $(0 \sigma, 1 \sigma)$, blue for $(1 \sigma, 2 \sigma)$, and violet for $(2 \sigma, 3 \sigma)$.}
\end{figure}

For each point, $10^6$ MC events were generated in the simulations, and the LHC analyses listed in Table~\ref{tab:analyses} were used to test it. In particular, the following LHC analyses were included in our study, which played a crucial role in constraining the scenario:
\begin{enumerate}
	\item The search with the ATLAS detector for chargino and slepton pair production with two lepton final states in $\sqrt{s} = 13~{\rm TeV}$ pp collisions (Report No. CERN-EP-2019-106)~\cite{Aad:2019vnb}.
	\item The search with the ATLAS detector for chargino-neutralino pair production with involved mass splittings near the electroweak scale in three-lepton final states in pp collisions at $\sqrt{s} = 13~{\rm TeV}$ (Report No. CERN-EP-2019-263)~\cite{Aad:2019vvi}.
	\item The search with the ATLAS detector for the direct production of electroweakinos in final states with one lepton, missing transverse momentum and with a Higgs boson decaying into two b-jets in pp collisions at $\sqrt{s} = 13~{\rm TeV}$ (Report No. CERN-EP-2019-188)~\cite{Aad:2019vvf}.
	\item The search with the ATLAS detector for chargino-neutralino pair production in final states with three leptons and missing transverse momentum in $\sqrt{s} = 13~{\rm TeV}$ pp collisions (Report No. CERN-EP-2021-059 )~\cite{ATLAS:2021moa}.
	\item The combined search with the CMS detector for charginos and neutralinos (Report No. CMS-SUS-17-004)~\cite{Sirunyan:2018ubx}.
	\item The search with the CMS setector for final states with two oppositely charged same-flavor leptons, jets, and missing transverse momentum in pp collisions at $\sqrt{s} = 13~{\rm TeV}$ (Report No. CMS-SUS-20-001)~\cite{CMS:2020bfa}.
\end{enumerate}

The quantity $R$ was used to describe the LHC's limitation on the samples in the discussion. It is defined by $R \equiv \max\{S_i/S_{{\rm obs},i}^{95}\}$, where $S_i$ stands for the number of simulated events in the $i$-th signal region (SR) of all the included analyses, and $S_{{\rm obs},i}^{95}$ represents corresponding observed 95$\%$ confidence level upper limit. Accordingly, without considering the involved uncertainties, $R > 1$ represents that the considered parameter point is excluded due to the inconsistency with the LHC limit. Otherwise, it is allowed by the LHC searches~\cite{Domingo:2018ykx}.

The collider simulation results given by \texttt{CheckMATE} implied that the LHC searches for SUSY set strong restrictions on the $h_2$ scenario of the $\mu$NMSSM. In order to display the analysis results clearly, the points were classified by NLSP's dominated component, which may be $\tilde{\mu}_R$, $\tilde{\nu}_\mu$, $\tilde{B}$, $\tilde{W}$, or $\tilde{H}$. The Histograms of R-value distribution for the different NLSP types were displayed in Fig.~{\ref{fig:Rck}} from left to right and top to bottom, respectively. Points colored by turquoise, orange, pink, green, blue, and violet correspond respectively to the cases that $a^{\rm SUSY}_{\mu}$ are in the range of $(-3 \sigma, -2 \sigma)$, $(-2 \sigma, -1 \sigma)$,  $(-1 \sigma, 0 \sigma)$, $(0 \sigma, 1 \sigma)$, $(1 \sigma, 2 \sigma)$ and $(2 \sigma, 3 \sigma)$. The results were also summarized in Table~{\ref{tab:mcsum}}, which includes the number of samples obtained by the scan (denoted $N_{\rm tot}$), that satisfying $R < 1$ (denoted by $N_{\rm pass}$), and the more detailed classification of $N_{tot}$ and $N_{\rm pass}$ by the ranges of $a_{\mu}^{\rm SUSY}$.

\begin{table}[t]
\centering
\begin{tabular}{|c|cc|cc|cc|cc|cc|}
\hline\hline
$\rm NLSP$ & \multicolumn{2}{|c|}{$\tilde{H}$} & \multicolumn{2}{|c|}{$\tilde{B}$}	& \multicolumn{2}{|c|}{$\tilde{W}$} & \multicolumn{2}{|c|}{$\tilde{\mu}_R$} & \multicolumn{2}{|c|}{$\tilde{\nu}_\mu$}   \\ \hline
$N_{\rm tot}$ \qquad$\mid$\qquad$N_{\rm pass}$& 15356 & 221 & 756 & 12 & 3116 & 4 & 3304 & 2 & 3408 & 9 \\
$N_{a_{\mu}^{\rm SUSY}} \in \left(-3 \sigma, -2 \sigma \right)$ & 1728 & 33 & 92 & 0 & 364 & 1 & 406 & 0 &  271 & 3 \\
$N_{a_{\mu}^{\rm SUSY}} \in \left(-2 \sigma, -1 \sigma \right)$ & 2370 & 17 & 126 & 0 & 518 & 0 & 527 & 0 & 449 & 2 \\
$N_{a_{\mu}^{\rm SUSY}} \in \left(-1 \sigma, ~~0 \sigma \right)$ & 4380 & 15 & 196 & 0 & 926 & 0 & 899 & 0 & 1059 & 1 \\
$N_{a_{\mu}^{\rm SUSY}} \in \left(~~0 \sigma, ~~1 \sigma \right)$ & 3497 & 6 & 135 & 0 & 721 & 0 & 632 & 0 & 851 & 1 \\
$N_{a_{\mu}^{\rm SUSY}} \in \left(~~1 \sigma, ~~2 \sigma \right)$ & 982 & 0 & 48 & 0 & 200 & 0 & 216 & 0 & 242 & 0 \\
$N_{a_{\mu}^{\rm SUSY}} \in \left(~~2 \sigma, ~~3 \sigma \right)$ & 462 & 0 & 21 & 0 & 79 & 0 & 100 & 0 & 122 & 0 \\ \hline \hline
\end{tabular}
\caption{\label{tab:mcsum} Numbers of the samples classified by NLSP's dominant component. $N_{\rm tot}$ denotes the total number for each type samples, which were obtained by the scan and sequently surveyed by MC simulations, $N_{\rm pass}$ represents the number of the points satisfying $R < 1$, and $N_{a_{\mu}^{\rm SUSY}}$ corresponds to further classifications of $N_{\rm tot}$ and $N_{\rm pass}$ by the magnitude of $a_\mu^{\rm SUSY}$. It was verified that the $R$-values were always larger than 0.4, which means that all the samples are to be tested at high-luminosity LHC, and that all samples passing the LHC constraints were characterized by $|m_{\tilde{\chi}_1^0}| > 110~{\rm GeV}$. }
\end{table}
	
According to Table~{\ref{tab:mcsum}} and Figs.~{\ref{fig:Rck}}, the following conclusions are inferred:
\begin{itemize}
\item Among the five types of NLSP, the $\tilde{H}$-dominated NLSP is the easiest one for explaining the discrepancy in the $h_2$ scenario, and by contrast, the $\tilde{B}$-dominated NLSP is the least preferred one (see $N_{\rm tot}$ in Table \ref{tab:mcsum}).  The LHC restrictions are extremely strong in excluding parameter points for any type of NLSP (see $N_{\rm pass}$ in the table). In particular, they are strengthened significantly  once the scenario is required to explain the discrepancy at $3 \sigma$ level (see $N_{a_{\mu}^{\rm SUSY}}$ in the table). Specifically, it takes  dozens of parameter points with $\tilde{H}$-dominated NLSP and only few points with $\tilde{\nu}_\mu$-dominated NLSP to  interpret the discrepancy at the $3 \sigma$ level. This situation reflects the difficulty of the $h_2$ scenario in explaining the discrepancy. One fundamental reason comes from the fact that the scenario prefers a relatively small $\rm tan \beta$ and $\mu_{tot}$, and hence moderately light sparticles are predicted to obtain a sizable $a_\mu^{\rm SUSY}$.

    It was verified that, among the experimental analyses, the analysis 4 usually sets the tightest constraints. In the case that the parameter points predict sizable signals with four or more leptons, analysis 5 could also impose the strongest restriction.

\item In the case of $\tilde{\mu}_R$- or $\tilde{\nu}_{\mu}$-dominated NLSP, Wino- and Higgsino-dominated electroweakinos will decay mainly into leptonic final states via slepton and/or sneutrino, which proliferates the lepton signals. As a result, $R$ can reach 100 for lots of points, which is shown on the top left and right panels in Fig.~{\ref{fig:Rck}}. In addition, the LHC constraints on the $\tilde{\mu}_R$-dominated NLSP point are usually tighter than those on the $\tilde{\nu}_{\mu}$-dominated NLSP point because neutralinos will decay by $\tilde{\chi}_i^0 \to \tilde{\mu}^\pm \mu^\mp \to \tilde{\chi}_1^0 \mu^+ \mu^-$ for the former case and by $\tilde{\chi}_i^0 \to \tilde{\mu}^\pm \mu^\mp, \tilde{\nu}_\mu \nu \to \tilde{\chi}_1^0 \mu^+ \mu^-, \tilde{\chi}_1^0 \nu \nu$ for the latter case. The former case can produce more $\mu$ leptons.

\item In the case of $\tilde{B}$-dominated NLSP, although most parameter points correspond to $R < 20$, there are still a few points that predict $R > 80$. It was verified that the dominant decays of $\tilde{\chi}_2^0$ include $\tilde{\chi}_2^0 \to \tilde{\chi}_1^0 Z^{(\ast)}, \tilde{\chi}_1^0 h_{1,2}, \tilde{\chi}_1^0 \mu^+ \mu^-$, and smuons decay mainly by $\tilde{\mu}_{1,2} \to \mu \tilde{\chi}_2$ for most points. If the kinetics is allowed, other heavy sparticles prefer to decay dominantly into $\tilde{\chi}_2^0$ since they couple to $\tilde{\chi}_2^0$ by non-suppressed gauge couplings. For points with $R > 50$, $\tilde{\chi}_2^0 \to \tilde{\chi}_1^0 \mu^+ \mu^-$ is usually the largest decay channel of $\tilde{\chi}_2^0$, and the strongest constraints come from the analysis of four or more lepton signals in analysis 5.

\item In the case of $\tilde{W}$-dominated NLSP, most points predict $R < 20$, but in very rare case $R$ may reach 70. Detailed study indicated that $\tilde{\chi}_2^0$ decays mainly by $\tilde{\chi}_2^0 \to \tilde{\chi}_1^0 Z^{(\ast)}, \tilde{\chi}_1^0 h_{1,2}$ for most points, and $\tilde{\chi}_2^0 \to \tilde{\chi}_1^0 \mu^+ \mu^-, \tilde{\chi}_1^0 \nu \nu$ are the dominant decay only for a small portion of the points. It also indicated that $\tilde{\chi}_1^\pm$ decays mainly by $\tilde{\chi}_1^\pm \to \tilde{\chi}_1^0 W^{(*)}$ for nearly all points, and smuons decay in a complex way, e.g., any of the channels $\tilde{\mu} \to \tilde{\chi}_i^0 \mu~(i=1,\cdots, 5), \tilde{\chi}_{1,2}^- \nu$  may be the dominant decay.

    It is notable that $R$ in the $\tilde{W}$-dominated NLSP case can not be exceedingly large. This conclusion comes from the fact that the $\tilde{W}$-dominated electroweakinos are forbidden to decay into sleptons directly, and thus, even in the optimum case, the lepton signal from the decay $\tilde{\chi}_2^0 \to \tilde{\mu}^{\pm \ast} \mu^\mp \to \tilde{\chi}_1^0 \mu^+ \mu^-$ is not much larger than the other final states. Consequently, $p p \to \tilde{\chi}_2^0 \tilde{\chi}_1^\pm$, which is the largest sparticle production process, can not generate tri-lepton signal events efficiently. This feature results in a smaller signal rate than the $\tilde{B}$-dominated NLSP case, where the Wino-dominated electroweakinos may decay far dominantly into leptons.

\begin{table}[t]
	\caption{\label{tab:points}Detailed information of two benchmark points consistent with the DM and Higgs experiments. The point P1 is allowed by the LHC search for SUSY, while the point P2 has been excluded. Both of them predict $a_{\mu}^{\text{SUSY}}\simeq2.51\times 10^{-9}$. }
	\vspace{0.3cm}
	\resizebox{1.1\textwidth}{!}{
		\begin{tabular}{llll|llll}
			\hline\hline
			\multicolumn{4}{l|}{\texttt{Benchmark Point P1}}& \multicolumn{4}{l}{\texttt{Benchmark Point P2}}                                                                        \\ \hline
			\multicolumn{1}{l}{$\lambda$}     & \multicolumn{1}{r}{0.059} & \multicolumn{1}{l}{$\text{m}_{\text{h}_s}$}        &\multicolumn{1}{r|}{30.4 GeV}     & \multicolumn{1}{l}{$\lambda$}     & \multicolumn{1}{r}{0.094} & \multicolumn{1}{l}{$\text{m}_{\text{h}_s}$}        &\multicolumn{1}{r}{91.0 GeV}     \\
			\multicolumn{1}{l}{$\kappa$}     & \multicolumn{1}{r}{-0.12} & \multicolumn{1}{l}{$\text{m}_{\text{A}_s}$}        &\multicolumn{1}{r|}{209.3 GeV}     & \multicolumn{1}{l}{$\kappa$}     & \multicolumn{1}{r}{-0.17} & \multicolumn{1}{l}{$\text{m}_{\text{A}_s}$}        &\multicolumn{1}{r}{180.2 GeV}     \\
			\multicolumn{1}{l}{$\text{tan}\beta$}      & \multicolumn{1}{r}{24.99} & \multicolumn{1}{l}{$\text{m}_\text{h}$}         &\multicolumn{1}{r|}{124.7 GeV}     & \multicolumn{1}{l}{$\text{tan}\beta$}      & \multicolumn{1}{r}{12.75} & \multicolumn{1}{l}{$\text{m}_\text{h}$}         &\multicolumn{1}{r}{124.8 GeV}     \\
			\multicolumn{1}{l}{$\mu$}        & \multicolumn{1}{r}{170.5 GeV} & \multicolumn{1}{l}{$\text{m}_\text{H}$}         &\multicolumn{1}{r|}{1053 GeV}     & \multicolumn{1}{l}{$\mu$}        & \multicolumn{1}{r}{157.6 GeV} & \multicolumn{1}{l}{$\text{m}_\text{H}$}         &\multicolumn{1}{r}{1023 GeV}     \\
			\multicolumn{1}{l}{$\mu+\mu_{\text{eff}}$} & \multicolumn{1}{r}{200.5 GeV} & \multicolumn{1}{l}{$\text{m}_{\text{A}_\text{H}}$}        &\multicolumn{1}{r|}{1052 GeV}     & \multicolumn{1}{l}{$\mu+\mu_{\text{eff}}$} & \multicolumn{1}{r}{195.7 GeV} & \multicolumn{1}{l}{$\text{m}_{\text{A}_\text{H}}$}        &\multicolumn{1}{r}{1023 GeV}     \\
			\multicolumn{1}{l}{$\text{A}_t$}        & \multicolumn{1}{r}{-2284 GeV} & \multicolumn{1}{l}{$\text{m}_{\tilde{\chi}_1^0}$}        &\multicolumn{1}{r|}{-126.6 GeV}     & \multicolumn{1}{l}{$\text{A}_t$}        & \multicolumn{1}{r}{2078 GeV} & \multicolumn{1}{l}{$\text{m}_{\tilde{\chi}_1^0}$}        &\multicolumn{1}{r}{-136.5 GeV}     \\
			\multicolumn{1}{l}{$\text{A}_{\kappa}$}        & \multicolumn{1}{r}{231.6 GeV} & \multicolumn{1}{l}{$\text{m}_{\tilde{\chi}_2^0}$}        &\multicolumn{1}{r|}{194.6 GeV}     & \multicolumn{1}{l}{$\text{A}_{\kappa}$}        & \multicolumn{1}{r}{158.4 GeV} & \multicolumn{1}{l}{$\text{m}_{\tilde{\chi}_2^0}$}        &\multicolumn{1}{r}{156.6 GeV}     \\
			\multicolumn{1}{l}{$\text{M}_1$}        & \multicolumn{1}{r}{-767.1 GeV} & \multicolumn{1}{l}{$\text{m}_{\tilde{\chi}_3^0}$}        &\multicolumn{1}{r|}{-210.0 GeV}     & \multicolumn{1}{l}{$\text{M}_1$}        & \multicolumn{1}{r}{957.8 GeV} & \multicolumn{1}{l}{$\text{m}_{\tilde{\chi}_3^0}$}        &\multicolumn{1}{r}{-209.6 GeV}     \\
			\multicolumn{1}{l}{$\text{M}_2$}        & \multicolumn{1}{r}{429.0 GeV} & \multicolumn{1}{l}{$\text{m}_{\tilde{\chi}_4^0}$}        &\multicolumn{1}{r|}{468.1 GeV}     & \multicolumn{1}{l}{$\text{M}_2$}        & \multicolumn{1}{r}{217.8 GeV} & \multicolumn{1}{l}{$\text{m}_{\tilde{\chi}_4^0}$}        &\multicolumn{1}{r}{281.1 GeV}     \\
			\multicolumn{1}{l}{$\text{m}_{\text{L}}$}        & \multicolumn{1}{r}{524.7 GeV} & \multicolumn{1}{l}{$\text{m}_{\tilde{\chi}_5^0}$}        &\multicolumn{1}{r|}{-770.6 GeV}     & \multicolumn{1}{l}{$\text{m}_{\text{L}}$}        & \multicolumn{1}{r}{322.6 GeV} & \multicolumn{1}{l}{$\text{m}_{\tilde{\chi}_5^0}$}        &\multicolumn{1}{r}{961.1 GeV}     \\
			\multicolumn{1}{l}{$\text{m}_{\text{E}}$}        & \multicolumn{1}{r}{525.9 GeV} & \multicolumn{1}{l}{$\text{m}_{\tilde{\chi}_1^{\pm}}$}        &\multicolumn{1}{r|}{198.1 GeV}     & \multicolumn{1}{l}{$\text{m}_{\text{E}}$}        & \multicolumn{1}{r}{306.6 GeV} & \multicolumn{1}{l}{$\text{m}_{\tilde{\chi}_1^{\pm}}$}      &\multicolumn{1}{r}{160.6 GeV}     \\
			\multicolumn{1}{l}{$a_{\mu}^{\text{SUSY}}$}       & \multicolumn{1}{r}{$1.99 \times 10^{-9}$} & \multicolumn{1}{l}{$\text{m}_{\tilde{\chi}_2^{\pm}}$}        &\multicolumn{1}{r|}{468.6 GeV}     & \multicolumn{1}{l}{$a_{\mu}^{\text{SUSY}}$}       & \multicolumn{1}{r}{$2.51 \times 10^{-9}$} & \multicolumn{1}{l}{$\text{m}_{\tilde{\chi}_2^{\pm}}$}       &\multicolumn{1}{r}{284.6 GeV}     \\
			\multicolumn{1}{l}{${\Omega h}^2$}     & \multicolumn{1}{r}{0.085} & \multicolumn{1}{l}{$\text{m}_{\tilde{\mu}_{\text{L}}}$}      &\multicolumn{1}{r|}{531.4 GeV}     & \multicolumn{1}{l}{${\Omega h}^2$}     & \multicolumn{1}{r}{0.099} & \multicolumn{1}{l}{$\text{m}_{\tilde{\mu}_{\text{L}}}$}    &\multicolumn{1}{r}{329.1 GeV}     \\
			\multicolumn{1}{l}{$\sigma_{p}^{\text{SI}}$}   & \multicolumn{1}{r}{$1.40 \times 10^{-47} \text{cm}^{2}$} & \multicolumn{1}{l}{$\text{m}_{\tilde{\mu}_{\text{R}}}$}      &\multicolumn{1}{r|}{532.7 GeV}     & \multicolumn{1}{l}{$\sigma_{p}^{\text{SI}}$}   &\multicolumn{1}{r}{$4.93 \times 10^{-47}\text{cm}^{2}$}& \multicolumn{1}{l}{$\text{m}_{\tilde{\mu}_{\text{R}}}$}    &\multicolumn{1}{r}{493.8 GeV}     \\
			\multicolumn{1}{l}{$\sigma_{n}^{\text{SD}}$}   & \multicolumn{1}{r}{$4.30 \times 10^{-42}\text{cm}^2$} & \multicolumn{1}{l}{$\text{m}_{\tilde{\nu}_{\mu}}$}    &\multicolumn{1}{r|}{525.5 GeV}     & \multicolumn{1}{l}{$\sigma_{n}^{\text{SD}}$}   & \multicolumn{1}{r}{$4.41 \times 10^{-43}\text{cm}^2$} & \multicolumn{1}{l}{$\text{m}_{\tilde{\nu}_{\mu}}$}    &\multicolumn{1}{r}{319.0 GeV}     \\ \hline
			\multicolumn{2}{l}{$\text{N}_{11},\text{N}_{12},\text{N}_{13},\text{N}_{14},\text{N}_{15}$}           & \multicolumn{2}{l|}{0.005, 0.010, 0.046, 0.077, -0.996}                 & \multicolumn{2}{l}{$\text{N}_{11},\text{N}_{12},\text{N}_{13},\text{N}_{14},\text{N}_{15}$}           & \multicolumn{2}{l}{-0.005, 0.029, 0.09, 0.141, -0.986}                 \\
			\multicolumn{2}{l}{$\text{N}_{21},\text{N}_{22},\text{N}_{23},\text{N}_{24},\text{N}_{25}$}           & \multicolumn{2}{l|}{-0.032, -0.216, 0.709, -0.670, -0.021}                 & \multicolumn{2}{l}{$\text{N}_{21},\text{N}_{22},\text{N}_{23},\text{N}_{24},\text{N}_{25}$}           & \multicolumn{2}{l}{0.03, -0.586, 0.631, -0.506, -0.032}                 \\
			\multicolumn{2}{l}{$\text{N}_{31},\text{N}_{32},\text{N}_{33},\text{N}_{34},\text{N}_{35}$}           & \multicolumn{2}{l|}{0.053, 0.082, 0.697, 0.705, 0.088}                 & \multicolumn{2}{l}{$\text{N}_{31},\text{N}_{32},\text{N}_{33},\text{N}_{34},\text{N}_{35}$}           & \multicolumn{2}{l}{-0.024, 0.117, 0.682, 0.702, 0.166}                 \\
			\multicolumn{2}{l}{$\text{N}_{41},\text{N}_{42},\text{N}_{43},\text{N}_{44},\text{N}_{45}$}           & \multicolumn{2}{l|}{-0.008, 0.973, 0.098, -0.209, -0.002}                 & \multicolumn{2}{l}{$\text{N}_{41},\text{N}_{42},\text{N}_{43},\text{N}_{44},\text{N}_{45}$}           & \multicolumn{2}{l}{0.033, 0.801, 0.359, -0.478, -0.012}                 \\
			\multicolumn{2}{l}{$\text{N}_{51},\text{N}_{52},\text{N}_{53},\text{N}_{54},\text{N}_{55}$}           & \multicolumn{2}{l|}{0.998, -0.004, -0.014, -0.06, -4.7E-05}                 & \multicolumn{2}{l}{$\text{N}_{51},\text{N}_{52},\text{N}_{53},\text{N}_{54},\text{N}_{55}$}           & \multicolumn{2}{l}{0.999, -0.005, -0.014, 0.049, -1.63E-05}                 \\
			\hline
			\multicolumn{2}{l}{Annihilations}                     & \multicolumn{2}{l|}{Fractions[\%]}        & \multicolumn{2}{l}{Annihilations}                     & \multicolumn{2}{l}{Fractions[\%]}        \\
			\multicolumn{2}{l}{$\tilde{\chi}_1^0\tilde{\chi}_1^0 \to h_{s} \text{A}_{s}/h_{s}h_{s}$}                  & \multicolumn{2}{l|}{98.1/1.3}                 & \multicolumn{2}{l}{$\tilde{\chi}_1^0\tilde{\chi}_1^0 \to h_{s} \text{A}_{s}/h_{s}h_{s}$}                  & \multicolumn{2}{l}{86.0/4.5}                 \\
			\hline
			\multicolumn{2}{l}{Decays}                            & \multicolumn{2}{l|}{Branching ratios[\%]} & \multicolumn{2}{l}{Decays}                            & \multicolumn{2}{l}{Branching ratios[\%]} \\
			\multicolumn{2}{l}{$\tilde{\chi}_2^0 \to \tilde{\chi}_1^0Z^{\star}$}& \multicolumn{2}{l|}{100}
			&\multicolumn{2}{l}{$\tilde{\chi}_2^0 \to \tilde{\chi}_1^0Z^{\star}$}& \multicolumn{2}{l}{100}                 \\
			\multicolumn{2}{l}{$\tilde{\chi}_3^0 \to \tilde{\chi}_1^0h_{s}/\tilde{\chi}_1^0Z^{\star}$}& \multicolumn{2}{l|}{95.8/3.9}                 & \multicolumn{2}{l}{$\tilde{\chi}_3^0 \to \tilde{\chi}_1^{\pm}W^{\star}/\tilde{\chi}_2^0Z^{\star}/\tilde{\chi}_1^0Z^{\star}$}                           & \multicolumn{2}{l}{64.0/34.1/1.9}                 \\
			\multicolumn{2}{l}{$\tilde{\chi}_4^0 \to \tilde{\chi}_1^{\pm}W^{\mp}/\tilde{\chi}_3^0 Z/\tilde{\chi}_2^0 h/\tilde{\chi}_2^0 Z/\tilde{\chi}_3^0 h$}& \multicolumn{2}{l|}{56.7/20.5/16.8/1.9}                 & \multicolumn{2}{l}{$\tilde{\chi}_4^0 \to \tilde{\chi}_1^{\pm}W^{\mp}/\tilde{\chi}_1^{0}Z$}                           & \multicolumn{2}{l}{96.8/2.4}                 \\
			\multicolumn{2}{l}{$\tilde{\chi}_5^0 \to \tilde{\chi}_1^{\pm}W^{\mp}/\tilde{\nu}_{\mu}\nu_{\mu}/\tilde{\mu}^{\pm}_{\text{L}}\mu^{\mp}/\tilde{\chi}_3^0 Z/\tilde{\chi}_2^0 h/\tilde{\chi}_2^0 Z/\tilde{\chi}_2^0 h_{s}$}                           & \multicolumn{2}{l|}{24.8/23.7/19.1/12.3/10/3.6/2.0}                 & \multicolumn{2}{l}{$\tilde{\chi}_5^0 \to \tilde{\mu}^{\pm}_{\text{R}}\mu^{\mp}/\tilde{\nu}_{\mu}\nu_{\mu}/\tilde{\mu}^{\pm}_{\text{L}}\mu^{\mp}/\tilde{\chi}_2^{\pm}W^{\mp}/\tilde{\chi}_1^{\pm}W^{\mp}/\tilde{\chi}_3^0 Z/\tilde{\chi}_4^0 h/\tilde{\chi}_2^0 h/\tilde{\chi}_3^0 h$}                           & \multicolumn{2}{l}{46.6/11.9/11.3/9.0/5.7/5.5/3.5/2.4/1.6}                 \\
			\multicolumn{2}{l}{$\tilde{\chi}_1^{\pm} \to \tilde{\chi}_1^0W^{\star}$}                           & \multicolumn{2}{l|}{100}                 & \multicolumn{2}{l}{$\tilde{\chi}_1^{\pm} \to \tilde{\chi}_1^0W^{\star}/\tilde{\chi}_2^0W^{\star}$}                           & \multicolumn{2}{l}{99.1/0.9}                 \\
			\multicolumn{2}{l}{$\tilde{\chi}_2^{\pm} \to \tilde{\chi}_2^0W^{\pm}/\tilde{\chi}_3^0W^{\pm}/\tilde{\chi}_1^{\pm}Z/\tilde{\chi}_1^{\pm}h$}                     & \multicolumn{2}{l|}{26.7/26.4/25.6/20.5}                 & \multicolumn{2}{l}{$\tilde{\chi}_2^{\pm} \to \tilde{\chi}_2^0W^{\pm}/\tilde{\chi}_1^{\pm}Z/\tilde{\chi}_1^0W^{\pm}$}                     & \multicolumn{2}{l}{57.5/36.2/5.2}                 \\
			\multicolumn{2}{l}{$\tilde{\mu}_{\text{L}}^{\pm} \to \tilde{\chi}_2^{\pm}\nu_{\mu}/\tilde{\chi}_4^0\mu^{\pm}/\tilde{\chi}_2^0\mu^{\pm}/\tilde{\chi}_3^0\mu^{\pm}/$}                              & \multicolumn{2}{l|}{41.5/20.1/14.9/3.9}                 & \multicolumn{2}{l}{$\tilde{\mu}_{\text{L}}^{\pm} \to \tilde{\chi}_1^{\pm}\nu_{\mu}/\tilde{\chi}_2^0\mu^{\pm}/\tilde{\chi}_2^{\pm}\nu_{\mu}/\tilde{\chi}_4^0\mu^{\pm}$}                              & \multicolumn{2}{l}{47.8/29.7/14.2/4.5}                 \\
			\multicolumn{2}{l}{$\tilde{\mu}_{\text{R}}^{\pm} \to \tilde{\chi}_2^{\pm}\nu_{\mu}/\tilde{\chi}_4^0\mu^{\pm}/\tilde{\chi}_2^0\mu^{\pm}/\tilde{\chi}_3^0\mu^{\pm}/\tilde{\chi}_1^{\pm}\nu_{\mu}$}                              & \multicolumn{2}{l|}{46.0/22.3/10.8/5.9}                 & \multicolumn{2}{l}{$\tilde{\mu}_{\text{R}}^{\pm} \to \tilde{\chi}_2^0\mu^{\pm}/\tilde{\chi}_1^{\pm}\nu_{\mu}/\tilde{\chi}_3^0\mu^{\pm}/\tilde{\chi}_4^0\mu^{\pm}/\tilde{\chi}_2^{\pm}\nu_{\mu}$} &\multicolumn{2}{l}{48.0/36.2/9.6/3.4/1.8}\\
			\multicolumn{2}{l}{$\tilde{\nu}_{\mu} \to \tilde{\chi}_1^{\pm}\mu^{\mp}/\tilde{\chi}_2^{\pm}\mu^{\mp}/\tilde{\chi}_4^0\tilde{\nu}_{\mu}/\tilde{\chi}_2^0\tilde{\nu}_{\mu}$}
			& \multicolumn{2}{l|}{47.2/27.2/14.5/10.4}                 & \multicolumn{2}{l}{$\tilde{\nu}_{\mu} \to \tilde{\chi}_1^{\pm}\mu^{\mp}/\tilde{\chi}_2^0\tilde{\nu}_{\mu}/\tilde{\chi}_2^{\pm}\mu^{\mp}/\tilde{\chi}_4^0\tilde{\nu}_{\mu}$}                             & \multicolumn{2}{l}{65.7/24.9/5.0/3.6}                 \\
			\hline
			\multicolumn{2}{l}{R value}                           & \multicolumn{2}{l|}{0.84}                 & \multicolumn{2}{l}{R value}                           & \multicolumn{2}{l}{4.14}                 \\ \hline\hline
	\end{tabular}}
\end{table}

\item By considering the $\tilde{H}$-dominated NLSP case it was found that the decay modes of $\tilde{\chi}_2^0$, $\tilde{\chi}_1^\pm$ and $\tilde{\mu}_{1,2}$ are similar to those of the  $\tilde{W}$-dominated NLSP case, and the LHC constraints tend to be weaker than the other cases. This observation may be understood from four aspects~\cite{Cao:2021tuh}. First, since the $\tilde{H}$-dominated $\tilde{\chi}_{2,3}^0$ and $\tilde{\chi}_1^\pm$ can not decay into sleptons, the leptonic signal rate is usually much smaller than the case where $\tilde{\mu}_R$ or $\tilde{\nu}_{\mu}$ acts as NLSP. Second, the collider sensitive signal events are often diluted by the complicated decay chains of sparticles, given that heavy sparticles prefer to decay into the NLSP or other non-singlet-dominated sparticles first. They are diluted also by the decays $\tilde{\chi}_{2,3}^0 \to \tilde{\chi}_1^0 h_s, \tilde{\chi}_1^0 h$, given that $Br(h_s/h \to \ell^\pm \ell^\mp)$ is much smaller than $Br(Z \to \ell^\pm \ell^\mp)$. Third, the interpretation of $\Delta a_\mu$ requires that all crucial sparticles are usually in several hundred GeVs for a not too large $\tan \beta$. Thus, for the parameter points surviving the LHC constraints, the mass splitting between sparticles is not large enough to produce high-$p_{\rm T}$ signal objects, which can be significantly distinguished from the background in the collider. Last, in some rare cases, the leptonic signal of SUSY may mainly come from the NLSP. For this situation, the discussion of the LHC constraints can be simplified by considering the system that only contains NLSP and LSP. From this, it is evident that the constraints on the $\tilde{H}$-dominated NLSP case are significantly weaker than those on the $\tilde{W}$-dominated NLSP case. In fact, we once scrutinized the property of all the samples surviving the LHC constraints. It was found that the above factors applied to these samples.
\end{itemize}

In order to emphasize the characteristics of the parameter point with $\tilde{H}$-dominated NLSP, two benchmark points, P1 and P2, are chosen to present their detailed information in Table~\ref{tab:points}. Both points satisfy the DM constraints and can explain the $a_\mu$ discrepancy at $2 \sigma$ level. The P1 point survives the LHC constraints, while the P2 point has been excluded by the LHC search for SUSY. These two benchmark points verify part of the discussions in this work.

\begin{figure}[t]
	\centering
     \includegraphics[width=0.50\textwidth]{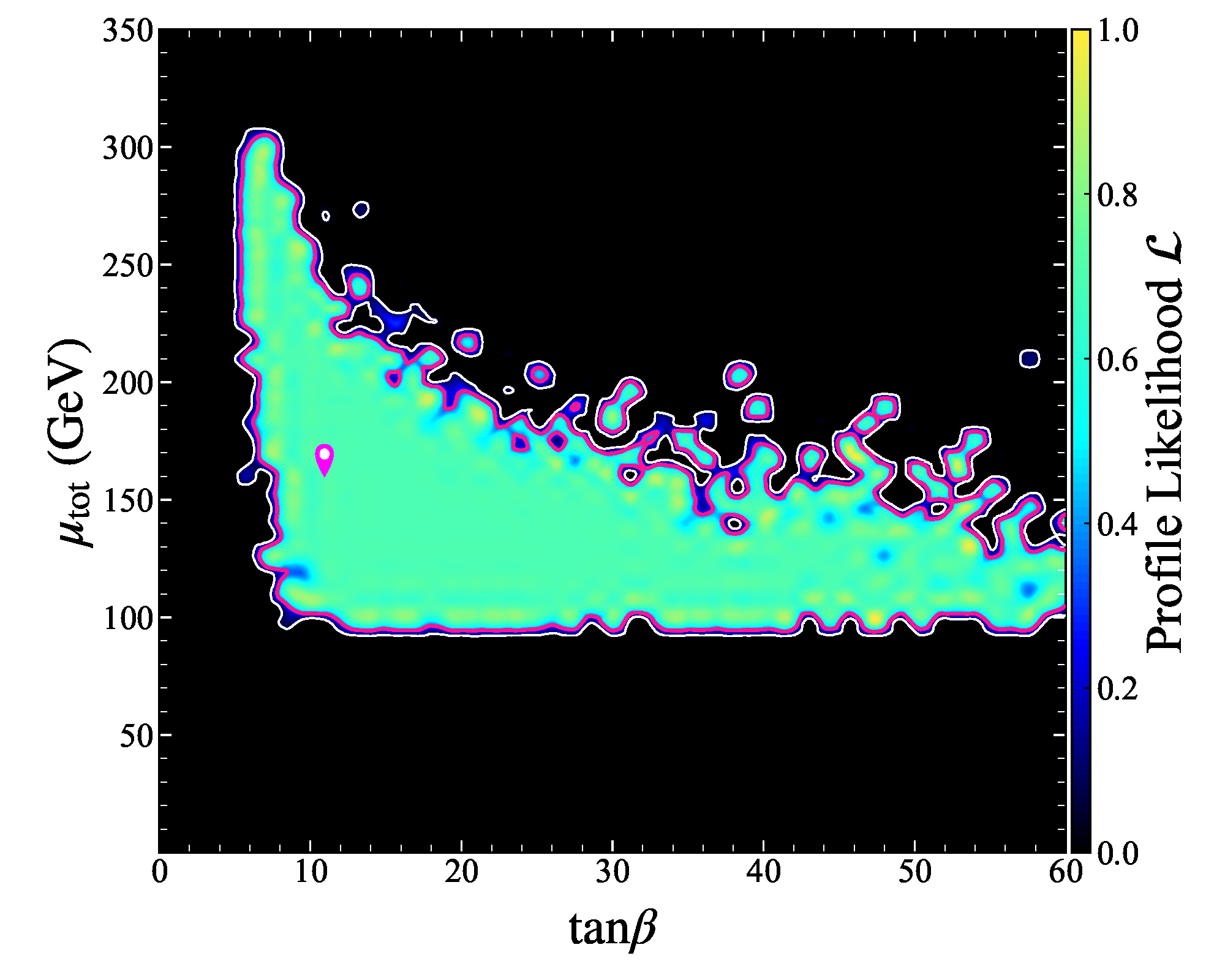} \hspace{-0.3cm}
	  \includegraphics[width=0.50\textwidth]{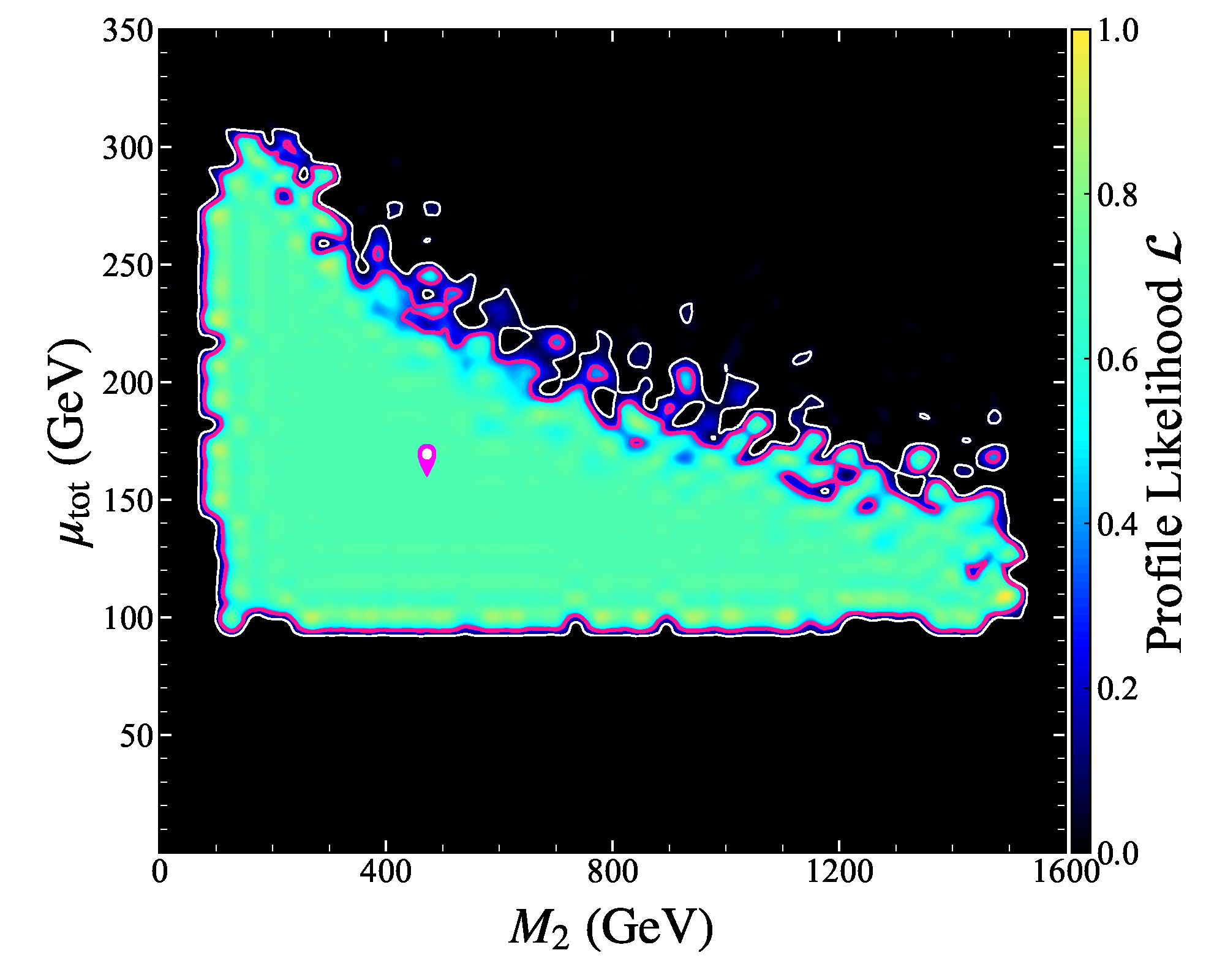}
	\caption{\label{fig:plamm-GNMSSM} Same as Fig.~\ref{fig:plamm}, but for the results of the GNMSSM. The best point is located at $\tan \beta \simeq 10.9$, $m_{\tilde{\chi}_1^0} \simeq 42.8~{\rm GeV}$, $\mu_{tot} \simeq 160~{\rm GeV}$, $M_2 \simeq 473~{\rm GeV}$, $m_{\tilde{\mu}_L} \simeq 164~{\rm GeV}$, and $m_{\tilde{\mu}_R} \simeq 143~{\rm GeV}$.}
\end{figure}

\begin{figure}[t]
	\centering
	\includegraphics[width=0.9\textwidth]{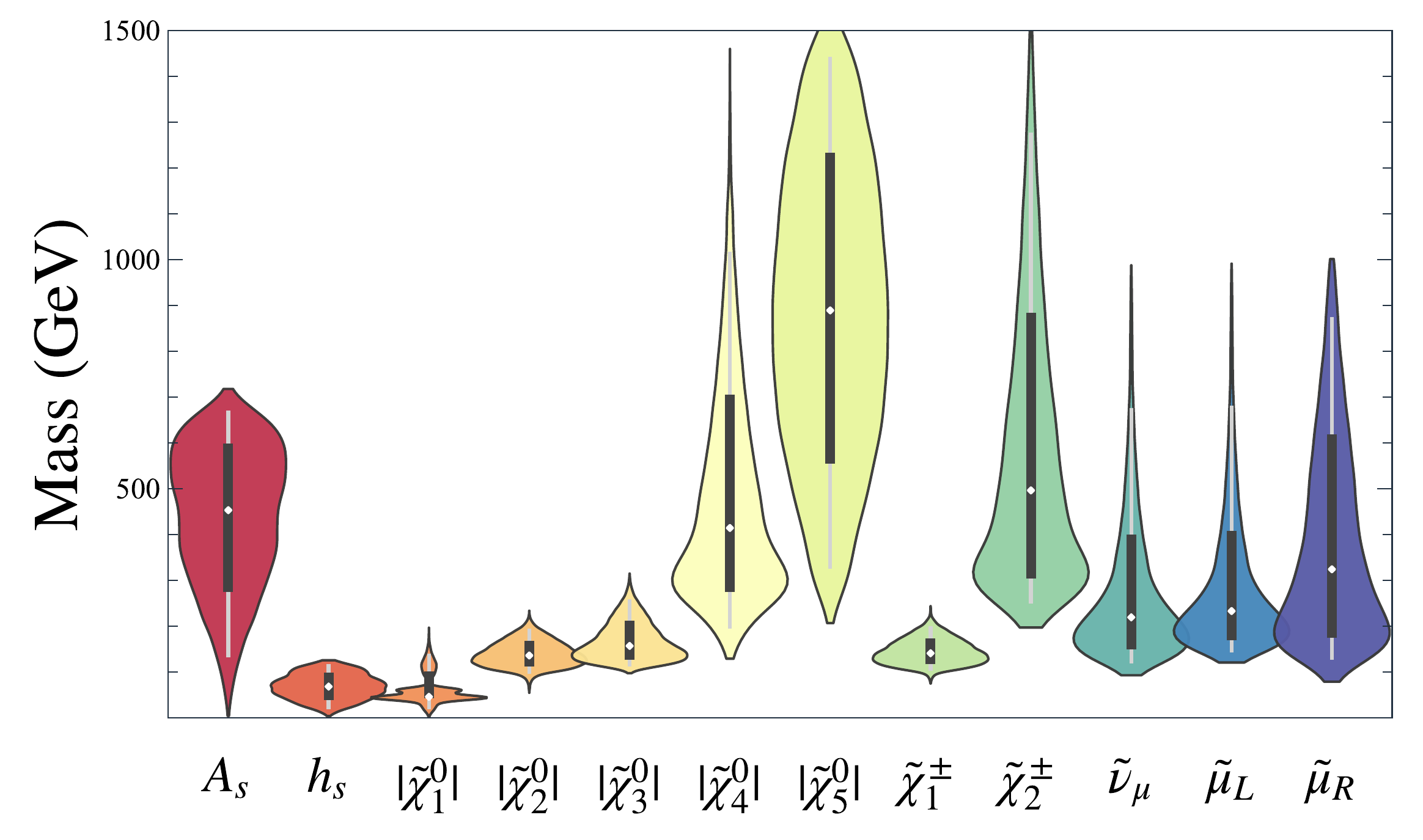}
	\caption{\label{fig:smassviolin-GNMSSM} Same as Fig.~\ref{fig:smassviolin}, but for the results of the GNMSSM.}
\end{figure}

\begin{table}[t]
\centering
\begin{tabular}{|c|cc|cc|cc|cc|cc|}
\hline\hline
$\rm NLSP$ & \multicolumn{2}{|c|}{$\tilde{H}$} & \multicolumn{2}{|c|}{$\tilde{B}$}	& \multicolumn{2}{|c|}{$\tilde{W}$} & \multicolumn{2}{|c|}{$\tilde{\mu}_R$} & \multicolumn{2}{|c|}{$\tilde{\nu}_\mu$}   \\ \hline
$N_{\rm tot}$ \qquad$\mid$\qquad$N_{\rm pass}$& 11753 & 21 & 470 & 0 & 1259 & 0 & 867 & 0 & 2041 & 0 \\
$N_{a_{\mu}^{\rm SUSY}} \in \left(-3 \sigma, -2 \sigma \right)$ & 1519 & 5 & 67 & 0 & 105 & 0 & 92 & 0 & 104  & 0 \\
$N_{a_{\mu}^{\rm SUSY}} \in \left(-2 \sigma, -1 \sigma \right)$ & 2113 & 3 & 97 & 0 & 180 & 0 & 154 & 0 & 270 & 0 \\
$N_{a_{\mu}^{\rm SUSY}} \in \left(-1 \sigma, ~~0 \sigma \right)$ & 3777 & 1 & 137 & 0 & 448 & 0 & 314 & 0 & 906 & 0 \\
$N_{a_{\mu}^{\rm SUSY}} \in \left(~~0 \sigma, ~~1 \sigma \right)$ & 2462 & 0 & 83 & 0 & 361 & 0 & 180 & 0 & 578 & 0 \\
$N_{a_{\mu}^{\rm SUSY}} \in \left(~~1 \sigma, ~~2 \sigma \right)$ & 499 & 0 & 27 & 0 & 65 & 0 & 26 & 0 & 117 & 0 \\
$N_{a_{\mu}^{\rm SUSY}} \in \left(~~2 \sigma, ~~3 \sigma \right)$ & 224 & 0 & 14 & 0 & 19 & 0 & 35 & 0 & 14 & 0 \\ \hline \hline
\end{tabular}
\caption{\label{tab:mcsum-GNMSSM} Same as Table \ref{tab:mcsum}, but for the results of the GNMSSM. All samples predict $R > 0.4$, thus they are to be tested at high-luminosity LHC. Moreover, all samples passing the LHC constraints are characterized by $|m_{\tilde{\chi}_1^0}| \gtrsim 150~{\rm GeV}$.   }
\end{table}

Finally, it should be noted that the LHC search for $\tau$-leptons plus missing momentum signal, such as the ATLAS analyses in~\cite{ATLAS:2017qwn} and~\cite{ATLAS:2019gti}, was not considered because of the massive $\tilde{\tau}$ assumption in this study. Specifically, the assumption implies that the $\tau$-leptons mainly come from the decay of the $W/Z$ or Higgs bosons, which are the decay products of parent sparticles. For the former case, the final states containing $e$ and/or $\mu$ are more efficient than the $\tau$ final state in restricting SUSY mass spectrum since all the lepton signal rates are roughly equal. For the latter case, the $\tau$ signal is usually less crucial in SUSY search because the branching ratio of the Higgs decay into $\tau \bar{\tau}$ is significantly small (in comparison with $\tilde{\tau}$ decay). With the codes for the analyses in~\cite{ATLAS:2017qwn} and~\cite{ATLAS:2019gti}, which were implemented in our previous work~\cite{Cao:2018iyk} and \texttt{CheckMATE-2.0.29}, respectively, we studied their prediction of $R$ for the two benchmark points. We found that the analyses do not affect the results in Table~\ref{tab:points}. As an alternative, if $\tilde{\tau}$ or all sleptons are assumed to the NLSP (see, e.g.,~\cite{Hagiwara:2017lse}), the production rate of the $e/\mu$ final states will be affected. In this case, $R$ should be recalculated. In particular, the experimental analysis of the $\tau$ final state must be included in the study. It is expected that the LHC constraints are still strong because the $h_2$ scenario is featured by moderately light Higgsinos.

\section{\label{numerical study2} Explaining $\Delta a_\mu$ in the $h_2$ scenario of GNMSSM}

The impact of the muon g-2 anomaly on the $h_2$ scenario of the GNMSSM is studied in this section. For this purpose, the parameter space including $|\mu^\prime| \leq 1~{\rm TeV}$, $-10^6~{\rm TeV}^2 \leq m_S^{\prime\ 2} \leq 10^6~{\rm TeV}^2$, and that in Eq.~(\ref{Parameter-region}), were scanned in a way similar to what we did in Section \ref{numerical study1}. It was found that the DM was Singlino-dominated for all the obtained samples, and it annihilated mainly by a resonant $Z$, $h$, or $h_s/A_s$ to obtain the measured abundance. These channels contributed to the total Bayesian evidence by about $43\%$, $19.6\%$, and $37\%$, respectively, before the MC simulations were implemented. The basic reason for such a behavior is that $\tilde{\chi}_1^0$, $m_{h_s}$ and $m_{A_s}$ in the GNMSSM can be changed freely by tuning $\mu^\prime$, $A_\kappa$, and $m_S^{\prime\ 2}$, respectively. Thus, the annihilations could easily happen. Given that the GMSSM might have different key features from the $\mu$NMSSM, various PL maps of the GNMSSM were surveyed in this study.

In Fig.~\ref{fig:plamm-GNMSSM}, the two-dimensional profile likelihood function was projected onto $\tan \beta-\mu_{tot}$ and $M_2-\mu_{tot}$ planes. They show that the GNMSSM results are quite similar to the $\mu$NMSSM predictions. In particular, $\mu_{tot}$ and $m_{\tilde{\chi}_1^\pm}$ have upper bounds of about $310~{\rm GeV}$ and $300~{\rm GeV}$, respectively\footnote{We inferred that the $2\sigma$ CI of the $\mu$NMSSM covers a broader region on the parameter planes than that of the GNMSSM by comparing Fig.~\ref{fig:plamm-GNMSSM} with Fig.~\ref{fig:plamm}. This is contrary to the common sense that the former should be narrower than the latter since the parameter space of the $\mu$NMSSM is only a subset of the GNMSSM's parameter space. This phenomenon originates from the MultiNest algorithm utilized in the scan, which mainly collects the samples contributing significantly to the Bayesian evidence~\cite{Feroz:2008xx}. The parameter points of the $\mu$NMSSM correspond to $\mu^\prime = 0$ and $m_S^{\prime\ 2} = 0$, and are relatively unimportant for the evidence, which was verified by the study of the two-dimensional posterior probability distribution function, $P(\mu^\prime,m_S^{\prime\ 2})$~\cite{Fowlie:2016hew}. Thus, only a few of them were considered in the sampling. It is expected that, with the increase of the setting $n_{live}$, more samples of the GNMSSM will be collected, which will broaden the CI regions~\cite{Cao:2018iyk}. This process, however, is very computationally expensive, since a high-dimensional parameter space is surveyed.}. The fundamental reason for this, as was emphasized before, is that the $h_2$ scenario prefers moderately small $\mu_{tot}$ and $\tan \beta$ to predict $m_{h_1} \lesssim 125~{\rm GeV}$ and $h_2$ to be SM-like, which was verified by the posterior probability distribution function of the samples obtained from the scan. This characteristic, once combined with the requirement to explain the anomaly, will entail certain moderately light sparticles. In Fig.~\ref{fig:smassviolin-GNMSSM}, the violin diagrams for the mass spectrum of the singlet-dominated Higgs bosons, the electroweakinos and $\mu$-type sleptons are shown. The profiles for the sparticles are quite similar to those in Fig.~\ref{fig:smassviolin} for the $\mu$NMSSM results, except that $|m_{\tilde{\chi}_1^0}|$ can be as low as several GeV. This difference mainly comes from the DM annihilation mechanisms, and it usually makes the LHC's constraints much stronger.

In Table \ref{tab:mcsum-GNMSSM}, the numbers of the samples surveyed by MC simulations and those passing the LHC constraints were presented in a way similar to Table \ref{tab:mcsum}. This table shows that the LHC analyses have strongly constrained the parameter space of the GNMSSM.

\section{\label{conclusion}Conclusion}

The recent measurement of $a_\mu$ by the FNAL  corroborates further the long-standing discrepancy of $a_\mu^{\rm Exp}$ from $a_\mu^{\rm SM}$. It can not only reveal useful information of the physics beyond the SM, but also place strong restrictions on certain theories. Recently, implications of the discrepancy were comprehensively discussed  with respect to the GNMSSM, which is a theory that has the following attractive features: it is free from the tadpole problem and the domain-wall problem of the $Z_3$-NMSSM, and it is capable of forming an economic secluded DM sector to naturally yield the DM experimental results~\cite{Cao:2021ljw}. It was found that the $h_1$ scenario of the GNMSSM could easily and significantly weaken the constraints from the LHC search for SUSY. It also predicted more stable vacuums than the $Z_3$-NMSSM. As a result, the scenario can explain the discrepancy in a broad parameter space that is consistent with all experimental results, and at same time keeps the electroweak symmetry breaking natural~\cite{Cao:2021tuh}. By contrast, it is difficult for the popular MSSM and $Z_3$-NMSSM to do this.

These theoretical advantages inspired us to consider the $h_2$ scenario of the GNMSSM, which is another well-known realization of the theory. It was shown by analytic formulae that, in order to obtain $m_{h_1} \lesssim 125~{\rm GeV}$ and an SM-like $h_2$ without significant tunings of relevant parameters, the scenario prefers a moderately light $\mu_{tot}$ and $\tan \beta \lesssim 30$. This characteristic, if combined with the requirement to account for the anomaly, will entail some light sparticles, and sequentially make the LHC constraints rather tight. In this work, this speculation was tested using numerical results. Specifically, a special case of the GNMSSM called $\mu$NMSSM was first studied by scanning its parameter space with the MultiNest algorithm and considering the constraints from the LHC Higgs data, the DM experimental results, the B-physics observations, and the vacuum stability. Then,
the samples obtained from the scan were surveyed by the LHC analyses in sparticle searches. Through sophisticated MC simulations, it was found that only a dozen of the samples, among about twenty thousand, passed the constraints, which corresponded to about $0.04\%$ of the total Bayesian evidence. Given that the scan results have statistical significance, we conclude that the $h_2$ scenario of the $\mu$NMSSM is tightly constrained if it is intended to explain the anomaly. A similar study was carried out for the GNMSSM, and it was found that a smaller portion of the samples (about $0.008\%$ of the total evidence) satisfied the LHC constraints. This difference arises from DM annihilation mechanisms: for the former case, the Singlino-dominated DM achieved the measured abundance mainly through the process $\tilde{\chi}_1^0 \tilde{\chi}_1^0 \to h_s A_s$, while for the latter case, it was through a resonant $Z$, $h$, or $h_s/A_s$ annihilation to obtain the abundance. Since the latter case usually predicts a relatively light DM, the LHC constraints are stronger.

This work extends the research in~\cite{Cao:2018rix} by considering a more general theoretical framework with more advanced and sophisticated research strategies. As a result, the conclusions obtained in this work are more robust than those of the previous work, and apply to any realizations of the NMSSM.

\section*{Acknowledgement}
J. Cao and Y. Yue thank Dr. H. J. Zhou for her great patience in reading the manuscript carefully and giving good suggestions. This work is supported by the National Natural Science Foundation of China (NNSFC) under grant No. 12075076.

\appendix

\section{DM-nucleon scattering in the MSSM}

In this section, we use analytic formulae to focus on Bino-dominated DM and study DM-nucleon scatterings for three typical cases.

We begin with the neutralino mass matrix given by~\cite{Pierce:2013rda}:
\begin{equation}
M_N=\left(
\begin{array}{cccc}
M_1 & 0 & -\frac{v g_1c_{\beta}}{2} & \frac{v g_1 s_{\beta }}{2} \\
0 & M_2 & \frac{v c_W g_1 c_{\beta}}{2 s_W } & -\frac{v c_W g_1 s_{\beta }}{2 s_W } \\
-\frac{v g_1c_{\beta}}{2} & \frac{v c_W g_1 c_{\beta}}{2 s_W } & 0 & -\mu  \\
\frac{v g_1 s_{\beta }}{2 } & -\frac{v c_W g_1 s_{\beta }}{2 s_W } & -\mu  & 0 \\
\end{array}
\right),
\end{equation}
where $g_1= 2 M_Z s_W/v$, $s_\beta \equiv \sin\beta$ and $c_\beta\equiv \cos\beta$. In terms of neutralino mass, $m_{\tilde{\chi}_i^0}$, the eigenvectors are then exactly formulated by
\begin{equation}
N_i=\frac{1}{\sqrt{C_i}}\left(
\begin{array}{c}
\left(\mu ^2-m_{\tilde{\chi}_i^0}^2\right) \left(M_2-m_{\tilde{\chi}_i^0}\right)- M_Z^2 c_W^2\left(m_{\tilde{\chi}_i^0}+2 \mu  s_\beta c_\beta\right)\\
\\
- M_Z^2 s_W c_W \left(m_{\tilde{\chi}_i^0}+2 \mu  s_\beta c_\beta\right)\\
\\
\left(M_2-m_{\tilde{\chi}_i^0}\right) \left(m_{\tilde{\chi}_i^0} c_\beta+\mu  s_\beta\right)M_Z s_W\\
\\
- \left(M_2-m_{\tilde{\chi}_i^0}\right) \left( m_{\tilde{\chi}_i^0} s_\beta+\mu c_\beta\right)M_Z s_W
\end{array}
\right)  \nonumber
\end{equation}
where
\begin{eqnarray}
C_i&&= M_Z^2c_W^2 \left(m_{\tilde{\chi}_i^0}+2 \mu  s_{\beta } c_\beta\right) \left[M_Z^2\left(m_{\tilde{\chi}_i^0}+2 \mu  s_{\beta } c_\beta\right)+2  \left(\mu ^2-m_{\tilde{\chi}_i^0}^2\right) \left(m_{\tilde{\chi} _i^0}-M_2\right)\right]\nonumber\\
&&\qquad \qquad+ \left(m_{\tilde{\chi}_i^0}-M_2\right)^2 \left\{M_Z^2 s_W^2 \left[ \left(m_{\tilde{\chi}_i^0}^2+\mu ^2\right)+4 \mu  m_{\tilde{\chi}_i^0} s_{\beta } c_\beta  \right]+ \left(m_{\tilde{\chi}_i^0}^2-\mu ^2\right)^2\right\}. \nonumber
\end{eqnarray}
Parameterizing the couplings of the DM to the SM-like Higgs boson, $h$, and Z boson as the following form~\cite{Haber:1984rc,Gunion:1984yn}
\begin{eqnarray}
{\cal{L}}_{\rm MSSM} \ni  C_{\tilde{\chi}_1^0 \tilde{\chi}_1^0 h} h \overline{\tilde{\chi}_1^0} \tilde{\chi}_1^0 + C_{\tilde{\chi}_1^0 \tilde{\chi}_1^0 Z} Z_\mu \overline{\tilde{\chi}_1^0} \gamma^\mu \gamma_5 \tilde{\chi}_1^0,  \nonumber
\end{eqnarray}
we obtain~\cite{Cao:2019qng}
\begin{eqnarray}
C_{\tilde{\chi}_1^0  \tilde{\chi}_1^0 h} &=& \frac{e~ \tan\theta_w}{C_1} \left( {\mu}^2 - m_{\tilde{\chi}_1^0}^2  \right) \left( M_2 - m_{\tilde{\chi}_1^0} \right)^2 M_Z     \left[ m_{\tilde{\chi}_1^0} \sin(\beta - \alpha) + \mu \cos(\beta + \alpha)     \right], \nonumber \\
C_{\tilde{\chi}_1^0  \tilde{\chi}_1^0 Z} &=& \frac{e~ \tan\theta_w ~ \cos 2\beta}{2 C_1} \left( {\mu}^2 - m_{\tilde{\chi}_1^0}^2  \right) \left( M_2 - m_{\tilde{\chi}_1^0} \right)^2 M_Z^2,
\end{eqnarray}
where $\alpha$ is the mixing angle of CP-even Higgs fields in forming mass eigenstates~\cite{Djouadi:2005gj}. In the decoupling limit of the Higgs sector, i.e., $m_A \gg v $, the SI and SD cross-sections of the DM with nucleons are approximated by~\cite{Cao:2019qng}:
\begin{eqnarray}
\sigma_{\tilde{\chi}_1^0-N}^{\rm SI} & \simeq & 5 \times 10^{-45} {\rm cm^2} \left ( \frac{C_{\tilde{\chi}_1^0 \tilde{\chi}_1^0 h}}{0.1} \right )^2 \left (\frac{m_h}{125 {\rm GeV}} \right )^2,   \\
\sigma_{\tilde{\chi}_1^0-N}^{\rm SD} & \simeq & C_N \times \left ( \frac{C_{\tilde{\chi}_1^0 \tilde{\chi}_1^0 Z}}{0.01} \right )^2,
\end{eqnarray}
with $C_p \simeq 2.9 \times 10^{-41}~{\rm cm^2} $ for protons and $C_n \simeq 2.3 \times 10^{-41}~{\rm cm^2} $ for neutrons.

In the following, we assume $\tan \beta \gg 1$ and $m_A \gg v$ so that $\alpha \simeq \beta - \pi/2$~\cite{Djouadi:2005gj}, and investigate the dependence of $\sigma_{\tilde{\chi}_1^0-N}^{\rm SI} $ and $\sigma_{\tilde{\chi}_1^0-N}^{\rm SD}$ on parameter $\mu$ for three cases.
\begin{itemize}
\item Case I: the DM co-annihilated with the Wino-dominated electroweakinos to obtain the measured abundance, $M_2$ and $m_{\tilde{\chi}_1^0}$ are of same sign, and $|\mu|$ is comparable with $|m_{\tilde{\chi}_1^0}|$. In this case, $M_2 \simeq 1.1 \times m_{\tilde{\chi}_1^0}$ to obtain the measured DM abundance, and $C_1$ is approximated by:
   \begin{equation}
 C_1  \simeq M_Z^4 c^2_w \left( m_{\tilde{\chi}_1^0}   + \mu \sin 2\beta \right) ^2 \simeq  M_Z^4 c^2_w m_{\tilde{\chi}_1^0}^2.
 \end{equation}
Consequently, $C_{h \tilde{\chi}_1^0  \tilde{\chi}_1^0}$ and  $C_{Z \tilde{\chi}_1^0  \tilde{\chi}_1^0}$ are given by:
 \begin{eqnarray}
 C_{\tilde{\chi}_1^0  \tilde{\chi}_1^0 h} &\simeq& 0.01 \times e \tan\theta_w  \dfrac{m_{\tilde{\chi}_1^0} \left( {\mu}^2 - m_{\tilde{\chi}_1^0}^2  \right) \left( \sin 2 \beta + m_{\tilde{\chi}_1^0}/\mu   \right)}{M_Z^3 c_w^2} \nonumber \\
 &\simeq & \pm 0.01 \times e \tan\theta_w  \dfrac{2 \delta_1 m_{\tilde{\chi}_1^0}^3 }{(1 + \delta_1) M_Z^3 c_w^2},  \\
 C_{\tilde{\chi}_1^0  \tilde{\chi}_1^0 Z} &\simeq& 0.01 \times e \tan\theta_w \cos 2 \beta \dfrac{\left( {\mu}^2 - m_{\tilde{\chi}_1^0}^2  \right) }{ 2 M_Z^2 c_w^2} \nonumber \\
 &\simeq & 0.01 \times e \tan\theta_w \cos 2 \beta \dfrac{\delta_1 m_{\tilde{\chi}_1^0}^2 }{M_Z^2 c_w^2},
 \end{eqnarray}
where $\mu$ is parameterized by $|\mu| = (1 + \delta_1) |m_{\tilde{\chi}_1^0}| $, with $\delta_1$ denoting a small positive dimensionless number. These approximations indicate that the DM-nucleon scattering rates increase monotonously as the DM becomes heavier and/or $|\mu|$ departures from $|m_{\tilde{\chi}_1^0}|$. Specifically, for $m_{\tilde{\chi}_1^0} = 210~{\rm GeV}$, which are the lower mass bound of the DM from the LHC search for Wino-dominated electroweakinos in the compressed mass spectrum case~\cite{ATLAS:2021moa}, we found that  $\delta_1$ must be less than about $0.26$ to be consistent with the XENON-1T data on SI cross-section. For $m_{\tilde{\chi}_1^0} = 300~{\rm GeV}$, it must be less than about $0.12$. The Higgsinos in this narrow mass region contribute significantly to lepton signals at the LHC and the DM relic abundance. They also affect $a_\mu$.
Consequently, such a situation needs tuning to satisfy all experimental constraints.

\item Case II: the DM co-annihilated with the Wino-dominated electroweakinos to obtain the measured abundance, and $|\mu|$ is much larger than $|m_{\tilde{\chi}_1^0}|$. This case predicts~\cite{Cao:2019qng}:
\begin{eqnarray}
 C_1  &\simeq& \left( {\mu}^2 - m_{\tilde{\chi}_1^0}^2  \right)^2 \left( {M}_2 - m_{\tilde{\chi}_1^0}  \right)^2, \\
 C_{\tilde{\chi}_1^0  \tilde{\chi}_1^0 h} &\simeq& e \tan\theta_w \dfrac{M_Z \left( \sin 2 \beta + m_{\tilde{\chi}_1^0}/\mu   \right)}{ \mu \left(   1- m_{\tilde{\chi}_1^0}^2/\mu^2 \right) }, \label{C_DMDMh} \\
  C_{\tilde{\chi}_1^0  \tilde{\chi}_1^0 Z} &\simeq& e \tan\theta_w \cos 2 \beta \dfrac{M_Z^2}{ 2\mu^2 \left(   1- m_{\tilde{\chi}_1^0}^2/\mu^2 \right) },  \label{C_DMDMZ}
\end{eqnarray}
and consequently the scattering rates decrease monotonously with the increase of $|\mu|$.

\item Case III: the DM co-annihilated with the Higgsino-dominated electroweakinos to obtain the measured abundance, and $|M_2|$ is much larger than $|m_{\tilde{\chi}_1^0}|$. For this case:
 \begin{eqnarray}
 C_1 &\simeq & \left( {\mu}^2 - m_{\tilde{\chi}_1^0}^2  \right)^2 \left( {M}_2 - m_{\tilde{\chi}_1^0}  \right)^2  +  \left( {M}_2 - m_{\tilde{\chi}_1^0}  \right)^2 M_Z^2 s_w^2 \left(   m_{\tilde{\chi}_1^0}^2 + \mu^2 + 2 m_{\tilde{\chi}_1^0} \mu \sin 2 \beta \right) \nonumber \\
 &\simeq& \left( {\mu}^2 - m_{\tilde{\chi}_1^0}^2  \right)^2 \left( {M}_2 - m_{\tilde{\chi}_1^0}  \right)^2,
 \end{eqnarray}
where the second approximation is obtained by assuming $|m_{\tilde{\chi}_1^0}| \gg m_Z$.  Thus, $C_{\tilde{\chi}_1^0  \tilde{\chi}_1^0 h}$ and  $C_{\tilde{\chi}_1^0  \tilde{\chi}_1^0 Z}$ have same approximations as those in Case-II. It should be noted that, because $m_{\tilde{\chi}_1^0}^2/\mu^2 \sim 1$, and consequently both $C_{\tilde{\chi}_1^0  \tilde{\chi}_1^0 h}$ and  $C_{\tilde{\chi}_1^0  \tilde{\chi}_1^0 Z}$ are enhanced by the factor  $1/(1- m_{\tilde{\chi}_1^0}^2/\mu^2)$, the DM must be as massive as several TeV to be consistent with the XENON-1T results.

\end{itemize}

\section{\label{Validation}Validation of CERN-EP-2021-059  (ATLAS\_2106\_01676)}
Within the framework of CheckMATE-2.0.29, our codes were validated for all Signal Regions (SRs) in ATLAS\_2106\_01676  by considering
$\tilde{\chi}_1^{\pm} \tilde{\chi}_2^{0}$ production at LHC.
The masses of other charginos and neutralinos apart from the bino-like $\tilde{\chi}_1^0$ and wino-like $\tilde{\chi}_2^0$, $\tilde{\chi}_1^{\pm}$  were set to be 2.5 TeV, and  $\tilde{\chi}_2^0$ and $\tilde{\chi}_1^{\pm}$ decay were set as follows: $\tilde{\chi}_2^0\rightarrow \tilde{\chi}_1^0 Z/h$, $\tilde{\chi}_1^{\pm}\rightarrow \tilde{\chi}_1^0 W^{\pm}$. 	
In the validation, $10^5$ events were generated by the package MG5\_aMC\_v3\_2\_0 for the parameter point $(m_{\tilde{\chi}_1^{\pm}}/m_{\tilde{\chi}_2^0}, m_{\tilde{\chi}_1^0}) = (300 {\rm GeV}, 200 {\rm GeV})$.

In the proc\_card.dat, the following setting was implemented:
\begin{lstlisting}[columns=fullflexible, backgroundcolor=\color{back},frame=trBL]
import model MSSM_SLHA2 --modelname
generate p p > n2 x1+, n2 > z > l- l+ n1, x1+ > w+ > l+ vl n1
add process p p > n2 x1-, n2 > z > l- l+ n1, x1- > w- > l- vl~ n1
add process p p > n2 x1+ j j, n2 > z > l- l+ n1, x1+ > w+ > l+ vl n1
add process p p > n2 x1- j j, n2 > l- l+ n1, x1- > w- > l- vl~ n1
\end{lstlisting}
In the run\_card.dat, the following setting was implemented and the others are kept default:
\begin{lstlisting}[backgroundcolor=\color{back},frame=trBL]
100000 = nevents ! Number of unweighted events requested.
0 = ickkw            ! 0 no matching, 1 MLM
75.0  =  ktdurham
\end{lstlisting}
In the param\_card.dat, we set the following information and others are kept default:
\begin{lstlisting}[backgroundcolor=\color{back},frame=trBL]
Block mass
1000022 2.000000e+02 # Mneu1
1000023 3.000000e+02 # Mneu2
1000024 3.000000e+02 # Mch1
1000025 -2.50000e+03 # Mneu3
1000037 2.500000e+03 # Mch2
Block nmix  # Neutralino Mixing Matrix^M
1  1     9.86364430E-01   # N_11^M
1  2    -5.31103553E-02   # N_12^M
1  3     1.46433995E-01   # N_13^M
1  4    -5.31186117E-02   # N_14^M
2  1     9.93505358E-02   # N_21^M
2  2     9.44949299E-01   # N_22^M
2  3    -2.69846720E-01   # N_23^M
2  4     1.56150698E-01   # N_24^M
Block umix  # Chargino Mixing Matrix U^M
1  1     9.16834859E-01   # U_11^M
1  2    -3.99266629E-01   # U_12^M
2  1     3.99266629E-01   # U_21^M
2  2     9.16834859E-01   # U_22^M
Block vmix  # Chargino Mixing Matrix V^M
1  1     9.72557835E-01   # V_11^M
1  2    -2.32661249E-01   # V_12^M
2  1     2.32661249E-01   # V_21^M
2  2     9.72557835E-01   # V_22^M
\end{lstlisting}
In the pythia8\_card.dat, we set:
\begin{lstlisting}[columns=fullflexible, backgroundcolor=\color{back},frame=trBL]
Merging:Process = pp>{ch1-,1000015}{ch1+,-1000015}{n2, 1000023}
Merging:mayRemoveDecayProducts=on
\end{lstlisting}
\begin{table}[t]
	\centering
	\caption{\label{tab:cf-wzonshell} Cutflow validation of the ATLAS analysis \texttt{atlas\_2106\_01676} for mass point $m(\tilde{\chi}_1^{\pm}/\tilde{\chi}_2^0, \tilde{\chi}_1^0) = (300, 200)~{\rm GeV}$ and $m(\tilde{\chi}_1^{\pm}/\tilde{\chi}_2^0, \tilde{\chi}_1^0) = (600, 100)~{\rm GeV}$.}
	\vspace{0.3cm}
	\resizebox{0.8\textwidth}{!}{%
\begin{tabular}{l|rrrr} 
			\hline\hline
			Process & \multicolumn{4}{c}{$pp \to \tilde{\chi}_2^0 \tilde{\chi}_1^{\pm}, \tilde{\chi}_1^\pm \to W^{\pm} \tilde{\chi}_1^0, \tilde{\chi}_2^0 \to Z \tilde{\chi}_1^0$} \\
			Point $m(\tilde{\chi}_2^0/\tilde{\chi}_1^\pm, \tilde{\chi}_1^0)$ & \multicolumn{2}{c}{(300, 200) GeV}&\multicolumn{2}{c}{(600, 100) GeV} \\
			Generated Events & \multicolumn{4}{c}{500,000} \\ \hline
			\multirow{1}{*}{} & \multicolumn{1}{c}{ATLAS} &\multicolumn{1}{c}{CheckMATE}&\multicolumn{1}{c}{ATLAS}& \multicolumn{1}{c}{CheckMATE}  \\
			\bf Selection & & & & \\
			$\mathcal{L}\times \sigma \times \text{BF}$ & 1760 &1760  &92 &92   \\
			$\mathcal{L}\times \sigma \times \text{BF} \times \text{filt.eff}$ & 1322 &1322  & 69 &69  \\
			3 isolated leptons & \multirow{3}{*}{227} & \multirow{3}{*}{221.55} &\multirow{3}{*}{ 23.9} &\multirow{3}{*}{22.64}   \\
\hline\hline
			
	\end{tabular}}
\end{table}

\begin{table}[t]
	\centering
	\caption{\label{tab:cf-whonshell} Cutflow validation of the ATLAS analysis \texttt{atlas\_2106\_01676} for mass point $m(\tilde{\chi}_1^{\pm}/\tilde{\chi}_2^0, \tilde{\chi}_1^0) = (190, 60)~{\rm GeV}$. }
	\vspace{0.3cm}
	\resizebox{0.8\textwidth}{!}{%
		\begin{tabular}{l|rr} 
			\hline\hline
			Process & \multicolumn{2}{c}{$pp \to \tilde{\chi}_2^0 \tilde{\chi}_1^{\pm}, \tilde{\chi}_1^\pm \to W^{\pm} \tilde{\chi}_1^0, \tilde{\chi}_2^0 \to h \tilde{\chi}_1^0$} \\
			Point $m(\tilde{\chi}_2^0/\tilde{\chi}_1^\pm, \tilde{\chi}_1^0)$ & \multicolumn{2}{c}{(190, 60) GeV}\\
			Generated Events & \multicolumn{2}{c}{500,000} \\ \hline
			\multirow{1}{*}{} & \multicolumn{1}{c}{ATLAS} &\multicolumn{1}{c}{CheckMATE}  \\
			\bf Selection & &  \\
			$\mathcal{L}\times \sigma \times \text{BF}$ & 10927 &10927   \\
			$\mathcal{L}\times \sigma \times \text{BF} \times \text{filt.eff}$ & 1174 &1174 \\
			3 isolated leptons & \multirow{3}{*}{192} & \multirow{3}{*}{172.23}   \\
\hline\hline
			
	\end{tabular}}
\end{table}	

\begin{table}[t]
	\centering
	\caption{\label{tab:cf-C250N170} Cutflow validation of the ATLAS analysis \texttt{atlas\_2106\_01676} for mass point $m(\tilde{\chi}_1^{\pm}/\tilde{\chi}_2^0, \tilde{\chi}_1^0) = (250, 170)~{\rm GeV}$. }
	\vspace{0.3cm}
	\resizebox{1.0\textwidth}{!}{%
		\begin{tabular}{l|lclllcllclcl}
			\hline
			$\tilde{\chi}_1^{\pm}\tilde{\chi}_2^0 \to WZ\tilde{\chi}_1^0\tilde{\chi}_1^0$ \quad wino/bino(+) &
			\multicolumn{12}{c}{m($\tilde{\chi}_1^{\pm}/\tilde{\chi}_2^{0}, \tilde{\chi}_1^0$)=(250, 170)GeV}                                                                                                                                                                                                                                     \\ \hline
			& \multicolumn{8}{c}{ATLAS}                                                                                                                                         & \multicolumn{4}{c}{CheckMATE}                                     \\
			$\mathcal{L}\times \sigma \times \text{BF}$                  &\multicolumn{1}{c}{}& \multicolumn{7}{c}{3559}                                                                                                                                              & \multicolumn{4}{c}{3559}                                       \\
			$\mathcal{L}\times \sigma \times \text{BF} \times \text{filt.eff}$           &\multicolumn{1}{c}{}& \multicolumn{7}{c}{1638}                                                                                                                                              & \multicolumn{4}{c}{1638}                                       \\
			3 isolated leptons       &\multicolumn{1}{c}{}& \multicolumn{7}{c}{384}                                                                                                                                              & \multicolumn{4}{c}{434}                                       \\
			b veto                   &\multicolumn{1}{c}{}& \multicolumn{7}{c}{367}                                                                                                                                              & \multicolumn{4}{c}{408}                                       \\
\hline
	\end{tabular}}
\end{table}

\begin{table}[t]
	\centering
	\caption{\label{tab:cf-C125N85} Cutflow validation of the ATLAS analysis \texttt{atlas\_2106\_01676} for mass point $m(\tilde{\chi}_1^{\pm}/\tilde{\chi}_2^0, \tilde{\chi}_1^0) = (125, 85)~{\rm GeV}$. }
	\vspace{0.3cm}
	\resizebox{1.0\textwidth}{!}{%
		\begin{tabular}{l|lclllcllclcl}
			\hline
			$\tilde{\chi}_1^{\pm}\tilde{\chi}_2^0 \to WZ\tilde{\chi}_1^0\tilde{\chi}_1^0$ \quad wino/bino(+) &
			\multicolumn{12}{c}{m($\tilde{\chi}_1^{\pm}/\tilde{\chi}_2^{0}, \tilde{\chi}_1^0$)=(125, 85)GeV}                                                                                                                                                                                                                                     \\ \hline
			& \multicolumn{8}{c}{ATLAS}                                                                                                                                         & \multicolumn{4}{c}{CheckMATE}                                     \\
			$\mathcal{L}\times \sigma \times \text{BF}$                  &\multicolumn{1}{c}{}& \multicolumn{7}{c}{45634}                                                                                                                                              & \multicolumn{4}{c}{45634}                                       \\
			$\mathcal{L}\times \sigma \times \text{BF} \times \text{filt.eff}$           &\multicolumn{1}{c}{}& \multicolumn{7}{c}{16811}                                                                                                                                              & \multicolumn{4}{c}{16811}                                       \\
			3 isolated leptons       &\multicolumn{1}{c}{}& \multicolumn{7}{c}{2660}                                                                                                                                              & \multicolumn{4}{c}{3110}                                       \\
			b veto                   &\multicolumn{1}{c}{}& \multicolumn{7}{c}{2550}                                                                                                                                              & \multicolumn{4}{c}{2961}                                       \\
\hline
	\end{tabular}}
\end{table}

\begin{table}[t]
	\centering
	\caption{\label{tab:cf-C185N125} Cutflow validation of the ATLAS analysis \texttt{atlas\_2106\_01676} for mass point $m(\tilde{\chi}_1^{\pm}/\tilde{\chi}_2^0, \tilde{\chi}_1^0) = (185, 125)~{\rm GeV}$. }
	\vspace{0.3cm}
	\resizebox{1.0\textwidth}{!}{%
		\begin{tabular}{l|lclllcllclcl}
			\hline
			$\tilde{\chi}_1^{\pm}\tilde{\chi}_2^0 \to WZ\tilde{\chi}_1^0\tilde{\chi}_1^0$ \quad higssino &
			\multicolumn{12}{c}{m($\tilde{\chi}_2^{0}, \tilde{\chi}_1^0$)=(185, 125)GeV}                                                                                                                                                                                                                    \\ \hline
			&\multicolumn{8}{c}{ATLAS}                                                                                                                                         & \multicolumn{4}{c}{CheckMATE}                                     \\
			$\mathcal{L}\times \sigma \times \text{BF}$                  &\multicolumn{1}{c}{}& \multicolumn{7}{c}{3912}                                                                                                                                              & \multicolumn{4}{c}{3912}                                       \\
			$\mathcal{L}\times \sigma \times \text{BF} \times \text{filt.eff}$           &\multicolumn{1}{c}{}& \multicolumn{7}{c}{1550}                                                                                                                                              & \multicolumn{4}{c}{1550}                                       \\
			3 isolated leptons       &\multicolumn{1}{c}{}& \multicolumn{7}{c}{392}                                                                                                                                              & \multicolumn{4}{c}{409}                                       \\
			b veto                   &\multicolumn{1}{c}{}& \multicolumn{7}{c}{374}                                                                                                                                              & \multicolumn{4}{c}{388}                                       \\
\hline
	\end{tabular}}
\end{table}

\begin{table}[t]
	\centering
	\caption{\label{tab:cf-C100N40} Cutflow validation of the ATLAS analysis \texttt{atlas\_2106\_01676} for mass point $m(\tilde{\chi}_1^{\pm}/\tilde{\chi}_2^0, \tilde{\chi}_1^0) = (100, 40)~{\rm GeV}$. }
	\vspace{0.3cm}
	\resizebox{1.0\textwidth}{!}{%
		\begin{tabular}{l|lclllcllclcl}
			\hline
			$\tilde{\chi}_1^{\pm}\tilde{\chi}_2^0 \to WZ\tilde{\chi}_1^0\tilde{\chi}_1^0$ \quad higssino &
			\multicolumn{12}{c}{m($\tilde{\chi}_2^{0}, \tilde{\chi}_1^0$)=(100, 40)GeV}                                                                                                                                                                                                                    \\ \hline
			&\multicolumn{8}{c}{ATLAS}                                                                                                                                         & \multicolumn{4}{c}{CheckMATE}                                     \\
			$\mathcal{L}\times \sigma \times \text{BF}$                  &\multicolumn{1}{c}{}& \multicolumn{7}{c}{48276}                                                                                                                                              & \multicolumn{4}{c}{48276}                                       \\
			$\mathcal{L}\times \sigma \times \text{BF} \times \text{filt.eff}$           &\multicolumn{1}{c}{}& \multicolumn{7}{c}{16084}                                                                                                                                              & \multicolumn{4}{c}{16084}                                       \\
			3 isolated leptons       &\multicolumn{1}{c}{}& \multicolumn{7}{c}{3760}                                                                                                                                              & \multicolumn{4}{c}{4247}                                       \\
			b veto                   &\multicolumn{1}{c}{}& \multicolumn{7}{c}{3620}                                                                                                                                              & \multicolumn{4}{c}{4027}                                       \\
\hline
	\end{tabular}}
\end{table}

\section*{Acknowledgement}
This work is supported by the National Natural Science Foundation of China (NNSFC) under grant No. 12075076.


\bibliographystyle{CitationStyle}
\bibliography{myrefs}

\providecommand{\href}[2]{#2}\begingroup\raggedright\begin{thebibliography}{100}

\bibitem{Abi:2021gix}
{\scshape Muon g-2} collaboration, B.~Abi et~al., \emph{{Measurement of the
  Positive Muon Anomalous Magnetic Moment to 0.46 ppm}},
  \href{https://doi.org/10.1103/PhysRevLett.126.141801}{\emph{Phys. Rev. Lett.}
  {\bfseries 126} (2021) 141801}
  [\href{https://arxiv.org/abs/2104.03281}{{\ttfamily 2104.03281}}].

\bibitem{Bennett:2006fi}
{\scshape Muon g-2} collaboration, G.~W. Bennett et~al., \emph{{Final Report of
  the Muon E821 Anomalous Magnetic Moment Measurement at BNL}},
  \href{https://doi.org/10.1103/PhysRevD.73.072003}{\emph{Phys. Rev. D}
  {\bfseries 73} (2006) 072003}
  [\href{https://arxiv.org/abs/hep-ex/0602035}{{\ttfamily hep-ex/0602035}}].

\bibitem{Aoyama:2020ynm}
T.~Aoyama et~al., \emph{{The anomalous magnetic moment of the muon in the
  Standard Model}},
  \href{https://doi.org/10.1016/j.physrep.2020.07.006}{\emph{Phys. Rept.}
  {\bfseries 887} (2020) 1} [\href{https://arxiv.org/abs/2006.04822}{{\ttfamily
  2006.04822}}].

\bibitem{Aoyama:2012wk}
T.~Aoyama, M.~Hayakawa, T.~Kinoshita and M.~Nio, \emph{{Complete Tenth-Order
  QED Contribution to the Muon $g-2$}},
  \href{https://doi.org/10.1103/PhysRevLett.109.111808}{\emph{Phys. Rev. Lett.}
  {\bfseries 109} (2012) 111808}
  [\href{https://arxiv.org/abs/1205.5370}{{\ttfamily 1205.5370}}].

\bibitem{Aoyama:2019ryr}
T.~Aoyama, T.~Kinoshita and M.~Nio, \emph{{Theory of the Anomalous Magnetic
  Moment of the Electron}},
  \href{https://doi.org/10.3390/atoms7010028}{\emph{Atoms} {\bfseries 7} (2019)
  28}.

\bibitem{Czarnecki:2002nt}
A.~Czarnecki, W.~J. Marciano and A.~Vainshtein, \emph{{Refinements in
  electroweak contributions to the muon anomalous magnetic moment}},
  \href{https://doi.org/10.1103/PhysRevD.67.073006}{\emph{Phys. Rev.}
  {\bfseries D67} (2003) 073006}
  [\href{https://arxiv.org/abs/hep-ph/0212229}{{\ttfamily hep-ph/0212229}}].

\bibitem{Gnendiger:2013pva}
C.~Gnendiger, D.~St{\"o}ckinger and H.~St{\"o}ckinger-Kim, \emph{{The
  electroweak contributions to $(g-2)_\mu$ after the Higgs boson mass
  measurement}}, \href{https://doi.org/10.1103/PhysRevD.88.053005}{\emph{Phys.
  Rev.} {\bfseries D88} (2013) 053005}
  [\href{https://arxiv.org/abs/1306.5546}{{\ttfamily 1306.5546}}].

\bibitem{Davier:2017zfy}
M.~Davier, A.~Hoecker, B.~Malaescu and Z.~Zhang, \emph{{Reevaluation of the
  hadronic vacuum polarisation contributions to the Standard Model predictions
  of the muon $g-2$ and ${\alpha (m_Z^2)}$ using newest hadronic cross-section
  data}}, \href{https://doi.org/10.1140/epjc/s10052-017-5161-6}{\emph{Eur.
  Phys. J.} {\bfseries C77} (2017) 827}
  [\href{https://arxiv.org/abs/1706.09436}{{\ttfamily 1706.09436}}].

\bibitem{Keshavarzi:2018mgv}
A.~Keshavarzi, D.~Nomura and T.~Teubner, \emph{{Muon $g-2$ and $\alpha(M_Z^2)$:
  a new data-based analysis}},
  \href{https://doi.org/10.1103/PhysRevD.97.114025}{\emph{Phys. Rev.}
  {\bfseries D97} (2018) 114025}
  [\href{https://arxiv.org/abs/1802.02995}{{\ttfamily 1802.02995}}].

\bibitem{Colangelo:2018mtw}
G.~Colangelo, M.~Hoferichter and P.~Stoffer, \emph{{Two-pion contribution to
  hadronic vacuum polarization}},
  \href{https://doi.org/10.1007/JHEP02(2019)006}{\emph{JHEP} {\bfseries 02}
  (2019) 006} [\href{https://arxiv.org/abs/1810.00007}{{\ttfamily
  1810.00007}}].

\bibitem{Hoferichter:2019gzf}
M.~Hoferichter, B.-L. Hoid and B.~Kubis, \emph{{Three-pion contribution to
  hadronic vacuum polarization}},
  \href{https://doi.org/10.1007/JHEP08(2019)137}{\emph{JHEP} {\bfseries 08}
  (2019) 137} [\href{https://arxiv.org/abs/1907.01556}{{\ttfamily
  1907.01556}}].

\bibitem{Davier:2019can}
M.~Davier, A.~Hoecker, B.~Malaescu and Z.~Zhang, \emph{{A new evaluation of the
  hadronic vacuum polarisation contributions to the muon anomalous magnetic
  moment and to $\mathbf{\boldsymbol\alpha(m_Z^2)}$}},
  \href{https://doi.org/10.1140/epjc/s10052-020-7792-2}{\emph{Eur. Phys. J.}
  {\bfseries C80} (2020) 241}
  [\href{https://arxiv.org/abs/1908.00921}{{\ttfamily 1908.00921}}].

\bibitem{Keshavarzi:2019abf}
A.~Keshavarzi, D.~Nomura and T.~Teubner, \emph{{The $g-2$ of charged leptons,
  $\alpha(M_Z^2)$ and the hyperfine splitting of muonium}},
  \href{https://doi.org/10.1103/PhysRevD.101.014029}{\emph{Phys. Rev.}
  {\bfseries D101} (2020) 014029}
  [\href{https://arxiv.org/abs/1911.00367}{{\ttfamily 1911.00367}}].

\bibitem{Kurz:2014wya}
A.~Kurz, T.~Liu, P.~Marquard and M.~Steinhauser, \emph{{Hadronic contribution
  to the muon anomalous magnetic moment to next-to-next-to-leading order}},
  \href{https://doi.org/10.1016/j.physletb.2014.05.043}{\emph{Phys. Lett.}
  {\bfseries B734} (2014) 144}
  [\href{https://arxiv.org/abs/1403.6400}{{\ttfamily 1403.6400}}].

\bibitem{Melnikov:2003xd}
K.~Melnikov and A.~Vainshtein, \emph{{Hadronic light-by-light scattering
  contribution to the muon anomalous magnetic moment revisited}},
  \href{https://doi.org/10.1103/PhysRevD.70.113006}{\emph{Phys. Rev.}
  {\bfseries D70} (2004) 113006}
  [\href{https://arxiv.org/abs/hep-ph/0312226}{{\ttfamily hep-ph/0312226}}].

\bibitem{Masjuan:2017tvw}
P.~Masjuan and P.~S{\'a}nchez-Puertas, \emph{{Pseudoscalar-pole contribution to
  the $(g_{\mu}-2)$: a rational approach}},
  \href{https://doi.org/10.1103/PhysRevD.95.054026}{\emph{Phys. Rev.}
  {\bfseries D95} (2017) 054026}
  [\href{https://arxiv.org/abs/1701.05829}{{\ttfamily 1701.05829}}].

\bibitem{Colangelo:2017fiz}
G.~Colangelo, M.~Hoferichter, M.~Procura and P.~Stoffer, \emph{{Dispersion
  relation for hadronic light-by-light scattering: two-pion contributions}},
  \href{https://doi.org/10.1007/JHEP04(2017)161}{\emph{JHEP} {\bfseries 04}
  (2017) 161} [\href{https://arxiv.org/abs/1702.07347}{{\ttfamily
  1702.07347}}].

\bibitem{Hoferichter:2018kwz}
M.~Hoferichter, B.-L. Hoid, B.~Kubis, S.~Leupold and S.~P. Schneider,
  \emph{{Dispersion relation for hadronic light-by-light scattering: pion
  pole}}, \href{https://doi.org/10.1007/JHEP10(2018)141}{\emph{JHEP} {\bfseries
  10} (2018) 141} [\href{https://arxiv.org/abs/1808.04823}{{\ttfamily
  1808.04823}}].

\bibitem{Gerardin:2019vio}
A.~G{\'e}rardin, H.~B. Meyer and A.~Nyffeler, \emph{{Lattice calculation of the
  pion transition form factor with $N_f=2+1$ Wilson quarks}},
  \href{https://doi.org/10.1103/PhysRevD.100.034520}{\emph{Phys. Rev.}
  {\bfseries D100} (2019) 034520}
  [\href{https://arxiv.org/abs/1903.09471}{{\ttfamily 1903.09471}}].

\bibitem{Bijnens:2019ghy}
J.~Bijnens, N.~Hermansson-Truedsson and A.~Rodr{\'i}guez-S{\'a}nchez,
  \emph{{Short-distance constraints for the HLbL contribution to the muon
  anomalous magnetic moment}},
  \href{https://doi.org/10.1016/j.physletb.2019.134994}{\emph{Phys. Lett.}
  {\bfseries B798} (2019) 134994}
  [\href{https://arxiv.org/abs/1908.03331}{{\ttfamily 1908.03331}}].

\bibitem{Colangelo:2019uex}
G.~Colangelo, F.~Hagelstein, M.~Hoferichter, L.~Laub and P.~Stoffer,
  \emph{{Longitudinal short-distance constraints for the hadronic
  light-by-light contribution to $(g-2)_\mu$ with large-$N_c$ Regge models}},
  \href{https://doi.org/10.1007/JHEP03(2020)101}{\emph{JHEP} {\bfseries 03}
  (2020) 101} [\href{https://arxiv.org/abs/1910.13432}{{\ttfamily
  1910.13432}}].

\bibitem{Blum:2019ugy}
T.~Blum, N.~Christ, M.~Hayakawa, T.~Izubuchi, L.~Jin, C.~Jung et~al.,
  \emph{{The hadronic light-by-light scattering contribution to the muon
  anomalous magnetic moment from lattice QCD}},
  \href{https://doi.org/10.1103/PhysRevLett.124.132002}{\emph{Phys. Rev. Lett.}
  {\bfseries 124} (2020) 132002}
  [\href{https://arxiv.org/abs/1911.08123}{{\ttfamily 1911.08123}}].

\bibitem{Colangelo:2014qya}
G.~Colangelo, M.~Hoferichter, A.~Nyffeler, M.~Passera and P.~Stoffer,
  \emph{{Remarks on higher-order hadronic corrections to the muon $g-2$}},
  \href{https://doi.org/10.1016/j.physletb.2014.06.012}{\emph{Phys. Lett.}
  {\bfseries B735} (2014) 90}
  [\href{https://arxiv.org/abs/1403.7512}{{\ttfamily 1403.7512}}].

\bibitem{Athron:2021iuf}
P.~Athron, C.~Bal\'azs, D.~H. Jacob, W.~Kotlarski, D.~St\"ockinger and
  H.~St\"ockinger-Kim, \emph{{New physics explanations of $a_\mu$ in light of
  the FNAL muon $g-2$ measurement}},
  \href{https://arxiv.org/abs/2104.03691}{{\ttfamily 2104.03691}}.

\bibitem{Fayet:1976cr}
P.~Fayet and S.~Ferrara, \emph{{Supersymmetry}},
  \href{https://doi.org/10.1016/0370-1573(77)90066-7}{\emph{Phys. Rept.}
  {\bfseries 32} (1977) 249}.

\bibitem{Haber:1984rc}
H.~E. Haber and G.~L. Kane, \emph{{The Search for Supersymmetry: Probing
  Physics Beyond the Standard Model}},
  \href{https://doi.org/10.1016/0370-1573(85)90051-1}{\emph{Phys. Rept.}
  {\bfseries 117} (1985) 75}.

\bibitem{Martin:1997ns}
S.~P. Martin, \emph{{A Supersymmetry primer}},
  \href{https://arxiv.org/abs/hep-ph/9709356}{{\ttfamily hep-ph/9709356}}.

\bibitem{Jungman:1995df}
G.~Jungman, M.~Kamionkowski and K.~Griest, \emph{{Supersymmetric dark matter}},
  \href{https://doi.org/10.1016/0370-1573(95)00058-5}{\emph{Phys. Rept.}
  {\bfseries 267} (1996) 195}
  [\href{https://arxiv.org/abs/hep-ph/9506380}{{\ttfamily hep-ph/9506380}}].

\bibitem{Martin:2001st}
S.~P. Martin and J.~D. Wells, \emph{{Muon Anomalous Magnetic Dipole Moment in
  Supersymmetric Theories}},
  \href{https://doi.org/10.1103/PhysRevD.64.035003}{\emph{Phys. Rev. D}
  {\bfseries 64} (2001) 035003}
  [\href{https://arxiv.org/abs/hep-ph/0103067}{{\ttfamily hep-ph/0103067}}].

\bibitem{Domingo:2008bb}
F.~Domingo and U.~Ellwanger, \emph{{Constraints from the Muon g-2 on the
  Parameter Space of the NMSSM}},
  \href{https://doi.org/10.1088/1126-6708/2008/07/079}{\emph{JHEP} {\bfseries
  07} (2008) 079} [\href{https://arxiv.org/abs/0806.0733}{{\ttfamily
  0806.0733}}].

\bibitem{Moroi:1995yh}
T.~Moroi, \emph{{The Muon anomalous magnetic dipole moment in the minimal
  supersymmetric standard model}},
  \href{https://doi.org/10.1103/PhysRevD.53.6565}{\emph{Phys. Rev. D}
  {\bfseries 53} (1996) 6565}
  [\href{https://arxiv.org/abs/hep-ph/9512396}{{\ttfamily hep-ph/9512396}}].

\bibitem{Hollik:1997vb}
W.~Hollik, J.~I. Illana, S.~Rigolin and D.~Stockinger, \emph{{One loop MSSM
  contribution to the weak magnetic dipole moments of heavy fermions}},
  \href{https://doi.org/10.1016/S0370-2693(97)01259-8}{\emph{Phys. Lett. B}
  {\bfseries 416} (1998) 345}
  [\href{https://arxiv.org/abs/hep-ph/9707437}{{\ttfamily hep-ph/9707437}}].

\bibitem{Athron:2015rva}
P.~Athron, M.~Bach, H.~G. Fargnoli, C.~Gnendiger, R.~Greifenhagen, J.-h. Park
  et~al., \emph{{GM2Calc: Precise MSSM prediction for $(g - 2)$ of the muon}},
  \href{https://doi.org/10.1140/epjc/s10052-015-3870-2}{\emph{Eur. Phys. J. C}
  {\bfseries 76} (2016) 62} [\href{https://arxiv.org/abs/1510.08071}{{\ttfamily
  1510.08071}}].

\bibitem{Endo:2021zal}
M.~Endo, K.~Hamaguchi, S.~Iwamoto and T.~Kitahara, \emph{{Supersymmetric
  interpretation of the muon g \textendash{} 2 anomaly}},
  \href{https://doi.org/10.1007/JHEP07(2021)075}{\emph{JHEP} {\bfseries 07}
  (2021) 075} [\href{https://arxiv.org/abs/2104.03217}{{\ttfamily
  2104.03217}}].

\bibitem{Stockinger:2006zn}
D.~Stockinger, \emph{{The Muon Magnetic Moment and Supersymmetry}},
  \href{https://doi.org/10.1088/0954-3899/34/2/R01}{\emph{J. Phys. G}
  {\bfseries 34} (2007) R45}
  [\href{https://arxiv.org/abs/hep-ph/0609168}{{\ttfamily hep-ph/0609168}}].

\bibitem{Czarnecki:2001pv}
A.~Czarnecki and W.~J. Marciano, \emph{{The Muon anomalous magnetic moment: A
  Harbinger for 'new physics'}},
  \href{https://doi.org/10.1103/PhysRevD.64.013014}{\emph{Phys. Rev. D}
  {\bfseries 64} (2001) 013014}
  [\href{https://arxiv.org/abs/hep-ph/0102122}{{\ttfamily hep-ph/0102122}}].

\bibitem{Cao:2011sn}
J.~Cao, Z.~Heng, D.~Li and J.~M. Yang, \emph{{Current experimental constraints
  on the lightest Higgs boson mass in the constrained MSSM}},
  \href{https://doi.org/10.1016/j.physletb.2012.03.052}{\emph{Phys. Lett. B}
  {\bfseries 710} (2012) 665}
  [\href{https://arxiv.org/abs/1112.4391}{{\ttfamily 1112.4391}}].

\bibitem{Kang:2016iok}
Z.~Kang, \emph{{$H_{u,d}$-messenger Couplings Address the $\mu/B_\mu$
  \textbackslash{}\& $A_t/m_{H_u}^2$ Problem and $(g-2)_\mu$ Puzzle}},
  \href{https://arxiv.org/abs/1610.06024}{{\ttfamily 1610.06024}}.

\bibitem{Zhu:2016ncq}
B.~Zhu, R.~Ding and T.~Li, \emph{{Higgs mass and muon anomalous magnetic moment
  in the MSSM with gauge-gravity hybrid mediation}},
  \href{https://doi.org/10.1103/PhysRevD.96.035029}{\emph{Phys. Rev. D}
  {\bfseries 96} (2017) 035029}
  [\href{https://arxiv.org/abs/1610.09840}{{\ttfamily 1610.09840}}].

\bibitem{Yanagida:2017dao}
T.~T. Yanagida and N.~Yokozaki, \emph{{Muon g \ensuremath{-} 2 in MSSM gauge
  mediation revisited}},
  \href{https://doi.org/10.1016/j.physletb.2017.07.002}{\emph{Phys. Lett. B}
  {\bfseries 772} (2017) 409}
  [\href{https://arxiv.org/abs/1704.00711}{{\ttfamily 1704.00711}}].

\bibitem{Hagiwara:2017lse}
K.~Hagiwara, K.~Ma and S.~Mukhopadhyay, \emph{{Closing in on the chargino
  contribution to the muon g-2 in the MSSM: current LHC constraints}},
  \href{https://doi.org/10.1103/PhysRevD.97.055035}{\emph{Phys. Rev. D}
  {\bfseries 97} (2018) 055035}
  [\href{https://arxiv.org/abs/1706.09313}{{\ttfamily 1706.09313}}].

\bibitem{Cox:2018qyi}
P.~Cox, C.~Han and T.~T. Yanagida, \emph{{Muon $g-2$ and dark matter in the
  minimal supersymmetric standard model}},
  \href{https://doi.org/10.1103/PhysRevD.98.055015}{\emph{Phys. Rev. D}
  {\bfseries 98} (2018) 055015}
  [\href{https://arxiv.org/abs/1805.02802}{{\ttfamily 1805.02802}}].

\bibitem{Tran:2018kxv}
H.~M. Tran and H.~T. Nguyen, \emph{{GUT-inspired MSSM in light of muon $g-2$
  and LHC results at $\sqrt{s}=13$ TeV}},
  \href{https://doi.org/10.1103/PhysRevD.99.035040}{\emph{Phys. Rev. D}
  {\bfseries 99} (2019) 035040}
  [\href{https://arxiv.org/abs/1812.11757}{{\ttfamily 1812.11757}}].

\bibitem{Padley:2015uma}
B.~P. Padley, K.~Sinha and K.~Wang, \emph{{Natural Supersymmetry, Muon $g-2$,
  and the Last Crevices for the Top Squark}},
  \href{https://doi.org/10.1103/PhysRevD.92.055025}{\emph{Phys. Rev. D}
  {\bfseries 92} (2015) 055025}
  [\href{https://arxiv.org/abs/1505.05877}{{\ttfamily 1505.05877}}].

\bibitem{Choudhury:2017fuu}
A.~Choudhury, L.~Darm\'e, L.~Roszkowski, E.~M. Sessolo and S.~Trojanowski,
  \emph{{Muon g \ensuremath{-} 2 and related phenomenology in constrained
  vector-like extensions of the MSSM}},
  \href{https://doi.org/10.1007/JHEP05(2017)072}{\emph{JHEP} {\bfseries 05}
  (2017) 072} [\href{https://arxiv.org/abs/1701.08778}{{\ttfamily
  1701.08778}}].

\bibitem{Okada:2016wlm}
N.~Okada and H.~M. Tran, \emph{{125 GeV Higgs boson mass and muon $g-2$ in 5D
  MSSM}}, \href{https://doi.org/10.1103/PhysRevD.94.075016}{\emph{Phys. Rev. D}
  {\bfseries 94} (2016) 075016}
  [\href{https://arxiv.org/abs/1606.05329}{{\ttfamily 1606.05329}}].

\bibitem{Du:2017str}
X.~Du and F.~Wang, \emph{{NMSSM From Alternative Deflection in Generalized
  Deflected Anomaly Mediated SUSY Breaking}},
  \href{https://doi.org/10.1140/epjc/s10052-018-5921-y}{\emph{Eur. Phys. J. C}
  {\bfseries 78} (2018) 431}
  [\href{https://arxiv.org/abs/1710.06105}{{\ttfamily 1710.06105}}].

\bibitem{Ning:2017dng}
X.~Ning and F.~Wang, \emph{{Solving the muon g-2 anomaly within the NMSSM from
  generalized deflected AMSB}},
  \href{https://doi.org/10.1007/JHEP08(2017)089}{\emph{JHEP} {\bfseries 08}
  (2017) 089} [\href{https://arxiv.org/abs/1704.05079}{{\ttfamily
  1704.05079}}].

\bibitem{Wang:2018vxp}
K.~Wang, F.~Wang, J.~Zhu and Q.~Jie, \emph{{The semi-constrained NMSSM in light
  of muon g-2, LHC, and dark matter constraints}},
  \href{https://doi.org/10.1088/1674-1137/42/10/103109}{\emph{Chin. Phys. C}
  {\bfseries 42} (2018) 103109}
  [\href{https://arxiv.org/abs/1811.04435}{{\ttfamily 1811.04435}}].

\bibitem{Yang:2018guw}
J.-L. Yang, T.-F. Feng, Y.-L. Yan, W.~Li, S.-M. Zhao and H.-B. Zhang,
  \emph{{Lepton-flavor violation and two loop electroweak corrections to
  $(g-2)_\mu$ in the B-L symmetric SSM}},
  \href{https://doi.org/10.1103/PhysRevD.99.015002}{\emph{Phys. Rev. D}
  {\bfseries 99} (2019) 015002}
  [\href{https://arxiv.org/abs/1812.03860}{{\ttfamily 1812.03860}}].

\bibitem{Liu:2020nsm}
C.-X. Liu, H.-B. Zhang, J.-L. Yang, S.-M. Zhao, Y.-B. Liu and T.-F. Feng,
  \emph{{Higgs boson decay $h\rightarrow Z\gamma$ and muon magnetic dipole
  moment in the $\mu\nu$SSM}},
  \href{https://doi.org/10.1007/JHEP04(2020)002}{\emph{JHEP} {\bfseries 04}
  (2020) 002} [\href{https://arxiv.org/abs/2002.04370}{{\ttfamily
  2002.04370}}].

\bibitem{Cao:2019evo}
J.~Cao, J.~Lian, L.~Meng, Y.~Yue and P.~Zhu, \emph{{Anomalous muon magnetic
  moment in the inverse seesaw extended next-to-minimal supersymmetric standard
  model}}, \href{https://doi.org/10.1103/PhysRevD.101.095009}{\emph{Phys. Rev.
  D} {\bfseries 101} (2020) 095009}
  [\href{https://arxiv.org/abs/1912.10225}{{\ttfamily 1912.10225}}].

\bibitem{Cao:2021lmj}
J.~Cao, Y.~He, J.~Lian, D.~Zhang and P.~Zhu, \emph{{Electron and muon anomalous
  magnetic moments in the inverse seesaw extended NMSSM}},
  \href{https://doi.org/10.1103/PhysRevD.104.055009}{\emph{Phys. Rev. D}
  {\bfseries 104} (2021) 055009}
  [\href{https://arxiv.org/abs/2102.11355}{{\ttfamily 2102.11355}}].

\bibitem{Ke:2021kgy}
W.~Ke and P.~Slavich, \emph{{Higgs-mass constraints on a supersymmetric
  solution of the muon g-2 anomaly}},
  \href{https://arxiv.org/abs/2109.15277}{{\ttfamily 2109.15277}}.

\bibitem{Lamborn:2021snt}
J.~L. Lamborn, T.~Li, J.~A. Maxin and D.~V. Nanopoulos, \emph{{Resolving the
  $(g-2)_{\mu}$ Discrepancy with $\mathcal{F}$-$SU$(5) Intersecting D-branes}},
   \href{https://arxiv.org/abs/2108.08084}{{\ttfamily 2108.08084}}.

\bibitem{Li:2021xmw}
S.~Li, Y.~Xiao and J.~M. Yang, \emph{{Constraining CP-phases in SUSY: an
  interplay of muon/electron $g-2$ and electron EDM}},
  \href{https://arxiv.org/abs/2108.00359}{{\ttfamily 2108.00359}}.

\bibitem{Nakai:2021mha}
Y.~Nakai, M.~Reece and M.~Suzuki, \emph{{Supersymmetric alignment models for (g
  \ensuremath{-} 2)$_{\mu}$}},
  \href{https://doi.org/10.1007/JHEP10(2021)068}{\emph{JHEP} {\bfseries 10}
  (2021) 068} [\href{https://arxiv.org/abs/2107.10268}{{\ttfamily
  2107.10268}}].

\bibitem{Li:2021koa}
S.~Li, Y.~Xiao and J.~M. Yang, \emph{{Can electron and muon $g-2$ anomalies be
  jointly explained in SUSY?}},
  \href{https://arxiv.org/abs/2107.04962}{{\ttfamily 2107.04962}}.

\bibitem{Kim:2021suj}
J.~S. Kim, D.~E. Lopez-Fogliani, A.~D. Perez and R.~R. de~Austri, \emph{{The
  new $(g-2)_\mu$ and Right-Handed Sneutrino Dark Matter}},
  \href{https://arxiv.org/abs/2107.02285}{{\ttfamily 2107.02285}}.

\bibitem{Li:2021pnt}
Z.~Li, G.-L. Liu, F.~Wang, J.~M. Yang and Y.~Zhang, \emph{{Gluino-SUGRA
  scenarios in light of FNAL muon g-2 anomaly}},
  \href{https://arxiv.org/abs/2106.04466}{{\ttfamily 2106.04466}}.

\bibitem{Altmannshofer:2021hfu}
W.~Altmannshofer, S.~A. Gadam, S.~Gori and N.~Hamer, \emph{{Explaining
  $(g-2)_{\mu}$ with Multi-TeV Sleptons}},
  \href{https://arxiv.org/abs/2104.08293}{{\ttfamily 2104.08293}}.

\bibitem{Baer:2021aax}
H.~Baer, V.~Barger and H.~Serce, \emph{{Anomalous muon magnetic moment,
  supersymmetry, naturalness, LHC search limits and the landscape}},
  \href{https://doi.org/10.1016/j.physletb.2021.136480}{\emph{Phys. Lett. B}
  {\bfseries 820} (2021) 136480}
  [\href{https://arxiv.org/abs/2104.07597}{{\ttfamily 2104.07597}}].

\bibitem{Chakraborti:2021bmv}
M.~Chakraborti, L.~Roszkowski and S.~Trojanowski, \emph{{GUT-constrained
  supersymmetry and dark matter in light of the new $(g-2)_\mu$
  determination}}, \href{https://doi.org/10.1007/JHEP05(2021)252}{\emph{JHEP}
  {\bfseries 05} (2021) 252}
  [\href{https://arxiv.org/abs/2104.04458}{{\ttfamily 2104.04458}}].

\bibitem{Aboubrahim:2021xfi}
A.~Aboubrahim, M.~Klasen and P.~Nath, \emph{{What the Fermilab muon $g-$2
  experiment tells us about discovering supersymmetry at high luminosity and
  high energy upgrades to the LHC}},
  \href{https://doi.org/10.1103/PhysRevD.104.035039}{\emph{Phys. Rev. D}
  {\bfseries 104} (2021) 035039}
  [\href{https://arxiv.org/abs/2104.03839}{{\ttfamily 2104.03839}}].

\bibitem{Iwamoto:2021aaf}
S.~Iwamoto, T.~T. Yanagida and N.~Yokozaki, \emph{{Wino-Higgsino dark matter in
  MSSM from the g-2 anomaly}},
  \href{https://doi.org/10.1016/j.physletb.2021.136768}{\emph{Phys. Lett. B}
  {\bfseries 823} (2021) 136768}
  [\href{https://arxiv.org/abs/2104.03223}{{\ttfamily 2104.03223}}].

\bibitem{Chakraborti:2021dli}
M.~Chakraborti, S.~Heinemeyer and I.~Saha, \emph{{The new ''MUON G-2'' Result
  and Supersymmetry}},  \href{https://arxiv.org/abs/2104.03287}{{\ttfamily
  2104.03287}}.

\bibitem{Cao:2021tuh}
J.~Cao, J.~Lian, Y.~Pan, D.~Zhang and P.~Zhu, \emph{{Improved $(g-2)_\mu$
  measurement and singlino dark matter in $\mu$-term extended
  $\mathbb{Z}_3$-NMSSM}},
  \href{https://doi.org/10.1007/JHEP09(2021)175}{\emph{JHEP} {\bfseries 09}
  (2021) 175} [\href{https://arxiv.org/abs/2104.03284}{{\ttfamily
  2104.03284}}].

\bibitem{Yin:2021mls}
W.~Yin, \emph{{Muon g \ensuremath{-} 2 anomaly in anomaly mediation}},
  \href{https://doi.org/10.1007/JHEP06(2021)029}{\emph{JHEP} {\bfseries 06}
  (2021) 029} [\href{https://arxiv.org/abs/2104.03259}{{\ttfamily
  2104.03259}}].

\bibitem{Zhang:2021gun}
H.-B. Zhang, C.-X. Liu, J.-L. Yang and T.-F. Feng, \emph{{Muon anomalous
  magnetic dipole moment in the $\mu\nu$SSM}},
  \href{https://arxiv.org/abs/2104.03489}{{\ttfamily 2104.03489}}.

\bibitem{Ibe:2021cvf}
M.~Ibe, S.~Kobayashi, Y.~Nakayama and S.~Shirai, \emph{{Muon $g-2$ in Gauge
  Mediation without SUSY CP Problem}},
  \href{https://arxiv.org/abs/2104.03289}{{\ttfamily 2104.03289}}.

\bibitem{Han:2021ify}
C.~Han, \emph{{Muon g-2 and CP violation in MSSM}},
  \href{https://arxiv.org/abs/2104.03292}{{\ttfamily 2104.03292}}.

\bibitem{Wang:2021bcx}
F.~Wang, L.~Wu, Y.~Xiao, J.~M. Yang and Y.~Zhang, \emph{{GUT-scale constrained
  SUSY in light of new muon g-2 measurement}},
  \href{https://doi.org/10.1016/j.nuclphysb.2021.115486}{\emph{Nucl. Phys. B}
  {\bfseries 970} (2021) 115486}
  [\href{https://arxiv.org/abs/2104.03262}{{\ttfamily 2104.03262}}].

\bibitem{Zheng:2021gug}
M.-D. Zheng and H.-H. Zhang, \emph{{Studying the $b \to s \ell + \ell$-
  anomalies and (g-2)\ensuremath{\mu} in R-parity violating MSSM framework with
  the inverse seesaw mechanism}},
  \href{https://doi.org/10.1103/PhysRevD.104.115023}{\emph{Phys. Rev. D}
  {\bfseries 104} (2021) 115023}.

\bibitem{Chakraborti:2021mbr}
M.~Chakraborti, S.~Heinemeyer, I.~Saha and C.~Schappacher, \emph{{$(g-2)_\mu$
  and SUSY Dark Matter: Direct Detection and Collider Search Complementarity}},
   \href{https://arxiv.org/abs/2112.01389}{{\ttfamily 2112.01389}}.

\bibitem{Aboubrahim:2021myl}
A.~Aboubrahim, M.~Klasen, P.~Nath and R.~M. Syed, \emph{{Tests of gluino-driven
  radiative breaking of the electroweak symmetry at the LHC}},  in \emph{{10th
  International Conference on New Frontiers in Physics}}, 12, 2021,
  \href{https://arxiv.org/abs/2112.04986}{{\ttfamily 2112.04986}}.

\bibitem{Ali:2021kxa}
M.~I. Ali, M.~Chakraborti, U.~Chattopadhyay and S.~Mukherjee, \emph{{Muon and
  Electron $(g-2)$ Anomalies with Non-Holomorphic Interactions in MSSM}},
  \href{https://arxiv.org/abs/2112.09867}{{\ttfamily 2112.09867}}.

\bibitem{Wang:2021lwi}
K.~Wang and J.~Zhu, \emph{{A smuon in the NMSSM confronted with the muon g-2
  and SUSY searches}},  \href{https://arxiv.org/abs/2112.14576}{{\ttfamily
  2112.14576}}.

\bibitem{Chakraborti:2020vjp}
M.~Chakraborti, S.~Heinemeyer and I.~Saha, \emph{{Improved $(g-2)_\mu$
  Measurements and Supersymmetry}},
  \href{https://doi.org/10.1140/epjc/s10052-020-08504-8}{\emph{Eur. Phys. J. C}
  {\bfseries 80} (2020) 984}
  [\href{https://arxiv.org/abs/2006.15157}{{\ttfamily 2006.15157}}].

\bibitem{Baum:2021qzx}
S.~Baum, M.~Carena, N.~R. Shah and C.~E.~M. Wagner, \emph{{The tiny (g-2) muon
  wobble from small-$\mu$ supersymmetry}},
  \href{https://doi.org/10.1007/JHEP01(2022)025}{\emph{JHEP} {\bfseries 01}
  (2022) 025} [\href{https://arxiv.org/abs/2104.03302}{{\ttfamily
  2104.03302}}].

\bibitem{XENON:2018voc}
{\scshape XENON} collaboration, E.~Aprile et~al., \emph{{Dark Matter Search
  Results from a One Ton-Year Exposure of XENON1T}},
  \href{https://doi.org/10.1103/PhysRevLett.121.111302}{\emph{Phys. Rev. Lett.}
  {\bfseries 121} (2018) 111302}
  [\href{https://arxiv.org/abs/1805.12562}{{\ttfamily 1805.12562}}].

\bibitem{XENON:2019rxp}
{\scshape XENON} collaboration, E.~Aprile et~al., \emph{{Constraining the
  spin-dependent WIMP-nucleon cross sections with XENON1T}},
  \href{https://doi.org/10.1103/PhysRevLett.122.141301}{\emph{Phys. Rev. Lett.}
  {\bfseries 122} (2019) 141301}
  [\href{https://arxiv.org/abs/1902.03234}{{\ttfamily 1902.03234}}].

\bibitem{PandaX-II:2021nsg}
{\scshape PandaX-II} collaboration, C.~Cheng et~al., \emph{{Search for Light
  Dark Matter-Electron Scatterings in the PandaX-II Experiment}},
  \href{https://doi.org/10.1103/PhysRevLett.126.211803}{\emph{Phys. Rev. Lett.}
  {\bfseries 126} (2021) 211803}
  [\href{https://arxiv.org/abs/2101.07479}{{\ttfamily 2101.07479}}].

\bibitem{PandaX-4T:2021bab}
{\scshape PandaX-4T} collaboration, Y.~Meng et~al., \emph{{Dark Matter Search
  Results from the PandaX-4T Commissioning Run}},
  \href{https://arxiv.org/abs/2107.13438}{{\ttfamily 2107.13438}}.

\bibitem{Aad:2019vvi}
{\scshape ATLAS} collaboration, G.~Aad et~al., \emph{{Search for
  chargino-neutralino production with mass splittings near the electroweak
  scale in three-lepton final states in $\sqrt {s}$=13 TeV $pp$ collisions with
  the ATLAS detector}},
  \href{https://doi.org/10.1103/PhysRevD.101.072001}{\emph{Phys. Rev. D}
  {\bfseries 101} (2020) 072001}
  [\href{https://arxiv.org/abs/1912.08479}{{\ttfamily 1912.08479}}].

\bibitem{Aad:2019vnb}
{\scshape ATLAS} collaboration, G.~Aad et~al., \emph{{Search for electroweak
  production of charginos and sleptons decaying into final states with two
  leptons and missing transverse momentum in $\sqrt{s}=13$ TeV $pp$ collisions
  using the ATLAS detector}},
  \href{https://doi.org/10.1140/epjc/s10052-019-7594-6}{\emph{Eur. Phys. J. C}
  {\bfseries 80} (2020) 123}
  [\href{https://arxiv.org/abs/1908.08215}{{\ttfamily 1908.08215}}].

\bibitem{Aad:2019vvf}
{\scshape ATLAS} collaboration, G.~Aad et~al., \emph{{Search for direct
  production of electroweakinos in final states with one lepton, missing
  transverse momentum and a Higgs boson decaying into two $b$-jets in $pp$
  collisions at $\sqrt{s}=13$ TeV with the ATLAS detector}},
  \href{https://doi.org/10.1140/epjc/s10052-020-8050-3}{\emph{Eur. Phys. J. C}
  {\bfseries 80} (2020) 691}
  [\href{https://arxiv.org/abs/1909.09226}{{\ttfamily 1909.09226}}].

\bibitem{Sirunyan:2018nwe}
{\scshape CMS} collaboration, A.~M. Sirunyan et~al., \emph{{Search for
  supersymmetric partners of electrons and muons in proton-proton collisions at
  $\sqrt{s}=$ 13 TeV}},
  \href{https://doi.org/10.1016/j.physletb.2019.01.005}{\emph{Phys. Lett. B}
  {\bfseries 790} (2019) 140}
  [\href{https://arxiv.org/abs/1806.05264}{{\ttfamily 1806.05264}}].

\bibitem{Sirunyan:2018ubx}
{\scshape CMS} collaboration, A.~M. Sirunyan et~al., \emph{{Combined search for
  electroweak production of charginos and neutralinos in proton-proton
  collisions at $\sqrt{s} =$ 13 TeV}},
  \href{https://doi.org/10.1007/JHEP03(2018)160}{\emph{JHEP} {\bfseries 03}
  (2018) 160} [\href{https://arxiv.org/abs/1801.03957}{{\ttfamily
  1801.03957}}].

\bibitem{ATLAS:2019lng}
{\scshape ATLAS} collaboration, G.~Aad et~al., \emph{{Searches for electroweak
  production of supersymmetric particles with compressed mass spectra in
  $\sqrt{s}=$ 13 TeV $pp$ collisions with the ATLAS detector}},
  \href{https://doi.org/10.1103/PhysRevD.101.052005}{\emph{Phys. Rev. D}
  {\bfseries 101} (2020) 052005}
  [\href{https://arxiv.org/abs/1911.12606}{{\ttfamily 1911.12606}}].

\bibitem{ATLAS:2021moa}
{\scshape ATLAS} collaboration, G.~Aad et~al., \emph{{Search for
  chargino--neutralino pair production in final states with three leptons and
  missing transverse momentum in $\sqrt{s} = 13$ TeV $pp$ collisions with the
  ATLAS detector}},  \href{https://arxiv.org/abs/2106.01676}{{\ttfamily
  2106.01676}}.

\bibitem{CMS:2020bfa}
{\scshape CMS} collaboration, A.~M. Sirunyan et~al., \emph{{Search for
  supersymmetry in final states with two oppositely charged same-flavor leptons
  and missing transverse momentum in proton-proton collisions at $\sqrt{s} =$
  13 TeV}}, \href{https://doi.org/10.1007/JHEP04(2021)123}{\emph{JHEP}
  {\bfseries 04} (2021) 123}
  [\href{https://arxiv.org/abs/2012.08600}{{\ttfamily 2012.08600}}].

\bibitem{Farrar:1978xj}
G.~R. Farrar and P.~Fayet, \emph{{Phenomenology of the Production, Decay, and
  Detection of New Hadronic States Associated with Supersymmetry}},
  \href{https://doi.org/10.1016/0370-2693(78)90858-4}{\emph{Phys. Lett. B}
  {\bfseries 76} (1978) 575}.

\bibitem{Gunion:1984yn}
J.~F. Gunion and H.~E. Haber, \emph{{Higgs Bosons in Supersymmetric Models.
  1.}}, \href{https://doi.org/10.1016/0550-3213(86)90340-8}{\emph{Nucl. Phys.
  B} {\bfseries 272} (1986) 1}.

\bibitem{Djouadi:2005gj}
A.~Djouadi, \emph{{The Anatomy of electro-weak symmetry breaking. II. The Higgs
  bosons in the minimal supersymmetric model}},
  \href{https://doi.org/10.1016/j.physrep.2007.10.005}{\emph{Phys. Rept.}
  {\bfseries 459} (2008) 1}
  [\href{https://arxiv.org/abs/hep-ph/0503173}{{\ttfamily hep-ph/0503173}}].

\bibitem{Planck:2018vyg}
{\scshape Planck} collaboration, N.~Aghanim et~al., \emph{{Planck 2018 results.
  VI. Cosmological parameters}},
  \href{https://doi.org/10.1051/0004-6361/201833910}{\emph{Astron. Astrophys.}
  {\bfseries 641} (2020) A6}
  [\href{https://arxiv.org/abs/1807.06209}{{\ttfamily 1807.06209}}].

\bibitem{Bagnaschi:2017tru}
E.~Bagnaschi et~al., \emph{{Likelihood Analysis of the pMSSM11 in Light of LHC
  13-TeV Data}},
  \href{https://doi.org/10.1140/epjc/s10052-018-5697-0}{\emph{Eur. Phys. J. C}
  {\bfseries 78} (2018) 256}
  [\href{https://arxiv.org/abs/1710.11091}{{\ttfamily 1710.11091}}].

\bibitem{Baer:2012uy}
H.~Baer, V.~Barger, P.~Huang and X.~Tata, \emph{{Natural Supersymmetry: LHC,
  dark matter and ILC searches}},
  \href{https://doi.org/10.1007/JHEP05(2012)109}{\emph{JHEP} {\bfseries 05}
  (2012) 109} [\href{https://arxiv.org/abs/1203.5539}{{\ttfamily 1203.5539}}].

\bibitem{CMS:2019ybf}
{\scshape CMS} collaboration, A.~M. Sirunyan et~al., \emph{{Searches for
  physics beyond the standard model with the $M_\mathrm{T2}$ variable in
  hadronic final states with and without disappearing tracks in proton-proton
  collisions at $\sqrt{s}=$ 13 TeV}},
  \href{https://doi.org/10.1140/epjc/s10052-019-7493-x}{\emph{Eur. Phys. J. C}
  {\bfseries 80} (2020) 3} [\href{https://arxiv.org/abs/1909.03460}{{\ttfamily
  1909.03460}}].

\bibitem{Cao:2018rix}
J.~Cao, Y.~He, L.~Shang, Y.~Zhang and P.~Zhu, \emph{{Current status of a
  natural NMSSM in light of LHC 13 TeV data and XENON-1T results}},
  \href{https://doi.org/10.1103/PhysRevD.99.075020}{\emph{Phys. Rev.}
  {\bfseries D99} (2019) 075020}
  [\href{https://arxiv.org/abs/1810.09143}{{\ttfamily 1810.09143}}].

\bibitem{ATLAS:2021upq}
{\scshape ATLAS} collaboration, G.~Aad et~al., \emph{{Search for charged Higgs
  bosons decaying into a top quark and a bottom quark at $ \sqrt{\mathrm{s}} $
  = 13 TeV with the ATLAS detector}},
  \href{https://doi.org/10.1007/JHEP06(2021)145}{\emph{JHEP} {\bfseries 06}
  (2021) 145} [\href{https://arxiv.org/abs/2102.10076}{{\ttfamily
  2102.10076}}].

\bibitem{Ellwanger:2009dp}
U.~Ellwanger, C.~Hugonie and A.~M. Teixeira, \emph{{The Next-to-Minimal
  Supersymmetric Standard Model}},
  \href{https://doi.org/10.1016/j.physrep.2010.07.001}{\emph{Phys. Rept.}
  {\bfseries 496} (2010) 1} [\href{https://arxiv.org/abs/0910.1785}{{\ttfamily
  0910.1785}}].

\bibitem{Maniatis:2009re}
M.~Maniatis, \emph{{The Next-to-Minimal Supersymmetric extension of the
  Standard Model reviewed}},
  \href{https://doi.org/10.1142/S0217751X10049827}{\emph{Int. J. Mod. Phys. A}
  {\bfseries 25} (2010) 3505}
  [\href{https://arxiv.org/abs/0906.0777}{{\ttfamily 0906.0777}}].

\bibitem{Cao:2016nix}
J.~Cao, Y.~He, L.~Shang, W.~Su and Y.~Zhang, \emph{{Natural NMSSM after LHC Run
  I and the Higgsino dominated dark matter scenario}},
  \href{https://doi.org/10.1007/JHEP08(2016)037}{\emph{JHEP} {\bfseries 08}
  (2016) 037} [\href{https://arxiv.org/abs/1606.04416}{{\ttfamily
  1606.04416}}].

\bibitem{Ellwanger:2016sur}
U.~Ellwanger, \emph{{Present Status and Future Tests of the Higgsino-Singlino
  Sector in the NMSSM}},
  \href{https://doi.org/10.1007/JHEP02(2017)051}{\emph{JHEP} {\bfseries 02}
  (2017) 051} [\href{https://arxiv.org/abs/1612.06574}{{\ttfamily
  1612.06574}}].

\bibitem{Xiang:2016ndq}
Q.-F. Xiang, X.-J. Bi, P.-F. Yin and Z.-H. Yu, \emph{{Searching for
  Singlino-Higgsino Dark Matter in the NMSSM}},
  \href{https://doi.org/10.1103/PhysRevD.94.055031}{\emph{Phys. Rev. D}
  {\bfseries 94} (2016) 055031}
  [\href{https://arxiv.org/abs/1606.02149}{{\ttfamily 1606.02149}}].

\bibitem{Baum:2017enm}
S.~Baum, M.~Carena, N.~R. Shah and C.~E. Wagner, \emph{{Higgs portals for
  thermal Dark Matter. EFT perspectives and the NMSSM}},
  \href{https://doi.org/10.1007/JHEP04(2018)069}{\emph{JHEP} {\bfseries 04}
  (2018) 069} [\href{https://arxiv.org/abs/1712.09873}{{\ttfamily
  1712.09873}}].

\bibitem{Ellwanger:2018zxt}
U.~Ellwanger and C.~Hugonie, \emph{{The higgsino\textendash{}singlino sector of
  the NMSSM: combined constraints from dark matter and the LHC}},
  \href{https://doi.org/10.1140/epjc/s10052-018-6204-3}{\emph{Eur. Phys. J. C}
  {\bfseries 78} (2018) 735}
  [\href{https://arxiv.org/abs/1806.09478}{{\ttfamily 1806.09478}}].

\bibitem{Domingo:2018ykx}
F.~Domingo, J.~S. Kim, V.~M. Lozano, P.~Martin-Ramiro and R.~Ruiz~de Austri,
  \emph{{Confronting the neutralino and chargino sector of the NMSSM with the
  multilepton searches at the LHC}},
  \href{https://doi.org/10.1103/PhysRevD.101.075010}{\emph{Phys. Rev. D}
  {\bfseries 101} (2020) 075010}
  [\href{https://arxiv.org/abs/1812.05186}{{\ttfamily 1812.05186}}].

\bibitem{Baum:2019uzg}
S.~Baum, N.~R. Shah and K.~Freese, \emph{{The NMSSM is within Reach of the LHC:
  Mass Correlations \textbackslash{}\& Decay Signatures}},
  \href{https://doi.org/10.1007/JHEP04(2019)011}{\emph{JHEP} {\bfseries 04}
  (2019) 011} [\href{https://arxiv.org/abs/1901.02332}{{\ttfamily
  1901.02332}}].

\bibitem{vanBeekveld:2019tqp}
M.~van Beekveld, S.~Caron and R.~Ruiz~de Austri, \emph{{The current status of
  fine-tuning in supersymmetry}},
  \href{https://doi.org/10.1007/JHEP01(2020)147}{\emph{JHEP} {\bfseries 01}
  (2020) 147} [\href{https://arxiv.org/abs/1906.10706}{{\ttfamily
  1906.10706}}].

\bibitem{Abdallah:2019znp}
W.~Abdallah, A.~Chatterjee and A.~Datta, \emph{{Revisiting singlino dark matter
  of the natural $Z_3$-symmetric NMSSM in the light of LHC}},
  \href{https://doi.org/10.1007/JHEP09(2019)095}{\emph{JHEP} {\bfseries 09}
  (2019) 095} [\href{https://arxiv.org/abs/1907.06270}{{\ttfamily
  1907.06270}}].

\bibitem{Cao:2019qng}
J.~Cao, L.~Meng, Y.~Yue, H.~Zhou and P.~Zhu, \emph{Suppressing the scattering
  of wimp dark matter and nucleons in supersymmetric theories},
  \href{https://doi.org/10.1103/PhysRevD.101.075003}{\emph{Phys. Rev. D}
  {\bfseries 101} (2020) 075003}.

\bibitem{Guchait:2020wqn}
M.~Guchait and A.~Roy, \emph{{Light Singlino Dark Matter at the LHC}},
  \href{https://doi.org/10.1103/PhysRevD.102.075023}{\emph{Phys. Rev. D}
  {\bfseries 102} (2020) 075023}
  [\href{https://arxiv.org/abs/2005.05190}{{\ttfamily 2005.05190}}].

\bibitem{Abdallah:2020yag}
W.~Abdallah, A.~Datta and S.~Roy, \emph{{A relatively light, highly bino-like
  dark matter in the Z$_{3}$-symmetric NMSSM and recent LHC searches}},
  \href{https://doi.org/10.1007/JHEP04(2021)122}{\emph{JHEP} {\bfseries 04}
  (2021) 122} [\href{https://arxiv.org/abs/2012.04026}{{\ttfamily
  2012.04026}}].

\bibitem{Zhou:2021pit}
H.~Zhou, J.~Cao, J.~Lian and D.~Zhang, \emph{{Singlino-dominated dark matter in
  Z3-symmetric NMSSM}},
  \href{https://doi.org/10.1103/PhysRevD.104.015017}{\emph{Phys. Rev. D}
  {\bfseries 104} (2021) 015017}
  [\href{https://arxiv.org/abs/2102.05309}{{\ttfamily 2102.05309}}].

\bibitem{Cao:2021ljw}
J.~Cao, D.~Li, J.~Lian, Y.~Yue and H.~Zhou, \emph{{Singlino-dominated dark
  matter in general NMSSM}},
  \href{https://doi.org/10.1007/JHEP06(2021)176}{\emph{JHEP} {\bfseries 06}
  (2021) 176} [\href{https://arxiv.org/abs/2102.05317}{{\ttfamily
  2102.05317}}].

\bibitem{Ellwanger:1983mg}
U.~Ellwanger, \emph{{Nonrenormalizable interactions from supergravity, quantum
  corrections and effecive low-energy theories}},
  \href{https://doi.org/10.1016/0370-2693(83)90557-9}{\emph{Phys. Lett. B}
  {\bfseries 133} (1983) 187}.

\bibitem{Abel:1996cr}
S.~A. Abel, \emph{{Destabilizing divergences in the NMSSM}},
  \href{https://doi.org/10.1016/S0550-3213(96)00470-1}{\emph{Nucl. Phys. B}
  {\bfseries 480} (1996) 55}
  [\href{https://arxiv.org/abs/hep-ph/9609323}{{\ttfamily hep-ph/9609323}}].

\bibitem{Kolda:1998rm}
C.~F. Kolda, S.~Pokorski and N.~Polonsky, \emph{{Stabilized singlets in
  supergravity as a source of the mu - parameter}},
  \href{https://doi.org/10.1103/PhysRevLett.80.5263}{\emph{Phys. Rev. Lett.}
  {\bfseries 80} (1998) 5263}
  [\href{https://arxiv.org/abs/hep-ph/9803310}{{\ttfamily hep-ph/9803310}}].

\bibitem{Panagiotakopoulos:1998yw}
C.~Panagiotakopoulos and K.~Tamvakis, \emph{{Stabilized NMSSM without domain
  walls}}, \href{https://doi.org/10.1016/S0370-2693(98)01493-2}{\emph{Phys.
  Lett. B} {\bfseries 446} (1999) 224}
  [\href{https://arxiv.org/abs/hep-ph/9809475}{{\ttfamily hep-ph/9809475}}].

\bibitem{Ross:2011xv}
G.~G. Ross and K.~Schmidt-Hoberg, \emph{{The Fine-Tuning of the Generalised
  NMSSM}}, \href{https://doi.org/10.1016/j.nuclphysb.2012.05.007}{\emph{Nucl.
  Phys. B} {\bfseries 862} (2012) 710}
  [\href{https://arxiv.org/abs/1108.1284}{{\ttfamily 1108.1284}}].

\bibitem{Lee:2010gv}
H.~M. Lee, S.~Raby, M.~Ratz, G.~G. Ross, R.~Schieren, K.~Schmidt-Hoberg et~al.,
  \emph{{A unique $\mathbb{Z}_4^R$ symmetry for the MSSM}},
  \href{https://doi.org/10.1016/j.physletb.2010.10.038}{\emph{Phys. Lett. B}
  {\bfseries 694} (2011) 491}
  [\href{https://arxiv.org/abs/1009.0905}{{\ttfamily 1009.0905}}].

\bibitem{Lee:2011dya}
H.~M. Lee, S.~Raby, M.~Ratz, G.~G. Ross, R.~Schieren, K.~Schmidt-Hoberg et~al.,
  \emph{{Discrete R symmetries for the MSSM and its singlet extensions}},
  \href{https://doi.org/10.1016/j.nuclphysb.2011.04.009}{\emph{Nucl. Phys. B}
  {\bfseries 850} (2011) 1} [\href{https://arxiv.org/abs/1102.3595}{{\ttfamily
  1102.3595}}].

\bibitem{Ross:2012nr}
G.~G. Ross, K.~Schmidt-Hoberg and F.~Staub, \emph{{The Generalised NMSSM at One
  Loop: Fine Tuning and Phenomenology}},
  \href{https://doi.org/10.1007/JHEP08(2012)074}{\emph{JHEP} {\bfseries 08}
  (2012) 074} [\href{https://arxiv.org/abs/1205.1509}{{\ttfamily 1205.1509}}].

\bibitem{Cao:2012fz}
J.-J. Cao, Z.-X. Heng, J.~M. Yang, Y.-M. Zhang and J.-Y. Zhu, \emph{{A SM-like
  Higgs near 125 GeV in low energy SUSY: a comparative study for MSSM and
  NMSSM}}, \href{https://doi.org/10.1007/JHEP03(2012)086}{\emph{JHEP}
  {\bfseries 03} (2012) 086} [\href{https://arxiv.org/abs/1202.5821}{{\ttfamily
  1202.5821}}].

\bibitem{Ferrara:2010yw}
S.~Ferrara, R.~Kallosh, A.~Linde, A.~Marrani and A.~Van~Proeyen, \emph{{Jordan
  Frame Supergravity and Inflation in NMSSM}},
  \href{https://doi.org/10.1103/PhysRevD.82.045003}{\emph{Phys. Rev. D}
  {\bfseries 82} (2010) 045003}
  [\href{https://arxiv.org/abs/1004.0712}{{\ttfamily 1004.0712}}].

\bibitem{Ferrara:2010in}
S.~Ferrara, R.~Kallosh, A.~Linde, A.~Marrani and A.~Van~Proeyen,
  \emph{{Superconformal Symmetry, NMSSM, and Inflation}},
  \href{https://doi.org/10.1103/PhysRevD.83.025008}{\emph{Phys. Rev. D}
  {\bfseries 83} (2011) 025008}
  [\href{https://arxiv.org/abs/1008.2942}{{\ttfamily 1008.2942}}].

\bibitem{Einhorn:2009bh}
M.~B. Einhorn and D.~R.~T. Jones, \emph{{Inflation with Non-minimal
  Gravitational Couplings in Supergravity}},
  \href{https://doi.org/10.1007/JHEP03(2010)026}{\emph{JHEP} {\bfseries 03}
  (2010) 026} [\href{https://arxiv.org/abs/0912.2718}{{\ttfamily 0912.2718}}].

\bibitem{Hollik:2018yek}
W.~G. Hollik, S.~Liebler, G.~Moortgat-Pick, S.~Paßehr and G.~Weiglein,
  \emph{{Phenomenology of the inflation-inspired NMSSM at the electroweak
  scale}}, \href{https://doi.org/10.1140/epjc/s10052-019-6561-6}{\emph{Eur.
  Phys. J. C} {\bfseries 79} (2019) 75}
  [\href{https://arxiv.org/abs/1809.07371}{{\ttfamily 1809.07371}}].

\bibitem{Hollik:2020plc}
W.~G. Hollik, C.~Li, G.~Moortgat-Pick and S.~Paasch, \emph{{Phenomenology of a
  Supersymmetric Model Inspired by Inflation}},
  \href{https://doi.org/10.1140/epjc/s10052-021-08869-4}{\emph{Eur. Phys. J. C}
  {\bfseries 81} (2021) 141}
  [\href{https://arxiv.org/abs/2004.14852}{{\ttfamily 2004.14852}}].

\bibitem{Cheung:2014lqa}
C.~Cheung, M.~Papucci, D.~Sanford, N.~R. Shah and K.~M. Zurek, \emph{{NMSSM
  Interpretation of the Galactic Center Excess}},
  \href{https://doi.org/10.1103/PhysRevD.90.075011}{\emph{Phys. Rev. D}
  {\bfseries 90} (2014) 075011}
  [\href{https://arxiv.org/abs/1406.6372}{{\ttfamily 1406.6372}}].

\bibitem{Badziak:2015exr}
M.~Badziak, M.~Olechowski and P.~Szczerbiak, \emph{{Blind spots for neutralino
  dark matter in the NMSSM}},
  \href{https://doi.org/10.1007/JHEP03(2016)179}{\emph{JHEP} {\bfseries 03}
  (2016) 179} [\href{https://arxiv.org/abs/1512.02472}{{\ttfamily
  1512.02472}}].

\bibitem{Badziak_2017}
M.~Badziak, M.~Olechowski and P.~Szczerbiak, \emph{Spin-dependent constraints
  on blind spots for thermal singlino-higgsino dark matter with(out) light
  singlets}, \href{https://doi.org/10.1007/jhep07(2017)050}{\emph{Journal of
  High Energy Physics} {\bfseries 2017} (2017) }.

\bibitem{Pospelov:2007mp}
M.~Pospelov, A.~Ritz and M.~B. Voloshin, \emph{{Secluded WIMP Dark Matter}},
  \href{https://doi.org/10.1016/j.physletb.2008.02.052}{\emph{Phys. Lett. B}
  {\bfseries 662} (2008) 53} [\href{https://arxiv.org/abs/0711.4866}{{\ttfamily
  0711.4866}}].

\bibitem{Feroz:2008xx}
F.~Feroz, M.~P. Hobson and M.~Bridges, \emph{{MultiNest: an efficient and
  robust Bayesian inference tool for cosmology and particle physics}},
  \href{https://doi.org/10.1111/j.1365-2966.2009.14548.x}{\emph{Mon. Not. Roy.
  Astron. Soc.} {\bfseries 398} (2009) 1601}
  [\href{https://arxiv.org/abs/0809.3437}{{\ttfamily 0809.3437}}].

\bibitem{Staub:2008uz}
F.~Staub, \emph{{SARAH}},  \href{https://arxiv.org/abs/0806.0538}{{\ttfamily
  0806.0538}}.

\bibitem{Staub:2012pb}
F.~Staub, \emph{{SARAH 3.2: Dirac Gauginos, UFO output, and more}},
  \href{https://doi.org/10.1016/j.cpc.2013.02.019}{\emph{Comput. Phys. Commun.}
  {\bfseries 184} (2013) 1792}
  [\href{https://arxiv.org/abs/1207.0906}{{\ttfamily 1207.0906}}].

\bibitem{Staub:2013tta}
F.~Staub, \emph{{SARAH 4 : A tool for (not only SUSY) model builders}},
  \href{https://doi.org/10.1016/j.cpc.2014.02.018}{\emph{Comput. Phys. Commun.}
  {\bfseries 185} (2014) 1773}
  [\href{https://arxiv.org/abs/1309.7223}{{\ttfamily 1309.7223}}].

\bibitem{Staub:2015kfa}
F.~Staub, \emph{{Exploring new models in all detail with SARAH}},
  \href{https://doi.org/10.1155/2015/840780}{\emph{Adv. High Energy Phys.}
  {\bfseries 2015} (2015) 840780}
  [\href{https://arxiv.org/abs/1503.04200}{{\ttfamily 1503.04200}}].

\bibitem{Porod:2003um}
W.~Porod, \emph{{SPheno, a program for calculating supersymmetric spectra, SUSY
  particle decays and SUSY particle production at e+ e- colliders}},
  \href{https://doi.org/10.1016/S0010-4655(03)00222-4}{\emph{Comput. Phys.
  Commun.} {\bfseries 153} (2003) 275}
  [\href{https://arxiv.org/abs/hep-ph/0301101}{{\ttfamily hep-ph/0301101}}].

\bibitem{Porod:2011nf}
W.~Porod and F.~Staub, \emph{{SPheno 3.1: Extensions including flavour,
  CP-phases and models beyond the MSSM}},
  \href{https://doi.org/10.1016/j.cpc.2012.05.021}{\emph{Comput. Phys. Commun.}
  {\bfseries 183} (2012) 2458}
  [\href{https://arxiv.org/abs/1104.1573}{{\ttfamily 1104.1573}}].

\bibitem{Porod:2014xia}
W.~Porod, F.~Staub and A.~Vicente, \emph{{A Flavor Kit for BSM models}},
  \href{https://doi.org/10.1140/epjc/s10052-014-2992-2}{\emph{Eur. Phys. J. C}
  {\bfseries 74} (2014) 2992}
  [\href{https://arxiv.org/abs/1405.1434}{{\ttfamily 1405.1434}}].

\bibitem{Belanger:2001fz}
G.~Belanger, F.~Boudjema, A.~Pukhov and A.~Semenov, \emph{{MicrOMEGAs: A
  Program for calculating the relic density in the MSSM}},
  \href{https://doi.org/10.1016/S0010-4655(02)00596-9}{\emph{Comput. Phys.
  Commun.} {\bfseries 149} (2002) 103}
  [\href{https://arxiv.org/abs/hep-ph/0112278}{{\ttfamily hep-ph/0112278}}].

\bibitem{Belanger:2005kh}
G.~Belanger, F.~Boudjema, C.~Hugonie, A.~Pukhov and A.~Semenov, \emph{{Relic
  density of dark matter in the NMSSM}},
  \href{https://doi.org/10.1088/1475-7516/2005/09/001}{\emph{JCAP} {\bfseries
  09} (2005) 001} [\href{https://arxiv.org/abs/hep-ph/0505142}{{\ttfamily
  hep-ph/0505142}}].

\bibitem{Belanger:2006is}
G.~Belanger, F.~Boudjema, A.~Pukhov and A.~Semenov, \emph{{MicrOMEGAs 2.0: A
  Program to calculate the relic density of dark matter in a generic model}},
  \href{https://doi.org/10.1016/j.cpc.2006.11.008}{\emph{Comput. Phys. Commun.}
  {\bfseries 176} (2007) 367}
  [\href{https://arxiv.org/abs/hep-ph/0607059}{{\ttfamily hep-ph/0607059}}].

\bibitem{Belanger:2010pz}
G.~Belanger, F.~Boudjema, A.~Pukhov and A.~Semenov, \emph{{micrOMEGAs: A Tool
  for dark matter studies}},
  \href{https://doi.org/10.1393/ncc/i2010-10591-3}{\emph{Nuovo Cim. C}
  {\bfseries 033N2} (2010) 111}
  [\href{https://arxiv.org/abs/1005.4133}{{\ttfamily 1005.4133}}].

\bibitem{Belanger:2013oya}
G.~Belanger, F.~Boudjema, A.~Pukhov and A.~Semenov, \emph{{micrOMEGAs$\_$3: A
  program for calculating dark matter observables}},
  \href{https://doi.org/10.1016/j.cpc.2013.10.016}{\emph{Comput. Phys. Commun.}
  {\bfseries 185} (2014) 960}
  [\href{https://arxiv.org/abs/1305.0237}{{\ttfamily 1305.0237}}].

\bibitem{Barducci:2016pcb}
D.~Barducci, G.~Belanger, J.~Bernon, F.~Boudjema, J.~Da~Silva, S.~Kraml et~al.,
  \emph{{Collider limits on new physics within micrOMEGAs$\_$4.3}},
  \href{https://doi.org/10.1016/j.cpc.2017.08.028}{\emph{Comput. Phys. Commun.}
  {\bfseries 222} (2018) 327}
  [\href{https://arxiv.org/abs/1606.03834}{{\ttfamily 1606.03834}}].

\bibitem{Aghanim:2018eyx}
{\scshape Planck} collaboration, N.~Aghanim et~al., \emph{{Planck 2018 results.
  VI. Cosmological parameters}},
  \href{https://doi.org/10.1051/0004-6361/201833910}{\emph{Astron. Astrophys.}
  {\bfseries 641} (2020) A6}
  [\href{https://arxiv.org/abs/1807.06209}{{\ttfamily 1807.06209}}].

\bibitem{Aprile:2018dbl}
{\scshape XENON} collaboration, E.~Aprile et~al., \emph{{Dark Matter Search
  Results from a One Ton-Year Exposure of XENON1T}},
  \href{https://doi.org/10.1103/PhysRevLett.121.111302}{\emph{Phys. Rev. Lett.}
  {\bfseries 121} (2018) 111302}
  [\href{https://arxiv.org/abs/1805.12562}{{\ttfamily 1805.12562}}].

\bibitem{Aprile:2019dbj}
{\scshape XENON} collaboration, E.~Aprile et~al., \emph{{Constraining the
  spin-dependent WIMP-nucleon cross sections with XENON1T}},
  \href{https://doi.org/10.1103/PhysRevLett.122.141301}{\emph{Phys. Rev. Lett.}
  {\bfseries 122} (2019) 141301}
  [\href{https://arxiv.org/abs/1902.03234}{{\ttfamily 1902.03234}}].

\bibitem{Ackermann:2015zua}
{\scshape Fermi-LAT} collaboration, M.~Ackermann et~al., \emph{{Searching for
  Dark Matter Annihilation from Milky Way Dwarf Spheroidal Galaxies with Six
  Years of Fermi Large Area Telescope Data}},
  \href{https://doi.org/10.1103/PhysRevLett.115.231301}{\emph{Phys. Rev. Lett.}
  {\bfseries 115} (2015) 231301}
  [\href{https://arxiv.org/abs/1503.02641}{{\ttfamily 1503.02641}}].

\bibitem{Carpenter:2016thc}
L.~M. Carpenter, R.~Colburn, J.~Goodman and T.~Linden, \emph{{Indirect
  Detection Constraints on s and t Channel Simplified Models of Dark Matter}},
  \href{https://doi.org/10.1103/PhysRevD.94.055027}{\emph{Phys. Rev. D}
  {\bfseries 94} (2016) 055027}
  [\href{https://arxiv.org/abs/1606.04138}{{\ttfamily 1606.04138}}].

\bibitem{Bechtle:2014ewa}
P.~Bechtle, S.~Heinemeyer, O.~Stål, T.~Stefaniak and G.~Weiglein,
  \emph{{Probing the Standard Model with Higgs signal rates from the Tevatron,
  the LHC and a future ILC}},
  \href{https://doi.org/10.1007/JHEP11(2014)039}{\emph{JHEP} {\bfseries 11}
  (2014) 039} [\href{https://arxiv.org/abs/1403.1582}{{\ttfamily 1403.1582}}].

\bibitem{Bechtle:2015pma}
P.~Bechtle, S.~Heinemeyer, O.~Stal, T.~Stefaniak and G.~Weiglein,
  \emph{{Applying Exclusion Likelihoods from LHC Searches to Extended Higgs
  Sectors}}, \href{https://doi.org/10.1140/epjc/s10052-015-3650-z}{\emph{Eur.
  Phys. J. C} {\bfseries 75} (2015) 421}
  [\href{https://arxiv.org/abs/1507.06706}{{\ttfamily 1507.06706}}].

\bibitem{PhysRevD.98.030001}
{\scshape Particle Data Group} collaboration, M.~Tanabashi, K.~Hagiwara, Hikasa
  et~al., \emph{Review of particle physics},
  \href{https://doi.org/10.1103/PhysRevD.98.030001}{\emph{Phys. Rev. D}
  {\bfseries 98} (2018) 030001}.

\bibitem{Khosa:2020zar}
C.~K. Khosa, S.~Kraml, A.~Lessa, P.~Neuhuber and W.~Waltenberger,
  \emph{{SModelS database update v1.2.3}},
  \href{https://arxiv.org/abs/2005.00555}{{\ttfamily 2005.00555}}.

\bibitem{Drees:2013wra}
M.~Drees, H.~Dreiner, D.~Schmeier, J.~Tattersall and J.~S. Kim,
  \emph{{CheckMATE: Confronting your Favourite New Physics Model with LHC
  Data}}, \href{https://doi.org/10.1016/j.cpc.2014.10.018}{\emph{Comput. Phys.
  Commun.} {\bfseries 187} (2015) 227}
  [\href{https://arxiv.org/abs/1312.2591}{{\ttfamily 1312.2591}}].

\bibitem{Dercks:2016npn}
D.~Dercks, N.~Desai, J.~S. Kim, K.~Rolbiecki, J.~Tattersall and T.~Weber,
  \emph{{CheckMATE 2: From the model to the limit}},
  \href{https://doi.org/10.1016/j.cpc.2017.08.021}{\emph{Comput. Phys. Commun.}
  {\bfseries 221} (2017) 383}
  [\href{https://arxiv.org/abs/1611.09856}{{\ttfamily 1611.09856}}].

\bibitem{Kim:2015wza}
J.~S. Kim, D.~Schmeier, J.~Tattersall and K.~Rolbiecki, \emph{{A framework to
  create customised LHC analyses within CheckMATE}},
  \href{https://doi.org/10.1016/j.cpc.2015.06.002}{\emph{Comput. Phys. Commun.}
  {\bfseries 196} (2015) 535}
  [\href{https://arxiv.org/abs/1503.01123}{{\ttfamily 1503.01123}}].

\bibitem{Camargo-Molina:2013qva}
J.~E. Camargo-Molina, B.~O'Leary, W.~Porod and F.~Staub,
  \emph{{$\mathbf{Vevacious}$: A Tool For Finding The Global Minima Of One-Loop
  Effective Potentials With Many Scalars}},
  \href{https://doi.org/10.1140/epjc/s10052-013-2588-2}{\emph{Eur. Phys. J. C}
  {\bfseries 73} (2013) 2588}
  [\href{https://arxiv.org/abs/1307.1477}{{\ttfamily 1307.1477}}].

\bibitem{Camargo-Molina:2014pwa}
J.~E. Camargo-Molina, B.~Garbrecht, B.~O'Leary, W.~Porod and F.~Staub,
  \emph{{Constraining the Natural MSSM through tunneling to color-breaking
  vacua at zero and non-zero temperature}},
  \href{https://doi.org/10.1016/j.physletb.2014.08.036}{\emph{Phys. Lett. B}
  {\bfseries 737} (2014) 156}
  [\href{https://arxiv.org/abs/1405.7376}{{\ttfamily 1405.7376}}].

\bibitem{Fowlie:2016hew}
A.~Fowlie and M.~H. Bardsley, \emph{{Superplot: a graphical interface for
  plotting and analysing MultiNest output}},
  \href{https://doi.org/10.1140/epjp/i2016-16391-0}{\emph{Eur. Phys. J. Plus}
  {\bfseries 131} (2016) 391}
  [\href{https://arxiv.org/abs/1603.00555}{{\ttfamily 1603.00555}}].

\bibitem{Cao:2018iyk}
J.~Cao, J.~Li, Y.~Pan, L.~Shang, Y.~Yue and D.~Zhang, \emph{{Bayesian analysis
  of sneutrino dark matter in the NMSSM with a type-I seesaw mechanism}},
  \href{https://doi.org/10.1103/PhysRevD.99.115033}{\emph{Phys. Rev. D}
  {\bfseries 99} (2019) 115033}
  [\href{https://arxiv.org/abs/1807.03762}{{\ttfamily 1807.03762}}].

\bibitem{hintze1998violin}
J.~L. Hintze and R.~D. Nelson, \emph{Violin plots: a box plot-density trace
  synergism}, {\emph{The American Statistician} {\bfseries 52} (1998) 181}.

\bibitem{Beenakker:1996ed}
W.~Beenakker, R.~Hopker and M.~Spira, \emph{{PROSPINO: A Program for the
  production of supersymmetric particles in next-to-leading order QCD}},
  \href{https://arxiv.org/abs/hep-ph/9611232}{{\ttfamily hep-ph/9611232}}.

\bibitem{Alwall:2011uj}
J.~Alwall, M.~Herquet, F.~Maltoni, O.~Mattelaer and T.~Stelzer, \emph{{MadGraph
  5 : Going Beyond}},
  \href{https://doi.org/10.1007/JHEP06(2011)128}{\emph{JHEP} {\bfseries 06}
  (2011) 128} [\href{https://arxiv.org/abs/1106.0522}{{\ttfamily 1106.0522}}].

\bibitem{Conte:2012fm}
E.~Conte, B.~Fuks and G.~Serret, \emph{{MadAnalysis 5, A User-Friendly
  Framework for Collider Phenomenology}},
  \href{https://doi.org/10.1016/j.cpc.2012.09.009}{\emph{Comput. Phys. Commun.}
  {\bfseries 184} (2013) 222}
  [\href{https://arxiv.org/abs/1206.1599}{{\ttfamily 1206.1599}}].

\bibitem{Sjostrand:2014zea}
T.~Sj\"ostrand, S.~Ask, J.~R. Christiansen, R.~Corke, N.~Desai, P.~Ilten
  et~al., \emph{{An introduction to PYTHIA 8.2}},
  \href{https://doi.org/10.1016/j.cpc.2015.01.024}{\emph{Comput. Phys. Commun.}
  {\bfseries 191} (2015) 159}
  [\href{https://arxiv.org/abs/1410.3012}{{\ttfamily 1410.3012}}].

\bibitem{deFavereau:2013fsa}
{\scshape DELPHES 3} collaboration, J.~de~Favereau, C.~Delaere, P.~Demin,
  A.~Giammanco, V.~Lema\^\i{}tre, A.~Mertens et~al., \emph{{DELPHES 3, A
  modular framework for fast simulation of a generic collider experiment}},
  \href{https://doi.org/10.1007/JHEP02(2014)057}{\emph{JHEP} {\bfseries 02}
  (2014) 057} [\href{https://arxiv.org/abs/1307.6346}{{\ttfamily 1307.6346}}].

\bibitem{ATLAS:2018ojr}
{\scshape ATLAS} collaboration, M.~Aaboud et~al., \emph{{Search for electroweak
  production of supersymmetric particles in final states with two or three
  leptons at $\sqrt{s}=13\,$TeV with the ATLAS detector}},
  \href{https://doi.org/10.1140/epjc/s10052-018-6423-7}{\emph{Eur. Phys. J. C}
  {\bfseries 78} (2018) 995}
  [\href{https://arxiv.org/abs/1803.02762}{{\ttfamily 1803.02762}}].

\bibitem{ATLAS:2018nud}
{\scshape ATLAS} collaboration, M.~Aaboud et~al., \emph{{Search for photonic
  signatures of gauge-mediated supersymmetry in 13 TeV $pp$ collisions with the
  ATLAS detector}},
  \href{https://doi.org/10.1103/PhysRevD.97.092006}{\emph{Phys. Rev. D}
  {\bfseries 97} (2018) 092006}
  [\href{https://arxiv.org/abs/1802.03158}{{\ttfamily 1802.03158}}].

\bibitem{ATLAS:2017vat}
{\scshape ATLAS} collaboration, M.~Aaboud et~al., \emph{{Search for electroweak
  production of supersymmetric states in scenarios with compressed mass spectra
  at $\sqrt{s}=13$ TeV with the ATLAS detector}},
  \href{https://doi.org/10.1103/PhysRevD.97.052010}{\emph{Phys. Rev. D}
  {\bfseries 97} (2018) 052010}
  [\href{https://arxiv.org/abs/1712.08119}{{\ttfamily 1712.08119}}].

\bibitem{CMS:2017moi}
{\scshape CMS} collaboration, A.~M. Sirunyan et~al., \emph{{Search for
  electroweak production of charginos and neutralinos in multilepton final
  states in proton-proton collisions at $\sqrt{s}=$ 13 TeV}},
  \href{https://doi.org/10.1007/JHEP03(2018)166}{\emph{JHEP} {\bfseries 03}
  (2018) 166} [\href{https://arxiv.org/abs/1709.05406}{{\ttfamily
  1709.05406}}].

\bibitem{CMS:2018kag}
{\scshape CMS} collaboration, A.~M. Sirunyan et~al., \emph{{Search for new
  physics in events with two soft oppositely charged leptons and missing
  transverse momentum in proton-proton collisions at $\sqrt{s}=$ 13 TeV}},
  \href{https://doi.org/10.1016/j.physletb.2018.05.062}{\emph{Phys. Lett. B}
  {\bfseries 782} (2018) 440}
  [\href{https://arxiv.org/abs/1801.01846}{{\ttfamily 1801.01846}}].

\bibitem{CMS:2016zvj}
{\scshape CMS} collaboration, \emph{{Search for new physics in the compressed
  mass spectra scenario using events with two soft opposite-sign leptons and
  missing momentum energy at 13 TeV}},
  \href{https://arxiv.org/abs/CMS-PAS-SUS-16-025}{{\ttfamily
  CMS-PAS-SUS-16-025}}.

\bibitem{ATLAS:2016uwq}
{\scshape ATLAS} collaboration, \emph{{Search for supersymmetry with two and
  three leptons and missing transverse momentum in the final state at
  \textbackslash{}sqrt{s}=13 TeV with the ATLAS detector}},
  \href{https://arxiv.org/abs/ATLAS-CONF-2016-096}{{\ttfamily
  ATLAS-CONF-2016-096}}.

\bibitem{ATLAS:2017qwn}
{\scshape ATLAS} collaboration, M.~Aaboud et~al., \emph{{Search for the direct
  production of charginos and neutralinos in final states with tau leptons in
  $\sqrt{s} = $ 13 TeV $pp$ collisions with the ATLAS detector}},
  \href{https://doi.org/10.1140/epjc/s10052-018-5583-9}{\emph{Eur. Phys. J. C}
  {\bfseries 78} (2018) 154}
  [\href{https://arxiv.org/abs/1708.07875}{{\ttfamily 1708.07875}}].

\bibitem{ATLAS:2019gti}
{\scshape ATLAS} collaboration, G.~Aad et~al., \emph{{Search for direct stau
  production in events with two hadronic $\tau$-leptons in $\sqrt{s} = 13$ TeV
  $pp$ collisions with the ATLAS detector}},
  \href{https://doi.org/10.1103/PhysRevD.101.032009}{\emph{Phys. Rev. D}
  {\bfseries 101} (2020) 032009}
  [\href{https://arxiv.org/abs/1911.06660}{{\ttfamily 1911.06660}}].

\bibitem{Pierce:2013rda}
A.~Pierce, N.~R. Shah and K.~Freese, \emph{{Neutralino Dark Matter with Light
  Staus}},  \href{https://arxiv.org/abs/1309.7351}{{\ttfamily 1309.7351}}.

\end{thebibliography}\endgroup
\end{document}